\documentclass[twoside,12pt]{book}
\usepackage[sort&compress,numbers]{natbib}
\usepackage[top=3cm, bottom=2.8cm, outer=3.5cm, inner=3.5cm, marginparwidth=2.5cm, marginparsep=0.5cm]{geometry}
\usepackage{afterpage}
\usepackage{titlesec}
\usepackage{titling}
\usepackage{wrapfig}
\usepackage[font=small,labelfont=bf]{caption}
\usepackage{import}
\usepackage{ragged2e}
\usepackage{tikz}
\usepackage{changepage}
\usepackage{datetime}
\usepackage{anyfontsize}
\usepackage{textcomp}
\usepackage{xcolor}
\usepackage{textcomp}
\usepackage{tabularx}
\usepackage{tabulary}
\usepackage[most]{tcolorbox}
\tcbuselibrary{breakable}
\usepackage{graphicx}
\usepackage[export]{adjustbox}
\usepackage{fancyhdr,color}
\usepackage{marginnote}
\usepackage[colorlinks]{hyperref} 
\definecolor{blue}{HTML}{0046FF}
\hypersetup{
    citecolor = red,
    linkcolor = blue,
    urlcolor = blue
}
\usepackage[chapter]{algorithm}
\usepackage{algorithmicx}
\usepackage{algpseudocode}
\usepackage{algcompatible,amsmath}
\algnewcommand\INPUT{\item[\textbf{Input:}]}%
\algnewcommand\OUTPUT{\item[\textbf{Output:}]}%

\newcommand{\mono}[1]{{\fontfamily{cmtt}\selectfont #1}}
\let\OLDitemize\itemize
\renewcommand\itemize{\OLDitemize\addtolength{\itemsep}{-20pt}}
\usepackage[hang,flushmargin,bottom]{footmisc}
\renewcommand{\footnoterule}{%
  \kern -3pt
  \hrule width \textwidth height 0.5pt
  \kern 5pt
}   

\definecolor{yellow}{HTML}{FFDB25}
\definecolor{red}{HTML}{FF006C}
\definecolor{mag}{HTML}{DF00FF}
\definecolor{blue}{HTML}{0303ff}
\definecolor{purple}{HTML}{7C00FF}

\definecolor{light_yellow}{HTML}{FFF6B3}
\definecolor{az}{HTML}{00B9E6}

\tcbuselibrary{listings,theorems,breakable}

\newtcolorbox{ch_abs}{center, blanker, breakable, width=0.99\textwidth, left= 20pt, right = 10pt, top = 3pt, bottom = 3pt,
borderline west={0.75mm}{0mm}{blue},
borderline west={0.75mm}{0.75mm}{az}}

\newtcolorbox{proofbox}[2]{fonttitle=\bfseries, coltitle = white, center, breakable, colback = white, arc=0mm, width=\textwidth, colframe = blue, pad at break=2mm,
break at=-\baselineskip/0pt, enhanced jigsaw,
height fixed for=middle, before upper={\parindent7pt\noindent},
title=#1, label=#2}

\newtcolorbox{eqbox}[2]{coltitle = black, center, breakable, colback = white, arc=5mm, width=\textwidth, colframe = yellow, pad at break=2mm,
break at=-\baselineskip/0pt, enhanced jigsaw,
height fixed for=middle, before upper={\parindent7pt\noindent},
title=#1, label=#2}

\newtcolorbox[auto counter,number within=chapter]{techbox}[2]{fonttitle=\bfseries\sffamily, coltitle = black, center, breakable, colback = white, arc=0mm, width=\textwidth, frame style = {top color=yellow,bottom color=yellow}, pad at break=2mm,
break at=-\baselineskip/0pt, enhanced jigsaw,
height fixed for=middle, before upper={\parindent7pt\noindent},
title={\textcolor{red}{Box~\thetcbcounter} #1}, label=#2}

\usepackage{titletoc}
\assignpagestyle{\chapter}{fancy}

\titlecontents{chapter}[0em]{\vskip12pt\bfseries\sffamily}
{\thecontentslabel\enspace}
{\hspace{1.05em}}
{ \hfill\contentspage}[\vskip 6pt]

\titlecontents{section}[1em]{\sffamily}
{\thecontentslabel\enspace}
{}
{\titlerule*[1pc]{.}\quad\contentspage}[\vskip 4pt]

\titlecontents{subsection}[2.7em]{\sffamily}
{\thecontentslabel\enspace}
{}
{\titlerule*[1pc]{.}\quad\contentspage}[\vskip 3pt]
\usepackage{etoolbox}
\pretocmd{\contentsname}{\sffamily}{}{}

\newcommand{\partitle}[1]{\noindent\textsf{\textbf{#1}} ---}

\usepackage{titlesec, blindtext, color}

\newcommand{\hsp}{\hspace{20pt}}

\titleformat{\chapter}[hang]{\Large\bfseries\fontfamily{phv}\selectfont}{\thechapter\hsp}{0pt}{\Large\bfseries}

\titleformat{\section}[hang]{\large\bfseries\fontfamily{phv}\selectfont}{{\textsf\thesection}\hsp}{0pt}{\large\bfseries}

\titleformat{\subsection}[hang]{\large\bfseries\fontfamily{phv}\selectfont}{{\textsf\thesubsection}\hsp}{0pt}{\large\bfseries}

\DeclareCaptionFont{blah}{\small \bf \fontfamily{phv}\selectfont}
\DeclareCaptionLabelSeparator*{twonl}{\\[0.5\baselineskip]}
\usepackage{tikz}
\usepackage{xcolor}
\DeclareRobustCommand*\circled[1]{\tikz[baseline=(char.base),every node/.style={scale=0.35}]{
            \node[shape=circle,fill] (char) {{{\Huge{{\textcolor{white}{#1}}}}}};}}
\usepackage[framemethod=tikz]{mdframed}
\usepackage{framed}

\def\tr{\texttt{tr}}
\def\bra#1{\langle{#1}|}
\def\ket#1{|{#1}\rangle}
\def\braket#1{\langle{#1}\rangle}
\newcommand{\ketbra}[2]{\ket{#1}\!\bra{#2}}

\usepackage{mathtools,amsfonts,amssymb,amsthm}
\usepackage[framemethod=tikz]{mdframed}
\usepackage{framed}
\usepackage{nameref}
\newcommand{\sgn}{\texttt{sign}}
\usepackage{bm}
\usepackage{bbm}
\renewcommand{\mathbb}[1]{\mathbbm{#1}}

\newcommand{\com}[1]{{\color{purple}{\fontfamily{phv}\selectfont #1}}}

\newcommand{\diag}{\texttt{diag}}

\newtheorem{definition}{Definition}[chapter]
\newtheorem{proposition}[definition]{Proposition}

\newtheorem{theorem}[definition]{Theorem}

\newtheorem{conjecture}[definition]{Conjecture}
\newtheorem{remark}[definition]{Remark}

\renewcommand{\qedsymbol}{\rule{0.7em}{0.7em}}

\author{Francesco Campaioli}
\title{
\textbf{\textsf{{\LARGE Tightening
Time-Energy Uncertainty Relations}}}}
\usepackage[font=small]{caption}

\begin{document}
\frontmatter
\begin{titlepage}
    \setlength\fboxsep{5pt}
	\newgeometry{left=2.2cm,right=2.2cm,top=2cm,bottom=2cm}
	\noindent
	\rule{\textwidth}{0.5pt}\\
	\noindent\colorbox{blue}{\textcolor{white}{\large\textsf{\textbf{PhD Thesis}}}}\hfill \textcolor{blue}{\large\textsf{\textbf{Monash Quantum Information Science}}}\vspace{-8.5pt}\\
	\noindent\rule{\textwidth}{0.5pt}\vspace{-11.5pt}\\
	\noindent\rule{\textwidth}{2.5pt}\vspace{10pt}\\
	{\noindent\LARGE\textbf{\textsf{Tightening Time-Energy Uncertainty Relations
	}}
	}\vspace{10pt}\\
	\noindent
	{\Large\textsf{Francesco Campaioli}}\\
	\noindent
	\noindent\rule{\textwidth}{2.5pt}\vspace{-13.5pt}\\
	\noindent\rule{\textwidth}{0.5pt}\vspace{60pt}
	
	\begin{center}
	\includegraphics[width=0.40\textwidth]{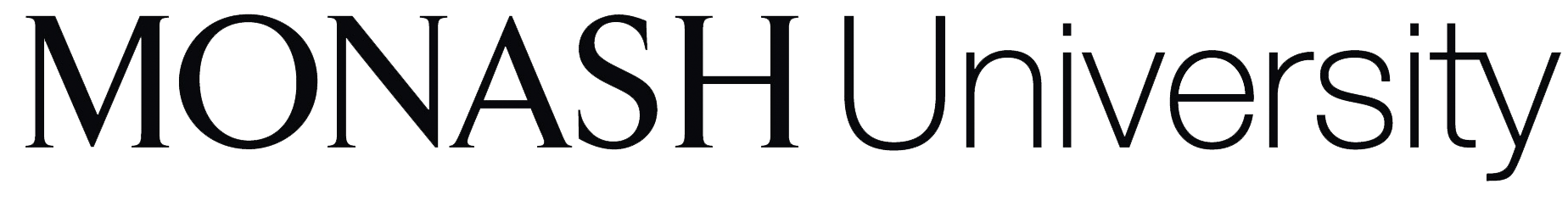}\\
	\includegraphics[width=0.35\textwidth]{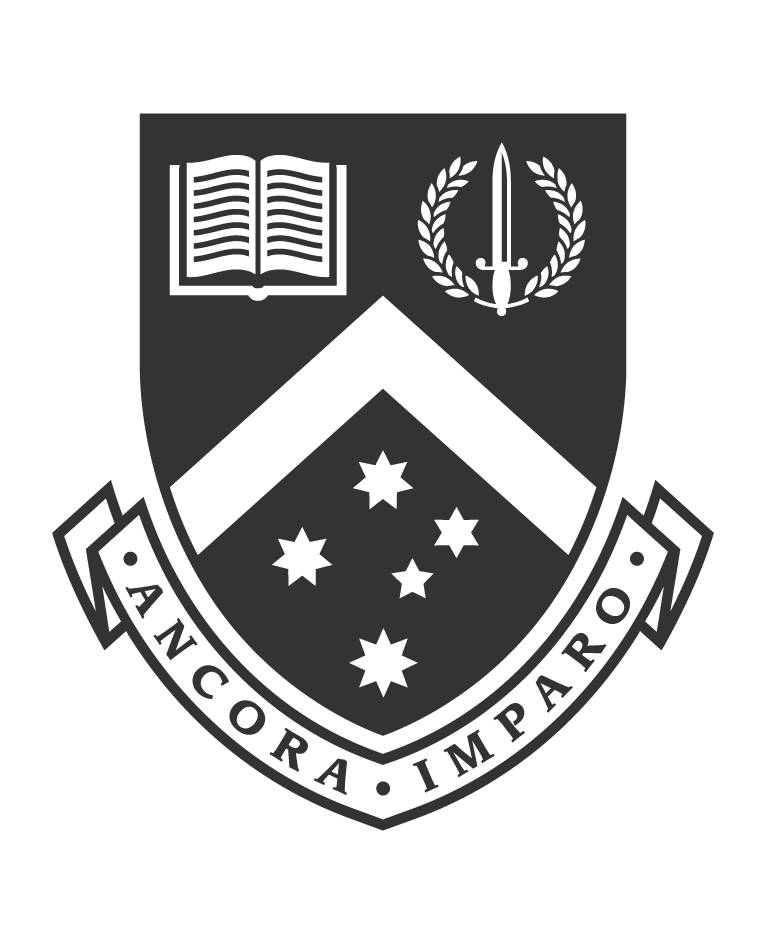}
	\end{center}
	
	\vfill
	\begin{center}
	   {\small A thesis submitted for the degree of\\
	\textit{Doctor of Philosophy}\\
	November 2019}

	\end{center}
	\vspace{5pt}
	\noindent\rule{\textwidth}{2.5pt}\vspace{-13.5pt}\\
	\noindent\rule{\textwidth}{0.5pt}\\
	\noindent\textcolor{blue}{\large\textsf{\textbf{School of Physics and Astronomy}}}\hfill\colorbox{blue}{\textcolor{white}{\large\textsf{\textbf{Monash University}}}}\vspace{-5pt}\\
	\noindent\rule{\textwidth}{0.5pt}
	\restoregeometry
\end{titlepage}
\thispagestyle{empty}
\noindent
{\textcopyright} Francesco Campaioli (2019) \vspace{10pt}

\newpage
\thispagestyle{empty}
\noindent
{\large \textsf{\textbf{Abstract}}}\vspace{-10pt}\\
\rule{\textwidth}{0.5pt}
\vspace{5pt}

\noindent
This thesis aims to study time-energy uncertainty relations, operationally interpreted as a bound on the minimal time for the evolution of quantum systems, thus known as quantum speed limits.\vspace{2pt} 

The \textit{first chapter} provides a brief review of the uncertainty principle and its interpretations in quantum mechanics, focusing on time and energy as pair of conjugate variables, and introducing how they can be considered as bounds on the minimal time of quantum evolution. \vspace{2pt} 

The \textit{second chapter} addresses the shortcomings of the traditional quantum speed limit for the case of unitary evolution, which is known to be loose for increasingly mixed states. The poor performance of the traditional bound is shown to arise from the notion of distance between states, that is used to derive it. A geometric approach is then adopted to obtain a new bound that that is provably tighter and easier to compute than the traditional one, thus improving the performance and efficacy of quantum speed limits applied as benchmarking tool for quantum optimal control, state preparation and gate design.\vspace{2pt}

In the \textit{third chapter}, these results are generalised to the case arbitrary quantum evolution, to extend their validity beyond the restrictive and idealised case of closed and isolated systems. A quantum speed limit is derived using the Euclidean notion of distance between the generalised Bloch vectors representing initial and final states.
This bound is shown to be robust under composition and mixing, tighter than other notable bounds, and generally easier to compute and measure experimentally. \vspace{2pt}

The converse problem of finding the time-optimal evolution between two states, is then addressed in the \textit{fourth chapter}, which provides an algorithmic method to look for Hamiltonians that drive a system saturating the quantum speed limit. The algorithm is based on the idea of progressively improving the efficiency of a Hamiltonian that connects the two states, by reducing its components that do not actively contribute to driving the system. This method can be interpreted as an optimisation over the geometric phases acquired by the initial state during the evolution, and can be applied to time-optimal state preparation and gate design problems.\vspace{2pt}

Finally, in the \textit{fifth chapter} time-energy uncertainty relations are applied to derive a bound on the achievable power of work extraction and energy deposition for isolated many-body quantum system. There, the notion of quantum battery is used to demonstrate the power advantage of using global operations over local ones. This advantage is proven to grow at most extensively with the number of subsystems while being strongly limited by the order of the interactions available for controlling them.
\pagestyle{empty}
\newpage
\noindent
{\large\textsf{\textbf{Declaration}}}\vspace{-10pt}\\
\rule{\textwidth}{0.5pt}
\vspace{10pt}

\noindent
I hereby declare that this thesis contains no material which has been accepted for the award of any other degree or diploma at any university or equivalent institution and that, to the best of my knowledge and belief, this thesis contains no material previously published or written by another person, except where due reference is made in the text of the thesis. 
\vspace{5pt}

\noindent
This thesis includes four original manuscripts published in peer reviewed journals and editions, and one submitted publication. The core theme of the thesis is the operational interpretation of time-energy uncertainty relations as a bound on the minimal time of evolution of quantum systems. The ideas, development and writing up of all the papers in the thesis were the principal responsibility of myself, the student, working within the School of Physics and Astronomy under the supervision of Kavan Modi.\vspace{10pt}

\noindent
I have rearranged and extended the sections of submitted and published papers in order to generate a consistent presentation within the thesis.\vspace{10pt}

\noindent
\textit{Melbourne, November 2019},\vspace{10pt}

\noindent
\hfill Francesco Campaioli
\vspace{30pt}

\noindent
I hereby certify that the above declaration correctly reflects the nature and extent of the student’s and co-authors’ contributions to this work. In instances where I am not the responsible author I have consulted with the responsible author to agree on the respective contributions of the authors. 
\vspace{10pt}

\noindent
\textit{Melbourne, November 2019},\vspace{10pt}

\noindent
\hfill Kavan Modi
\vspace{30pt}
\newpage
\thispagestyle{empty}
\noindent
{\large \textsf{\textbf{Publications}}}\vspace{-10pt}\\
\rule{\textwidth}{0.5pt}
\vspace{-5pt}

\noindent
Many of the results, ideas and figures contained in this thesis have appeared in the following publications and preprints: \cite{Campaioli2018,Campaioli2018b,Campaioli2019,Campaioli2017,Campaioli2018a}.
\vspace{10pt}

\noindent
\href{https://doi.org/10.1103/PhysRevLett.118.150601}{\textbf{Enhancing the Charging Power of Quantum Batteries}}
\vspace{4pt} \\
{\small \textit{Francesco Campaioli, Felix A. Pollock, Felix C. Binder, Lucas Céleri, John Goold, \\ Sai Vinjanampathy, and Kavan Modi}\vspace{4pt}\\
Phys. Rev. Lett. \textbf{118}, 150601 --- Published 12 April 2017\vspace{4pt} \\
Synopsis in \textbf{Physics}: \href{https://physics.aps.org/synopsis-for/10.1103/PhysRevLett.118.150601}{Speeding Up Battery Charging with Quantum Physics}}\vspace{10pt} \\
\href{https://doi.org/10.1103/PhysRevLett.120.060409}{\textbf{Tightening Quantum Speed Limits for Almost All States}}\vspace{4pt}\\
{\small \textit{Francesco Campaioli, Felix A. Pollock, Felix C. Binder, and Kavan Modi}\vspace{4pt}\\
Phys. Rev. Lett. \textbf{120}, 060409 --- Published 9 February 2018}\vspace{10pt} \\
\noindent
\href{https://doi.org/10.1007/978-3-319-99046-0_8}{\textbf{Quantum Batteries}}\vspace{4pt}\\
{\small \textit{Francesco Campaioli, Felix A. Pollock, Sai Vinjanampathy}\vspace{4pt}\\
\textit{Thermodynamics in the Quantum Regime}, pp 207-225 --- First Online 2 April 2019\\ Part of the \textit{Fundamental Theories of Physics} book series (FTPH, volume 195)}\vspace{10pt} \\
\noindent
\href{https://doi.org/10.22331/q-2019-08-05-168}{\textbf{Tight, robust, and feasible quantum speed limits for open dynamics}}\vspace{4pt}\\
{\small \textit{Francesco Campaioli, Felix A. Pollock, Kavan Modi}\vspace{5pt}\\
Quantum \textbf{3}, 168 (2019) --- Published 5 August 2019}\vspace{10pt} \\
\noindent
\href{https://journals.aps.org/pra/abstract/10.1103/PhysRevA.100.062328}{\textbf{Algorithm for solving unconstrained unitary quantum brachistochrone problems}}\vspace{4pt}\\
{\small \textit{Francesco Campaioli, William Sloan, Kavan Modi, Felix Alexander Pollock}\vspace{4pt}\\
Phys. Rev. A \textbf{100}, 062328 --- Published 20 December 2019}\vspace{10pt} \\
\noindent

\noindent
Original results and existing literature have been accurately specified and referenced through the thesis. In particular, \textit{Chapter~\ref{ch:intro}} is an overview of the uncertainty principle and its interpretation, and does not contain original results.
The original results of \textit{Chapter~\ref{ch:qsl_mixed}} are based on Ref.~\cite{Campaioli2018}, and can be found in Secs.~\ref{s:generatlised_Bloch_angle}~--~\ref{s:performance_qsl_mixed}.
The original results of \textit{Chapter~\ref{ch:qsl_open}} are based on Ref.~\cite{Campaioli2018b}, and can be found in Secs.~\ref{s:euclidean_open}~--~\ref{s:tightness}. The original results of \textit{Chapter~\ref{ch:fast_state_preparation}} are based on Ref.~\cite{Campaioli2019}, and can be found in Secs.~\ref{p:method}~--~\ref{s:non-mono}. The original results of \textit{Chapter~\ref{ch:quantum_batteries}} are based on Ref.~\cite{Campaioli2017}, and can be found in Sec.~\ref{s:enhancing}.

\newpage
\thispagestyle{empty}
\noindent
{\large \textsf{\textbf{Acknowledgments}}}\vspace{-10pt}\\
\rule{\textwidth}{0.5pt}
\vspace{10pt}

\noindent
My life as a PhD student has been much brighter than I initially anticipated, and most of the credit goes to the wonderful people that I had around me during this time.

My supervisors Kavan Modi and Felix A. Pollock deserve the greatest praise, for teaching me with patience and passion, motivating me professionally, and providing me with new challenges and responsibilities. 

As part of the Monash Quantum Information Science group I felt, and will always feel, a sense of belonging to something special, which I am extremely grateful for. I would like to thank Kavan and Felix for creating such an enjoyable and uplifting work environment, and the \textit{MonQIS} altogether for being the best team I could have hoped for.

In particular, to Simon Milz and Philip Taranto, who have taught me how to build an unsinkable friendship, I salute you. Thanks to Pedro, for teaching me that \textit{yes} means \textit{maybe}, and \textit{maybe} means \textit{no}, and to Josh, for being so imperceptibly always there. Thanks also to Mathias, Roberto, Graeme, and Top: I wish you all the best for the years to come. 
A special thanks goes to the honorary member Felix C. Binder, who I am obliged to for his intellectual acumen, work ethic, and unrivalled conviviality. 

I would like to express my sincere gratitude to the School of Physics and Astronomy for being a welcoming and inspiring space. A special recognition goes to Jean Pettigrew, for her extraordinary kindness and care, and to Karen Lee, for being a fantastic human being and dedicated basketball player, and for teaching me the essential Australian vocabulary. 
My studies at Monash University were possible thanks to the International Postgraduate Research Scholarship, Australian Postgraduate Award, and J. L. William Scholarship, which I kindly acknowledge for the support.

Moving to Australia for my postgraduate studies was not easy, and the distance from home seemed almost insurmountable at times. Fortunately, that burden was much easier to bear thanks to all the love that I received from my family every day. Thank you, Mum, Dad and Giulia for your encouragement and affection.

The deepest appreciation also goes to my hometown friends Ste, Pavi, Ane, Dami, Lele, and Des, fellows of adventures, whose presence was received, loud and clear, day by day. 

Once in Australia, I was lucky to be surrounded by amazing people, who made me feel at home from the very first days. I would like to thank Simon, Lochy, and Dale, for making me feel loved and included, even though I burnt their kettle and ate their ice cream, and for embracing my distinctly Italian manners with an open heart and amusing sense of humour. A special thank goes to Cody, brother in arms, and comrade through the good and the hard times: Your endeavour will not be forgotten.

Finally, I would like to express my gratitude and admiration for Vanessa, my loving partner, who has changed my life forever, and whose presence gave me all I needed to be happy every day. 
\thispagestyle{empty}
\newpage
    \hfill\textit{Per le mie nonne, Nella e Graziella}
{\hypersetup{hidelinks}
\thispagestyle{empty}
\tableofcontents}
\thispagestyle{empty}
\mainmatter
\thispagestyle{empty}
\pagestyle{empty}
\begin{center}
\includegraphics[width=\textwidth]{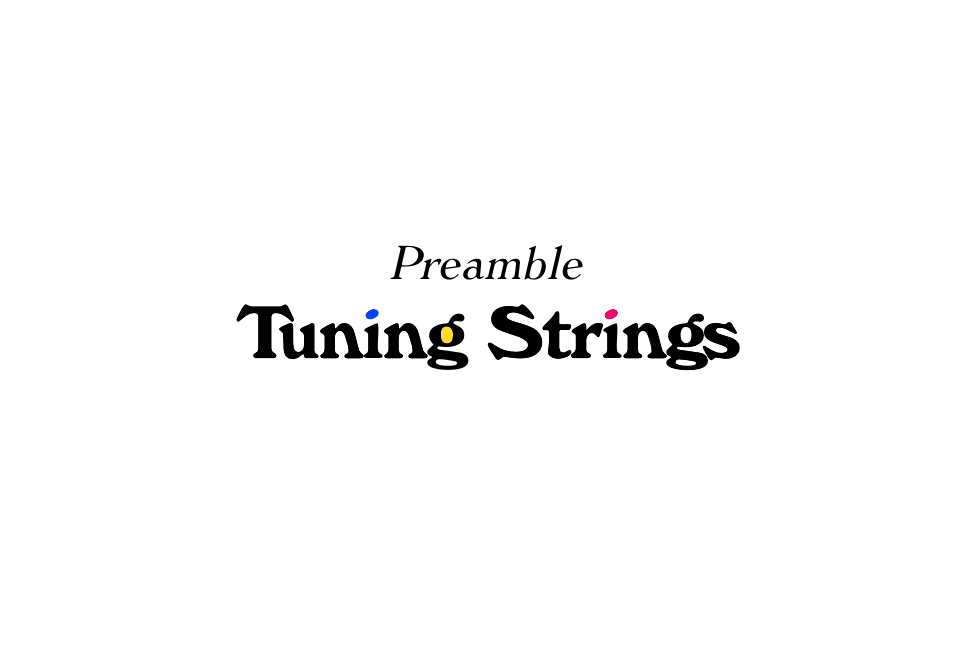}
\end{center}

\phantomsection
\addcontentsline{toc}{section}{\textit{Preamble}}
\noindent
Alice and Bob are flamenco guitarists, and before every set they meticulously tune their instruments. Alice is gifted with absolute pitch\footnote{Absolute pitch, or perfect pitch, is the ability to identify or reproduce a given pitch without the benefit of a reference tone.} and she tunes her classical guitar by ear. Bob does not have such innate talent, but his ears are well trained, and can confidently determine intervals between pitches.
To tune his guitar, he prefers not to resort to an electronic device, and instead he uses Alice's guitar as a reference. 

For an initial, coarse tuning, Alice plucks the first string and lets it oscillate, allowing for Bob to play the corresponding string of his guitar along with it. Bob then quickly adjusts the first tuning key in order to reach unison. He is experienced and able to get an excellent tuning already at this step, however, he is committed to get the best tuning he possibly can. Instead of relying on his almost intuitive ability to distinguish between pitch intervals, he exploits the interference between acoustic waves. 

Bob asks Alice to play a \textit{natural harmonic} on her first string. She proceeds plucking the first string of her guitar, roughly one-quarter of the string's length away from the bridge, while gently hovering a finger over the twelfth fret of the neck, located at the midpoint of the string. The sound produced by Alice's guitar this time is rounder, and higher than the one obtained by simply plucking the free string. Bob plays the natural harmonic on his first string too, which interferes with Alice's one to generate a distinctly audible pulsating sound, known as \textit{acoustic beats}.

The beats are the result of the interference pattern produced by two tones with almost identical pitches. The pulses that Bob and Alice hear occur with a frequency that is proportional to the \textit{detuning} between the tones they played. Due to the initial coarse tuning the period of the pulses is long enough for Alice and Bob to resolve them. This makes it possible for them to quantitatively compare almost identical tones, instead of relying on their natural ability to discern intervals between pitches. 

Bob knows that if the two strings were perfectly tuned, the period of the pulses would be so long to make them virtually inaudible. All he has to do to tune his string, is adjust the tuning key to gradually slow down the pulses' tempo, until they seemingly disappear. After finely tuning the first string, Alice and Bob proceed repeating the process for each string, and the duo is ready to play\footnote{For all that matters, Alice could be lying about her absolute pitch, as long as her guitar is tuned within the precision offered by the acoustic beats method. She and Bob will eventually still agree on their tuning, and could play a set together.}.\vspace{10pt}
\begin{figure}[h]
    \centering
    \includegraphics[width=\textwidth]{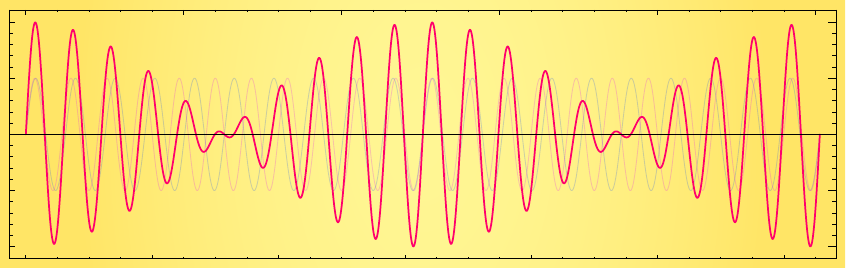}
    \captionsetup{labelformat=empty}
    \caption{\partitle{Acoustic beats} Approximating the natural harmonics played by Alice and Bob with sinusoidal waves of frequency $\nu_A$ and $\nu_B$, the period $T$ of the acoustic beats is inversely proportional to the detuning $\nu_A-\nu_B$, therefore, finer tuning requires longer time.}
    \label{fig:tuning_strings}
\end{figure}

For Alice and Bob, the advantage of using a natural harmonic instead of plucking the free string is that the former is a much better approximation of an acoustic wave with a unique frequency. Hovering the finger over the midpoint of the string forces it to oscillate mostly in its second harmonic mode\footnote{The second harmonic mode is the first overtone of the fundamental.}, softening the higher overtones and the transient frequencies produced by a sudden strum of the string. The acoustic beats that Bob uses to tune his instruments benefit from this filtering, since the closer the tones are to having a unique frequency, the more audible the pulses are. 

So, how well and how fast can the two guitarists tune their strings?
Under the assumption of acoustic waves characterised by a unique frequency, the fact that the period of the acoustic beats increases with the inverse of the string's detuning establishes a \textit{trade-off relation} between the error $\epsilon(\nu)$ on the frequency of the string to be tuned and the length $\Delta t$ of the time interval during which the pulses are listened to. In other words, finer tuning requires longer time.

At the same time, neither Alice's nor Bob's natural harmonics exactly correspond to waves characterised by a unique frequency. 
When looking at the frequency and time domains of these signal with Fourier analysis one can appreciate another well-known fundamental trade-off relation, i.e., that between the spread $\sigma(\nu)$ in the frequency domain of a wave, and the spread $\sigma(t)$ in its time domain.
In signal analysis, the latter is often interpreted as the duration of the signal.
Accordingly, a perfect harmonic characterised by a unique frequency, appears to be characterised by \emph{indefinite} duration.

The trade-off relations between frequency and time discussed in this informal example are intended to be a preparatory illustration of the time-energy uncertainty relations discussed in this Thesis. In particular, the bound on the time required to achieve a certain precision when tuning strings is a anticipation of the minimal time of quantum evolution, also known as \textit{quantum speed limit}. Similarly, the trade off between the spread of a wave in its time and frequency domain is analog to Robertson inequalities between canonically conjugate observables, which will be discussed in the Introduction.
From a purely mathematical perspective, uncertainty relations are often just seen as consequence of Fourier transform between conjugate domains. At the same time, exploring and understanding the physical meaning and implications of these trade-off relations is a fundamentally rich, complex and interesting task.

\chapter{Introduction}
\label{ch:intro} 
\pagestyle{fancy}
\section{Uncertainty relations and their interpretations}
\label{s:uncertainty_priciple}
\begin{figure}[htbp]
    \centering
    \includegraphics[width=0.9\textwidth]{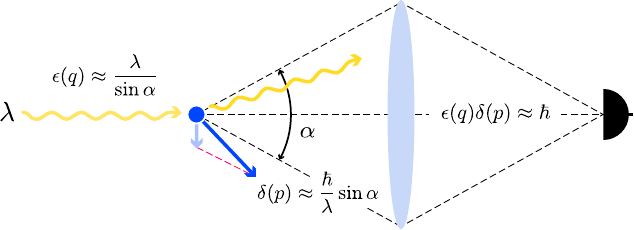}
    \caption{In Heisenberg thought-experiment a $\gamma$-ray with wavelength $\lambda$ is used to resolved the position $q$ of an electron. The error $\epsilon(q)\approx\lambda/\sin\alpha$ committed on this measurement is bounded by the limitations of diffractive optics. The interaction between the photon and the electron gives the latter a \emph{Compton kick} that disturbs its momentum $p$ of the order of the photon's momentum, $\delta(p)\approx \hbar \sin\alpha/\lambda$.}
    \label{fig:heisenberg_microscope}
\end{figure}

\noindent
In his seminal paper published in 1927, Werner Heisenberg presented the first formulation of the \emph{uncertainty principle}, proposing a relation between the disturbance in the momentum of a mass caused by a measurement of its position~\cite{Heisenberg}. His statement can be essentially expressed as a trade-off relation between the error $\epsilon(q)$ on a measurement of position $Q$, and the disturbance $\delta(p)$ of the momentum $P$ caused by the interaction with the measurement apparatus, 
\begin{equation}
\label{eq:heisenberg_tradeoff}
    \epsilon(q) \delta(p) \geq \hbar.
\end{equation}
He interpreted Eq.~\eqref{eq:heisenberg_tradeoff} as a consequence of the \textit{observer-effect}, for which any insightful measurement of a system is bound to affect its state, there represented by the famous $\gamma$-ray microscope \emph{Gedankenexperiment}. In his example, a $\gamma$-ray is used to resolve the position of an electron, as schematically depicted in Fig.~\ref{fig:heisenberg_microscope}. The disturbance of the electron's momentum caused by the scattering and the precision on the position measurement are bounded by the limits of diffraction.

Heisenberg suggested that this relation was a direct consequence of the canonical commutation relations between position and momentum operators,
\begin{equation}
\label{eq:canical_comm_rel}
    [Q,P] = i\hbar,
\end{equation}
showing that Eq.~\eqref{eq:heisenberg_tradeoff} is saturated for Gaussian states, and
drawing a similar conclusion for other canonically conjugate variables of classical mechanics, such as time and energy. 

However, it is now well known that Heisenberg's interpretation of the uncertainty principle does not correspond to a universally valid relation, and that it is in fact possible to achieve arbitrary precision on the measurement of position without disturbing the total momentum of a system~\cite{Ozawa1988,Ozawa2003}.
Nevertheless, Heisenberg's intuition on non-commuting operators is indeed at the core of a more formal expression of the uncertainty principle, represented by the Robertson inequality between the bounded standard deviation of non-commuting Hermitian operators~\cite{Robertson1929}
\begin{equation}
    \label{eq:robertson}
    \Delta A\:\Delta B\geq\frac{1}{2}|\langle\psi|[A,B]|\psi\rangle|,
\end{equation}
and obtained from the notion of standard deviation of an observable $A$ with respect to some state $\ket{\psi}$,
\begin{equation}
    \label{eq:pure_standard_dev}
    \Delta A:=\sqrt{\braket{\psi|A^2|\psi}-|\braket{\psi|A|\psi}|^2}.
\end{equation}

\begin{figure}[htbp]
    \centering
    \includegraphics[width=0.8\textwidth]{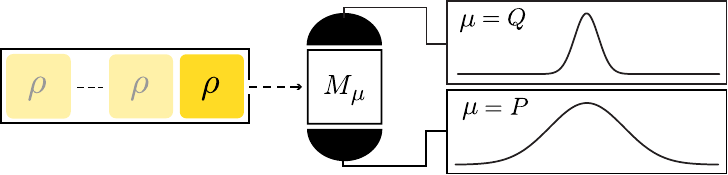}
    \caption{A system in state $\rho=\ketbra{\psi}{\psi}$ undergoes either a measurement of position $M_Q$, or of momentum $M_P$, after which the state of the system is disregarded. Several measurements $M_\mu$ are performed on identical copies of the state $\rho$, to obtain the probability distributions of the states' \textit{position} and \textit{momentum}, here represented by two Gaussian for the sake of simplicity. The standard deviations of these two distributions are bounded as in Eq.~\eqref{eq:robertson}.}
    \label{fig:robertson}
\end{figure}

The universally valid relation expressed by Eq.~\eqref{eq:robertson} does not directly correspond to Heisenberg's interpretation of Eq.~\eqref{eq:heisenberg_tradeoff}, when evaluated for position and momentum operators, since the standard deviation of an operator is a state-dependent quantity, and cannot account for the precision limits of the measurement apparatus, such as the resolution power of the $\gamma$-ray microscope described in Heisenberg's thought-experiment~\cite{Ozawa2004}.
Instead, Eq.~\eqref{eq:robertson} is correctly interpreted as a limitation on the statistical distribution of independent position and momentum measurements of an ensemble of identically prepared systems (see Fig.~\ref{fig:robertson}), or equivalently, on the preparation of such ensemble of identical states.

The discrepancy between Heisenberg and Robertson's interpretations of the uncertainty principle has led to extensive work in the attempt to reconcile the physical content of Eq.~\eqref{eq:heisenberg_tradeoff} with the formal mathematical statement expressed by Eq.~\eqref{eq:robertson}~\cite{Ozawa2003,Ozawa2004,Werner2004,Ozawa2005,Busch2014}. 
In the work by M. Ozawa, for example, the measurement apparatus and the system are described as a composite system that undergoes unitary evolution~\cite{Ozawa2004}. Ozawa uses the measurement operator formalism, widely accepted as the most general description of measurement in quantum mechanics, in order to obtain a universally valid uncertainty relation for the joint measurement of non-commuting observables, which reduces to Heisenberg's relation for observables affected by uncorrelated noise~\cite{Ozawa2004,Ozawa2005}.

\begin{figure}[htbp]
    \centering
    \includegraphics[width=0.95\textwidth]{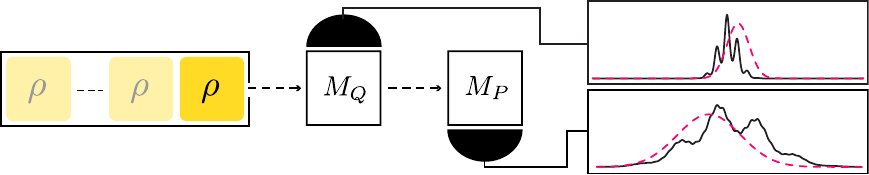}
    \caption{A system in state $\rho=\ketbra{\psi}{\psi}$ undergoes a \textit{joint measurement} of position  and momentum, here obtained by means sequential measurements $M_Q$ and  $M_P$, as described in Ref.~\cite{Werner2004}, after which the state of the system is disregarded. Several joint measurements are performed on identical copies of the state $\rho$, to obtain the statistics of the joint measurement apparatus. As shown in Ref.~\cite{Werner2004}, the difference between the \textit{marginals} of this joint measurement (\textit{solid black lines}) and the \textit{ideal distributions} of position and momentum (\textit{dashed magenta lines}) satisfy a trade-off inequality, which can be used to reconcile Heisenberg interpretation of Eq.~\eqref{eq:heisenberg_tradeoff} with the statistical interpretation of Robertson inequality.}
    \label{fig:werner}
\end{figure}

The measurement aspect of position and momentum uncertainty relations as been studied with the measurement operator formalism also by R. F. Werner, who, in contrast with Ozawa, derives a trade-off between the deviations of the marginals of a joint measurement from the ideal
position and momentum observables (see Fig.~\ref{fig:werner}). Werner's approach to the measurement problem is in essence operational, in the sense that it provides an interpretation of the uncertainty principle that is based solely on the information accessible to the experimenter, i.e., the statistics of the joint measurement apparatus~\cite{Werner2004}.

Remarkably, most rigorous formulations of the uncertainty principle, from Robertson inequalities to Werner's and Ozawa's formulations, are drawn from the commutation relations between conjugate operators.
As pointed out already by Heisenberg in his original work, the connection between the commutation relations of conjugate operators and the \textit{Poisson brackets} of canonically conjugate variables strongly suggests a similar derivation for uncertainty relations between energy and time. Although the conserved energy of the system is directly associated with the Hamiltonian operator $H$, there is not a universally accepted time-operator in quantum mechanics, and thus, time-energy uncertainty relations have required even more care for their formal definition and physical interpretation. 

While time is typically treated as a dynamical parameter in non-relativistic quantum mechanics, \textit{time measurements} are carried out on a regular basis in experiments, and thus the absence of a time operator is often considered a shortcoming of the theory. For this reason, there has been a great deal of effort put into formulating a convincing notion of time operator~\cite{Kijowski1974,Olkhovsky1974,Hilgevoord2002}. In Ref.~\cite{Delgado1997}, the authors show how simplistic formulations of time operators end up being either non-Hermitian, not canonically conjugate with the Hamiltonian operator, or not well behaved under time-translations. To circumvent this problem, they then introduce an effective, yet laboriously constructed \textit{time-of-arrival operator}, from which they derive time-energy uncertainty relations. 

Another notable approach that allows for the derivation of time-energy uncertainty relations consists in defining time as the state of an ancillary system that serves as clock, and that evolves unitarily with the state of the system, as presented in the recent work by Maccone and Sacha, in Ref.~\cite{Maccone2019}.
Using a system as reference clock also allows to derive entropic (dimensionless) time-energy uncertainty relations~\cite{Grabowski1987,Hall2008,Boette2016,Maccone2019,Hall2018,Maccone2015}, as done by the authors of Ref.~\cite{Coles2019}, who discuss their application to information-processing task and quantum technology, such as quantum key distribution. 

Interestingly, it is not necessary to introduce a time operator in order to obtain an operationally meaningful time-energy uncertainty relation. In a groundbreaking work published in 1945, Mandelstam and Tamm combine the dynamics of an observable,
\begin{equation}
    \label{eq:liouville_von_neumann}
    \partial_t A = \frac{i}{\hbar}[H,A],
\end{equation}
with the Robertson inequality of Eq.~\eqref{eq:robertson}, to obtain the relation
\begin{equation}
\label{eq:mandelstam}
    \Delta A_t\;\Delta E_t\geq \frac{1}{2} |\braket{\psi_t|[H,A]|\psi_t}|,
\end{equation}
where, with a minor breach of the notation used so far, 
\begin{equation}
\label{eq:standard_deviation}
    \Delta E_t = \sqrt{\braket{\psi_t | H^2_t |\psi_t}-|\braket{\psi_t|H_t|\psi_t}|^2},
\end{equation}
stands for the time-dependent standard deviation of the Hamiltonian $H$.
In order to obtain an explicit relation between time and energy, Eq.~\eqref{eq:mandelstam} is then integrated over a time interval $\Delta t$, and rearranged as
\begin{equation}
    \label{eq:mandelsta_explicit}
    \Delta t\;\overline{\Delta E} \geq \frac{\hbar}{2}\frac{1}{\overline{\Delta A}}\int_0^{\tau}dt |\partial_t \braket{\psi_t|A|\psi_t}|,
\end{equation}
where the horizontal bar corresponds to the time average in the interval $\tau=\Delta t$, such as for the time-averaged standard deviation
\begin{equation}
\label{eq:time_averaged_standard_deviation}
    \overline{\Delta E} = \frac{1}{\tau} \int_0^\tau dt \sqrt{\braket{\psi_t | H^2_t |\psi_t}-|\braket{\psi_t|H_t|\psi_t}|^2}.
\end{equation}

Mandelstam and Tamm suggest an interpretation of $\Delta t$ as the minimal time required to witness a variation of the expectation value of some observable $A$ comparable with its standard deviation $\Delta A$,
\begin{equation}
\label{eq:mandelstam_tamm_standard_time}
    \Delta t \geq \frac{\hbar}{2}\frac{1}{\overline{\Delta E}}.
\end{equation}
Their result is recognised as the first step towards the operational interpretation of time-energy uncertainty relations, now widely accepted as a bound on the \textit{minimal time} for the evolution of quantum systems.

\phantomsection\section{Minimal time of quantum evolution}
\label{s:minimal_time_quantum_evo}

\noindent
As seen in the previous section, Mandelstam and Tamm' result of Eq.~\eqref{eq:mandelstam_tamm_standard_time} was originally interpreted as the minimal time to witness a \textit{standard} variation in the expectation value of some operator $A$, associated with a physical observable. Naturally, one can choose this operator to be the projector $A=\ketbra{\psi_0}{\psi_0}$ on the initial state $\ket{\psi_0}$, in order to obtain
\begin{equation}
    \label{eq:mandelstam_tamm_projector}
    \Delta E_t \sqrt{\braket{A_t}-|\braket{A_t}|^2} \geq \frac{\hbar}{2}|\partial_t \braket{A_t}|,
\end{equation}
where $\braket{A_t}:=\braket{\psi_t|A|\psi_t}$, which is then integrated in the time interval $\tau =\Delta t$~\cite{Mandelstam1945,Deffner2017}, to obtain
\begin{equation}
    \label{eq:mandelstam_tamm_differential}
    \overline{\Delta E}\Delta t \geq \hbar \bigg( \frac{\pi}{2} -\arcsin{\sqrt{\braket{A_\tau}}} \bigg).
\end{equation}

If we consider a unitary evolution from the initial state $\ket{\psi_0}$ to some orthogonal state $\ket{\psi_\tau}$, with $\tau =\Delta t$ and $\braket{\psi_0|\psi_\tau} = 0$, and for which $\braket{A_\tau} = 0$ we can rearrange Eq.~\eqref{eq:mandelstam_tamm_differential} into
\begin{eqbox}{\textit{Mandelstam-Tamm bound}}{eqbox_mt}
\begin{equation}
    \label{eq:qsl_MT}
    \tau \geq \frac{\pi}{2} \frac{\hbar}{\overline{\Delta E}},
\end{equation}
\end{eqbox}
\noindent
which is thus interpreted as a bound on the minimal time required to unitarily evolve between two orthogonal states, and is often referred to as a \textit{Quantum Speed Limit} (QSL)~\cite{Aharonov1961, Eberly1973, Bauer1978, Uffink1985, Gislason1985,Anandan1990,Vaidman1992,Deffner2017}.

A alternative derivation for the QSL was proposed later by Margolus and Levitin, in Ref.~\cite{Margolus1998}, mainly to circumvent the problem of dealing with potentially diverging standard deviation of the Hamiltonian. 
Their approach proceeds by representing $\ket{\psi_0}$ and $\ket{\psi_t}$ in the eigenbasis of the driving Hamiltonian,
\begin{align}
\label{eq:initial_eigen}
&\ket{\psi_0}=\sum_k c_k \ket{E_k}, \\
\label{eq:evolved_eigen}
&\ket{\psi_t}=\sum_k c_k \exp(-i E_k t/\hbar)\ket{E_k},
\end{align}
in order to look at the real part of the time-dependent inner product $\braket{\psi_0|\psi_t}$ between initial and evolved state, and obtain
\begin{align}
\label{eq:ML}
\texttt{Re}[\braket{\psi_0|\psi_t}] &= \sum_k |c_k|^2 \cos\big(E_k t/\hbar\big), \\
\label{eq:inequality_ML}
& \geq \sum_k |c_k|^2\bigg[1 - \frac{2}{\pi}\bigg(\frac{E_k t}{\hbar}+\sin\Big(\frac{E_k t}{\hbar}\Big)\bigg)\bigg], \\
\label{eq:final_inequality}
& = 1 - \frac{2}{\pi}\frac{\braket{H}}{\hbar}t+\frac{2}{\pi}\texttt{Im}[\braket{\psi_0|\psi_t}],
\end{align}
where the inequality in Eq.~\eqref{eq:inequality_ML} holds since $\cos x \geq 1 - (2/\pi)(x+\sin x)$ for $x\geq 0$.
When considering positive average energy $\braket{H}$ and orthogonal initial and final states $\braket{\psi_\tau|\psi_0}=0$, Eq.~\eqref{eq:final_inequality} can be rearranged to obtain
\begin{equation}
    \tau \geq \frac{\pi}{2}\frac{\hbar}{\braket{H}}.
\end{equation}
The average energy $\braket{H}$ was later  replaced with the time-averaged energy of the Hamiltonian relative to its ground state energy~\cite{Giovannetti2003,Giovannetti2003b,Okuyama2018},
\begin{equation}
\label{eq:time_averaged_energy}
    \overline{E} = \frac{1}{\tau} \int_0^\tau dt E_t,
\end{equation}
where $E_t = \braket{\psi_t | H_t |\psi_t}-h_t^{(0)}$, and
where $h_t^{(0)}$ is the instantaneous ground state energy of $H_t$. The resulting bound
\begin{eqbox}{\textit{Margolus-Levitin bound}}{eqbox_ml}
\begin{equation}
\label{eq:qsl_ML}
    \tau \geq \frac{\pi}{2}\frac{\hbar}{\overline{E}},
\end{equation}
\end{eqbox}
\noindent
is valid for arbitrary unitary evolution between orthogonal pure states, generated by any time-dependent Hamiltonian~\cite{Deffner2017}.

As we will seen in detail in Chs.~\ref{ch:qsl_mixed} and~\ref{ch:qsl_open}, these bounds, originally derived for the transition between pure orthogonal states, have been extended to the unitary evolution of arbitrary pairs of pure states~\cite{Fleming1973, Uffink1993, Margolus1998}, and generalized to the case of mixed states~\cite{Uhlmann1992b,Deffner2013,Zhang2014,Mondal2016}, non-unitary evolution~\cite{Deffner2013, DelCampo2013, Taddei2013}, and multi-partite systems~\cite{Giovannetti2004, Zander2007, Borras2008, Batle2005, Batle2006}.
Extending their original scope, the significance of QSLs has evolved from fundamental physics to practical relevance. Now, QSLs are regularly used in theoretical and applied quantum mechanics, to study the limits of the rate of information transfer~\cite{Bekenstein1981,Murphy2010} and processing~\cite{Lloyd2000,Giovannetti2003}, the precision in quantum metrology~\cite{Alipour2014,Giovannetti2011,Chin2012,Demkowicz-Dobrzanski2012,Chenu2017}, the performance in quantum optimal control~\cite{Goerz2011,Reich2012,Caneva2009,DelCampo2012,Hegerfeldt2013,Murphy2010,An2016,Deffner2017,Campbell2017,Funo2017a,Basilewitsch2017}, the rate of entropy production~\cite{Deffner2010,deffner2019a}, and the power in quantum thermodynamics~\cite{Binder2015,Binder2016,Campaioli2017, Campaioli2018a,Andolina2018a}. For these reasons, they have received particular attention from the quantum information community in recent years~\cite{Kupferman2008, Uzdin2012,Santos2015,Santos2016, Goold2016, Uzdin2016,Mondal2016, Mondal2016b, Mirkin2016, Pires2016, Marvian2016, Friis2016, Epstein2017, Ektesabi2017, Russell2017, Garcia-Pintos2018, Berrada2018, Santos2018, Hu2018, Volkoff2018,Okuyama2018,Shanahan2018}.

\phantomsection\section{Overcoming the limitations of traditional bounds}
\label{s:limitations_of_qsl}

\noindent
A fundamental step forwards in the development and understanding of QSLs consists in using the notion of distinguishability between states in order to draw these bounds from the geometry of the state space.  
The idea is to exploit a suitable notion of distance $D$ between quantum states in order to cast the problem of minimal-time evolution as a \textit{geodesic problem}, i.e., that of finding a path of minimal length. In this formulation, often referred to as the \emph{geometric approach} to QSLs~\cite{Deffner2017}, the time $\tau$ required to evolve between two states $\rho$ and $\sigma$ is naturally bounded from below by the ratio between the distance $D(\rho,\sigma)$, which corresponds to the length of the geodesic, and average speed $\overline{v}$ of the evolution,
\begin{equation}
    \tau \geq \frac{D(\rho,\sigma)}{\overline{v}}.
\end{equation}

At the beginning of Ch.~\ref{ch:qsl_mixed} we discuss precisely how Mandelstam and Tamm's bound is generalised for arbitrary pairs of pure states, using the overlap $|\braket{\psi|\phi}|$ between state vectors as a measure of distinguishability, from which one obtains the Fubini-Study distance~\cite{Bengtsson2008}. This allows us to rigorously identify the standard deviation of the Hamiltonian along the orbit as the speed of the evolution. 

In this thesis we address the main obstacles found in the derivation of QSLs, identify the critical limitations of traditional bounds, and propose methods to overcome them. 
First, we address the \textit{tightness} of QSLs, i.e., their ability to estimate the actual minimal time of evolution, which is in some way also a measure of their performance. 
In Ch.~\ref{ch:qsl_mixed} we show that the traditional QSL for the unitary evolution of density operators is extremely loose when increasingly mixed states are considered, and explain why this looseness arises from the notion of distance used to derived it. We then show how a tight bound can be derived from a more suitable notion of distance for the considered dynamics, and present an improved QSL for unitary evolution, as well as for open quantum systems, in Ch.~\ref{ch:qsl_open}.

Another key element to consider when studying QSLs is to make sure that the distance from which they are derived corresponds to a physically meaningful and accessible measure of distinguishability. When pure states are considered, the overlap $|\braket{\psi|\phi}|$ between state vectors induces a tight distance while being relatively easy to measure experimentally~\cite{Mondal2016}. When mixed states are considered, the overlap is often replaced by the \textit{quantum fidelity}~\cite{Bengtsson2008},
\begin{equation}
    \label{eq:fidelity}
    \mathcal{F}(\rho,\sigma) = \tr{\bigg[\sqrt{\sqrt{\rho}\sigma\sqrt{\rho}}\bigg]},
\end{equation}
which also appears in many QSLs as the preferred choice for defining a distance on the state space. While it is certainly possible to measure $\mathcal{F}$ with an experimental method similar to that used to measure $|\braket{\psi|\phi}|$~\cite{Mondal2016}, in Ch.~\ref{ch:qsl_open} we show that this generally requires a large number of measurements and preparations. In Ch.~\ref{ch:qsl_mixed} and~\ref{ch:qsl_open} we show how to overcome this obstacle by considering a distance defined on the overlap $\tr[\rho\sigma]$ between density operators, which is easier to measure experimentally and to handle analytically.

While the overlap $\tr[\rho\sigma]$ is easy to handle for either theoretical and experimental applications, it is important to carefully tailor the distance and the bounds derived upon it to guarantee their physical meaningfulness. In Ch.~\ref{ch:qsl_open} we provide an extensive discussion of the properties that have to be satisfied by a QSL, with particular emphasis on its robustness under composition, the distance's behavior under physical maps, and other invariance properties. There, we show that it is possible to relax some requirements typically imposed on the distance between states and still obtain QSLs that behave well under physical maps. Surprisingly, such bounds are proven to be robust under such maps, outperforming many other QSLs, when it comes to their tightness and feasibility.

A clear conclusion from Chs.~\ref{ch:qsl_mixed} and~\ref{ch:qsl_open} is that, despite major progress in the field~\cite{DelCampo2013,Deffner2013,Pires2016,Deffner2017,Campaioli2018,Campaioli2019}, a generally attainable QSL is yet to be found and its search is a hard problem. In order to probe the absolute tightness of the QSLs derived in in this thesis, we address the converse problem of finding the fastest evolution between pairs of states. In Ch.~\ref{ch:fast_state_preparation} we formulate this problem with a constructive approach and obtain an iterative method to look for time-optimal unitary orbits between arbitrary pairs of density operators. We then discuss the practical relevance of this result and show its synergy with the QSLs derived in Ch.~\ref{ch:qsl_mixed}. By comparing the achievable time of evolution provided by the solutions of the iterative method with the inviolable bound offered by the QSL we can precisely study the performance of both results.

Finally, in Ch.~\ref{ch:quantum_batteries} we apply QSLs to study the limits on the achievable power of work extraction and deposition for quantum systems. There, we exploit the notion of \textit{quantum batteries} to obtain a bound on the achievable advantage of using global operation over local ones, and discuss our results with emphasis on entanglement and other quantum correlations. When considering the evolution of a composite many-body quantum system, such as a spin chain or an array of charged qubits, we show how the traditional QSL does not account for the dynamical constraints given by physical limitations, which are instead known to drastically influence the speed of the evolution and the time required to drive a system. To overcome the limitations of the traditional QSL we derive a bound that takes into account the physical restrictions on range and order of the interactions, and discuss their role for obtaining fast unitary driving.

Beyond the results examined in this thesis, there are many open problems of fundational relevance and practical impact related to time-energy uncertainty relation. In the conclusions we review some of these questions, and discuss their connection to other field of quantum information and quantum technology.

\chapter[The minimal time of unitary evolution]{The minimal time of unitary evolution}
\label{ch:qsl_mixed} 

\begin{ch_abs}
\textsf{\textbf{Chapter abstract}}\vspace{5pt}\\
\small\textsf{Traditional quantum speed limits for the unitary evolution of quantum systems perform poorly for mixed states, often underestimating the time required to perform the evolution. We show that the looseness of the traditional bound originates from an incongruous notion of distance between the initial and final states when their evolution has to preserve their spectrum. We then discuss more suitable  metrics and derive their corresponding speed limits. These bounds are shown to outperform the traditional one while being significantly simpler to compute and to experimentally measure.\vspace{15pt}\\
\emph{This chapter is based on publication \cite{Campaioli2018}.}}
\end{ch_abs}\vspace{20pt}

\section[The traditional unified quantum speed limit]{The traditional unified quantum speed limit}
\label{s:traditional_unified_qsl}

\noindent
As seen in the introduction, time-energy uncertainty relations are correctly interpreted as a bound on the minimal time required to evolve between two quantum states. In particular, we have seen how the Mandelstam-Tamm and Margolus-Levitin inequalities, given in Eqs.~\eqref{eq:qsl_MT} and~\eqref{eq:qsl_ML}, apply to orthogonal pure states. This result is generalised to the case of arbitrary pairs of pure states $\ket{\psi}$, $\ket{\phi}$ by the unified bound,
\begin{equation}
    \label{eq:qsl_pfifer} 
    \tau \geq \hbar\: \frac{\arccos\big(|\braket{\psi|\phi}|\big)}{\min\big\{\overline{E},\overline{\Delta E}\big\}},
\end{equation}
where the time $\tau$ required to unitarily evolve between the considered states is bounded by a quantity proportional to the angle associated with their overlap $|\braket{\psi|\phi}|$, and inversely proportional to the energy scale expressed by $\overline{E}$ and $\overline{\Delta E}$, given in Eqs.~\eqref{eq:time_averaged_energy} and~\eqref{eq:time_averaged_standard_deviation}~\cite{Pfeifer1993,Deffner2017}. Just like Eqs.~\eqref{eq:qsl_MT} and~\eqref{eq:qsl_ML}, this result is also obtained by combining the Robertson inequality with the Liouville-von-Neumann equation, given in Eqs.~\eqref{eq:robertson} and~\eqref{eq:liouville_von_neumann}, respectively. 

\begin{figure}[htbp]
    \centering
    \includegraphics[width=0.7\textwidth]{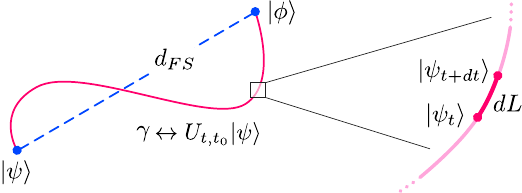}
    \caption{In order to derive the Mandelstam-Tamm bound with a geometric approach, we use the overlap between two state vectors to obtain the distance $d_{FS}$ between states, which correspond to the length of the shortest path that connects them. Other unitary orbits $U_{t,t_0}\ket{\psi}$ draw longer paths $\gamma$ between the considered states. Their length is calculated using the infinitesimal distance $dL$ induced by $d_{FS}$.}
    \label{fig:geodesic_approach}
\end{figure}

However, it is also possible to derive QSLs using a geometric approach, starting from a notion of distance between states, which is then used to measure the length of a path $\gamma$ associated with some unitary orbit $\ket{\psi_t} = U_{t,t_0}\ket{\psi}$, as depicted in Fig.~\ref{fig:geodesic_approach}. 
Exploiting the inner product of the Hilbert space, we can calculate the distance between pairs of states using the angle between their state vectors $d_{FS}(\ket{\psi},\ket{\phi}):=\arccos(|\braket{\psi|\phi}|)$, known as the Fubini-Study distance~\cite{Bengtsson2008,Fubini1904,Study1905}.
Since, by definition, a distance corresponds to the length of the shortest path between the pair of states, i.e., the geodesic induced by the corresponding metric, the length $L[\gamma]$ of any path $\gamma$ connecting the considered pair of states is either equal to or larger than the distance between them, 
\begin{equation}
    \label{eq:length_of_the_geodesic}
   d_{FS}(\ket{\psi},\ket{\phi}) \leq L[\gamma].
\end{equation}
We can now derive a Mandelstam-Tamm type of QSL, combining Eq.~\eqref{eq:length_of_the_geodesic} with the Schr\"odinger equation, and some notions of differential geometry.

First, we express $L[\gamma]$ as the path integral of its line element $dL$,
\begin{equation}
    \label{eq:infinitesimal_distance}
    L[\gamma] = \int_\gamma \: dL,
\end{equation}
then, we express $\gamma$ in terms of a general unitary orbit that connects $\ket{\psi}$ and $\ket{\phi}$, as $\ket{\psi_t}=U_{t,t_0} \ket{\psi}$, with the condition that, at the final time $t=\tau$, the target state is reached, i.e. $\ket{\psi_\tau}=\ket{\phi}$.
We can now write the line element as the infinitesimal distance between $\ket{\psi_t}$ and $\ket{\psi_{t+dt}}$,
\begin{equation}
    \label{eq:infinitesimal_distance_state}
    L[\gamma] = \int_0^\tau \: \arccos(|\braket{\psi_t|\psi_{t+dt}}|),
\end{equation}
where $\ket{\psi_{t+dt}}=U_{t+dt,t}\ket{\psi}$, and where we also fixed $t_0=0$. Note that in Eq.~\eqref{eq:infinitesimal_distance_state} the differential is $dt$, implicitly embedded in $\ket{\psi_{t+dt}}$. Considering the most general unitary evolution generated by a time-dependent Hamiltonian $H_t$, 
\begin{equation}
    \label{eq:dyson_series}
    U_{t,t_0} = \mathcal{T}\bigg\{\exp\bigg(-\frac{i}{\hbar}\int_{t_0}^t ds H(s)\bigg)\bigg\},
\end{equation}
where the \emph{Dyson series} defining the unitary is conveniently expressed with the time-ordering operator $\mathcal{T}$~\cite{Breuer2002},
we expand the infinitesimal unitary evolution $U_{t+dt,t}$ in powers of $dt$ around the Hamiltonian $H_t$ at time $t$,
\begin{equation}
    \label{eq:series_U}
    U_{t+dt,t} \approx \mathbb{1}-\frac{i}{\hbar} H_t dt-\frac{1}{2\hbar^2}H^2_t dt^2+\mathcal{O}(dt^3),
\end{equation}
and calculate the line element, neglecting the terms of order $\mathcal{O}(dt^3)$,
\begin{equation}
    \label{eq:line_element_expanded}
    dL = \arccos(\sqrt{1-x^2}), \:\:\: \textrm{and} \:\:\: x= \frac{\Delta E_t}{\hbar}\:dt,
\end{equation}
where $\Delta E_t$ is the standard deviation of the Hamiltonian at time $t$, given in Eq.~\eqref{eq:standard_deviation}. Expanding $\arccos(\sqrt{1-x^2})$ around $x\sim 0$, we obtain $dL = x$, from which we derive the expression for the length of a unitary orbit for pure states,
\begin{equation}
    \label{eq:metric_unitary_evolution_pure}
    L[\gamma]=\int_0^\tau dt\: \frac{\Delta E_t}{\hbar}.
\end{equation}
We then combine Eqs.~\eqref{eq:length_of_the_geodesic} and~\eqref{eq:metric_unitary_evolution_pure}, 
\begin{equation}
    d_{FS}(\ket{\psi},\ket{\phi}) \leq \int_0^\tau dt \: \frac{\Delta E_t}{\hbar},
\end{equation}
and obtain the Mandelstam-Tamm bound, by explicitly expressing the previous inequality in $\tau$,
\begin{equation}
    \label{eq:mt_qsl}
    \tau \geq \hbar \:\frac{d_{FS}(\ket{\psi},\ket{\phi})}{\overline{\Delta E}}.
\end{equation}
Like the unified bound of Eq.~\eqref{eq:qsl_pfifer}, this bound is provably tight~\cite{Levitin2009}, in the sense that, in the absence of constraints\footnote{As we will see in Ch.~\ref{ch:quantum_batteries}, many-body problems are characterised by stringent constraints on the form of the driving Hamiltonian, which strongly affect the attainability of traditional QSLs.}, such that the Hamiltonian can be proportional to any element of the Lie Algebra $\mathfrak{su}(d)$ of the special unitary group $SU(d)$, it is always possible to find a time-independent Hamiltonian that generates the optimal unitary that saturates the bound, for any Hilbert space dimension $d$~\cite{Bengtsson2008}.

\section{The unified bound is tight for pure states}
\label{s:attainable_pure_qsl}
\begin{figure}[t]
    \centering
    \includegraphics[width=0.75\textwidth]{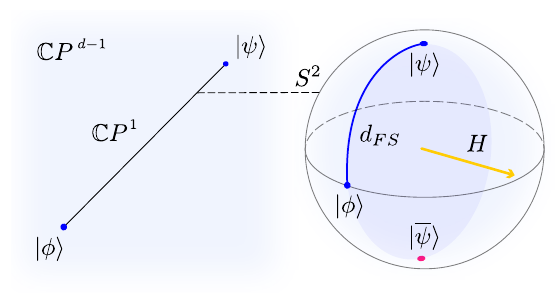}
    \caption{Any two pure states $\ket{\psi}$, $\ket{\phi}\in\mathcal{H}$ of dimension $d$ can be connected by a complex projective line $\mathbb{C}P^1 \subset \mathbb{C}P^{d-1}$, spanned by two orthogonal states $\ket{\psi}$, $\ket{\overline{\psi}}$, and isomorphic to a 2-sphere $S^2$. The Fubini-Study distance $d_{FS}$ between the two states is directly proportional to the length of the arc of great circle that connects them on $S^2$. The unitary orbit that connects them along that arc is generated by Hamiltonians of the form $H =  \ketbra{\psi}{\overline{\psi}}+h.c.$, where $\ket{\overline{\psi}}=(\ket{\phi}-\braket{\psi|\phi}\ket{\psi})/\sqrt{1-|\braket{\psi | \phi}|^2} $ is obtained by Gram-Schmidt orthonormalisation of $\ket{\psi}$ and $\ket{\phi}$.}
    \label{fig:saturate_qsl_pure}
\end{figure}

The attainability of the QSL for the unitary evolution of pure states is directly linked to the Fubini-Study distance, whose metric is the unique, unitary-invariant Riemannian metric on the space of pure states~\cite{Bengtsson2008}. The latter consists in the complex projective space $\mathbb{C}P^{d-1}$ obtained from the Hilbert space via the equivalence $\ket{\psi} \sim c \ket{\psi}$ for any $c\in\mathbb{C}$, $c\neq 0$.
Any initial and final states, $\ket{\psi}$ and $\ket{\phi}$, elements of this complex projective space, can always be connected on a complex projective line $\mathbb{C}P^1$, which is isomorphic to the 2-sphere $S^2$, and defined by the linear combinations of $\ket{\psi}$ and $\ket{\phi}$~\cite{Bengtsson2008}. As depicted in Fig.~\ref{fig:saturate_qsl_pure}, an optimal Hamiltonian, constructed via Gram-Schmidt orthonormalisation of $\ket{\psi}$ and $\ket{\phi}$,
\begin{align}
    \label{eq:optimal_H}
    &H = \omega \big( \ketbra{\psi}{\overline{\psi}}+\ketbra{\overline{\psi}}{\psi}\big), \\
    \label{eq:orthonormal_state}
    &\ket{\overline{\psi}} = \frac{\ket{\phi}-\braket{\psi|\phi}\ket{\psi}}{\sqrt{1-|\braket{\psi|\phi}|^2}},
\end{align}
drives the initial state to the final one along an arc of a great circle.

However, a complete description of quantum states and processes requires the use of density operators, necessary to account for the effect of environmental noise and witness the signature of quantum correlations~\cite{Nielsen2000,Breuer2002,Bengtsson2008}. For this reason, the unified bound of Eq.~\eqref{eq:qsl_pfifer} has been generalised to the case of density operators in Refs.~\cite{Uhlmann1976,Deffner2013}, by replacing the Fubini-Study distance $d_{FS}$ with the Bures angle,
\begin{equation}
    \label{eq:bures_angle}
    \mathcal{L} (\rho,\sigma) =\arccos(\mathcal{F}(\rho,\sigma)),
\end{equation}
a distance between density operators which consists of the angle associated with the fidelity $\mathcal{F}$, given in Eq.~\eqref{eq:fidelity}, and introduced by Uhlmann in Ref.~\cite{Uhlmann1976}.
Since the fidelity is a generalisation of the overlap between state vectors~\cite{Wootters1981, Uhlmann1992a}, the Bures angle $\mathcal{L}$ reduces to the Fubini-Study distance $d_{FS}$ when pure states are considered~\cite{Bengtsson2008}. The Mandelstam-Tamm type of bound that is obtained in this way is given by,
\begin{gather}
    \label{eq:QSL_mixed}
    \tau \geq T_{\mathcal{L}}(\rho,\sigma) := \hbar\: \frac{\mathcal{L}(\rho,\sigma)}{\overline{\Delta E}},
\end{gather}
where now
\begin{equation}
    \label{eq:mixed_average_standard_deviation}
    \overline{\Delta E} = \frac{1}{\tau}\int_0^\tau dt \sqrt{\tr[\rho_t H^2]-\tr[\rho_t H_t]^2},
\end{equation}
is the time-averaged standard deviation of $H_t$ calculated along the trajectory defined by $\rho_t = U_t \rho\: U_t^\dagger$. While this bound reduces to that of Eq.~\eqref{eq:mt_qsl} when $\rho=\ketbra{\psi}{\psi}$ and $\sigma=\ketbra{\phi}{\phi}$, it is not always attainable, in the sense that the existence of a Hamiltonian that saturates the bound via unitary evolution is not guaranteed for all states. On the contrary, $T_{\mathcal{L}}$ is known to be particularly loose, performing poorly for increasingly mixed states, as we will show in the next section.

\section{Attainability of QSL for mixed states}
\label{s:attainable_qsl}
\begin{figure}[htpb]
\centering
\includegraphics[width=0.8\textwidth]{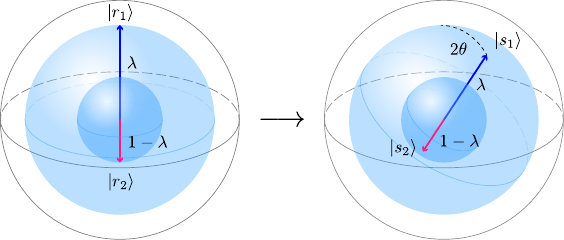}
\caption[width=0.7\textwidth]{Let $\rho = \lambda \ketbra{r_1}{r_1} + (1-\lambda) \ketbra{r_2}{r_2}$ and $\sigma = \lambda \ketbra{s_1}{s_1} + (1-\lambda) \ketbra{s_2}{s_2}$ be two mixed qubit states ($d=2$) with the same spectrum, with $\lambda \in (0,1)$, excluding the maximally mixed state ($\lambda=1/2$), where $\{\ket{r_1},\ket{r_2}\}$ and $\{\ket{s_1},\ket{s_2}\}$ are two orthonormal bases. The problem of unitarily evolving $\rho$ to $\sigma$ can be mapped to evolving $\ket{r_1}$ to $\ket{s_1}$ (or, equivalently, $\ket{r_2}$ to $\ket{s_2}$). Eq.~\eqref{eq:QSL_mixed} is tight for pure states, thus any Hamiltonian that takes $\ket{r_1}$ to $\ket{s_1}$, will also take $\rho$ to $\sigma$ in the same time. For any Hamiltonian, with bounded standard deviation $\Delta E \le \mathcal{E}$, this time is bounded from below by $\theta/\mathcal{E}$, where $\theta = d(\ket{r_1},\ket{s_1})$ is the distance between $\ket{r_1}$ and $\ket{s_1}$, i.e., half of the angle between the Bloch vectors associated with $\ket{r_1}$ and $\ket{s_1}$. However, Eq.~\eqref{eq:QSL_mixed} for the same constraint on the Hamiltonian suggests that $T_{QSL}=  \mathcal{L} (\rho,\sigma)/\mathcal{E}$, with $\mathcal{L}(\rho,\sigma)< \theta$ for every choice of $\lambda \neq 0,1$ (see Eq.~\eqref{eq:analytic_bures}), making the QSL unattainable for all mixed states.}
\label{fig:qubit_mixed}
\end{figure}

Let us consider two mixed states, $\rho$ and $\sigma$, with the same spectrum,
\begin{equation}
    \label{eq:rho_sigma}
    \rho=\sum_k \lambda_k \ketbra{r_k}{r_k}, \;\;\; \sigma=\sum_k \lambda_k \ketbra{s_k}{s_k}.
\end{equation}
Let $\rho'=\sum_k \lambda'_k \ketbra{r_k}{r_k}$ and $\sigma'=\sum_k \lambda'_k \ketbra{s_k}{s_k}$ be another pair of mixed states with the same degeneracy structure as $\rho$ and $\sigma$, but different eigenvalues $\lambda'_k$. Any driving Hamiltonian that maps $\rho$ to $\sigma$ will map $\rho'$ to $\sigma'$ in the same amount of time, independent of their spectrum. On the other hand, the Bures angle is a continuous function of the spectrum of a mixed state,  \textit{i.e.}, for some choices of $\{\lambda_k\}$ and $\{\lambda'_k\}$ one could have $\mathcal{L} (\rho,\sigma) \approx 1$, while $\mathcal{L} (\rho',\sigma') \approx 0$.
This observation alone is not enough to demonstrate the looseness of $T_{\mathcal{L}}$, since the denominator of Eq.~\eqref{eq:QSL_mixed} may in principle also differ between these two scenarios, due to its dependence on the state\footnote{Some authors have suggested quantifying the driving resource independently of the state, for instance in terms of norms of the driving Hamiltonian \cite{Uzdin2012,Binder2015, Binder2016, Campaioli2017, Deffner2017b, Russell2017}.}. However, as shown in the example of Fig.~\ref{fig:qubit_mixed}, the variation of the latter does not compensate for that of the former, implying that $T_{\mathcal{L}}$ cannot be tight for the case of mixed states. To understand the reason for the looseness of this bound we must understand the meaning of the Bures angle.

\begin{figure}[htbp]
    \centering
    \includegraphics[width=\textwidth]{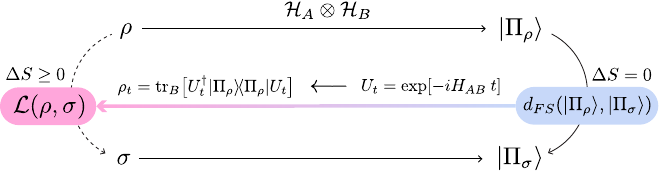}
    \caption{Any density operator $\rho$, $\sigma$ can be purified to a state vector $\ket{\Pi_\rho}$, $\ket{\Pi_\sigma}$ in a lager Hilber space $\mathcal{H}_A\otimes\mathcal{H}_B$. The Bures angle $\mathcal{L}$ is then defined as the minimal Fubini-Study distance over all the possible purifications of $\rho$ and $\sigma$. The geodesic between the purified states is given by a global unitary orbit of constant von Neumann entropy ($\Delta S=0$). The orbit $\rho_t$ in the reduced space is in general non unitary ($\Delta S\geq 0$) even when $\rho$ and $\sigma$ are isospectral states.}
    \label{fig:dilated_to_reduced_qsl}
\end{figure}

The poor performance of the bound in Eq.~\eqref{eq:QSL_mixed} stems from the construction of the Bures angle $\mathcal{L}(\rho,\sigma)$, which corresponds to the minimal Fubini-Study distance between purifications of $\rho$ and $\sigma$~\cite{Bengtsson2008}. Indeed,
any mixed state $\rho\in\mathcal{S}(\mathcal{H}_A)$ can be \emph{purified} to a state vector $\ket{\Pi_\rho}$ embedded in a larger Hilbert space $\mathcal{H}_A \otimes \mathcal{H}_B$, such that
\begin{equation}
    \label{eq:purified}
    \mathrm{tr}_B \big[\ketbra{\Pi_\rho}{\Pi_\rho}\big] = \rho,
\end{equation}
where $\tr_B[\cdot]$ denotes the partial trace over $\mathcal{H}_B$. Since neither the purification $\ket{\Pi_\rho}$ nor the dimension $d_B$ of the additional Hilbert space $\mathcal{H}_B$ are unique\footnote{However, $d_B = d_A$ is sufficient for a purification to always exist.}, the Fubini-Study distance between $\ket{\Pi_\rho}$ and $\ket{\Pi_\sigma}$ strongly depends on the choice of purifications.
A consistent notion of distance is found minimising $d_{FS}$ over all the possible purifications\footnote{The fidelity is obtained as the maximum overlap between all the possible purifications of the considered density operators. Via monotonicity of $\arccos$, the Bures angle is obtained as the minimum of the Fubini-Study distance.}~\cite{Bengtsson2008}, to obtain 
\begin{align}
    \label{eq:Bures_from_definition}
    \mathcal{L}(\rho,\sigma):=\min_{\ket{\Pi_\rho},\ket{\Pi_\sigma}}\big\{ d_{FS}(\ket{\Pi_\rho},\ket{\Pi_\sigma}) \big\}.
\end{align}
Since the speed limit for the unitary evolution of pure states is attainable, it is always possible to find a \emph{global} Hamiltonian $H_{AB}$ over the elements of $\mathcal{H}_A\otimes\mathcal{H}_B$ that saturates the bound of Eq.~\eqref{eq:QSL_mixed}.
However, the corresponding unitary dynamics between $\ket{\Pi_\rho}$ and $\ket{\Pi_\sigma}$ turns, in general, into non-unitary evolution between $\rho$ and $\sigma$, when tracing over $\mathcal{H}_B$, as depicted in Fig.~\ref{fig:dilated_to_reduced_qsl}. In other words, the Bures metric does not necessarily select geodesics generated by unitary operations in the system's state space, even if $\rho$ and $\sigma$ have the same spectrum.

We now propose two distance measures for mixed states with the same fixed spectrum, that do not suffer from the problems outlined above. 

\section{The generalised Bloch angle}
\label{s:generatlised_Bloch_angle}
Bloch vectors provide a representation of the states of quantum two-level systems, by associating any density operator $\rho$ with a real vector $\bm{r}\in V(\mathbb{R}^3)$ in a three-dimensional Euclidean vector space, 
\begin{equation}
    \label{eq:bloch_vector}
    \rho = \frac{\mathbb{1}+\bm{r}\cdot \bm{\Lambda}}{2},
\end{equation}
where the components of $\mathbf{\Lambda}=(\Lambda_x,\Lambda_y,\Lambda_z)$ are the Pauli operators. In this representation, the difference between pure and mixed states is evident, with the former consisting of the unit sphere $S^2$, and the latter forming the bulk of the ball delimited by the surface $S^2$. 

Similarly, any state $\rho \in \mathcal{S}(\mathcal{H})$ of Hilbert space dimension $d$ can be associated with a \emph{generalised} Bloch vector (GBV) $\bm{r}\in V(\mathbb{R}^{d^2-1})$, and represented as
\begin{gather}
\label{eq:rho_bloch}
\rho=\frac{\mathbb{1}+c\: \bm{r}\cdot \bm{\Lambda}}{d},
\end{gather}
with $c=\sqrt{d(d-1)/2}$, 
where now $\bm{\Lambda} = (\Lambda_1,\dots, \Lambda_{d^2-1})$ is a set of traceless Hermitian operators that form a Lie algebra for $SU(d)$, and that satisfies
\begin{equation}
    \label{eq:normalisation_su}
    \tr[\Lambda_i\Lambda_j]=2 \delta_{ij}.
\end{equation}
Such set always exists and it is easy to obtain; an example is given by the Gell-Mann matrices for $SU(3)$, and by their generalisation for $SU(d)$~\cite{Kimura2003}.
In order to represent a state, a generalised Bloch vector $\bm{r}$ has to satisfy a set of relations to reflect the invariance of $\tr[\rho^k]$, for $k=2,\dots, d$, therefore the geometry of the generalised Bloch sphere is vastly richer than that of the standard Bloch sphere. However, for now we will limit ourselves to calculating angles between GBVs in order to derive a tight speed limit for the unitary evolution of mixed states.

The generalised Bloch angle (GBA),
\begin{gather}
    \label{eq:gba}
    \Theta = \arccos{\big(\hat{\bm{r}}\cdot\hat{\bm{s}}\big)},
\end{gather}
is a distance between any pair of isospectral states\footnote{In fact, it is more generally a distance between states with the same purity.} $\rho$, $\sigma\in\mathcal{S}(\mathcal{H})$, where $\hat{\bm{r}}$ and $\hat{\bm{s}}$ are the generalised Bloch vectors associated with states $\rho$ and $\sigma$, respectively, normalised by their length $\lVert \bm{r}\rVert_2 = \lVert \bm{s}\rVert_2$. 
The angle $\Theta$ can be expressed as a function of $\rho$ and $\sigma$, independently of the chosen basis $\bm{\Lambda}$ for the Lie algebra $\mathfrak{su}(d)$, 
\begin{equation}
    \label{eq:gba_rho_sigma}
    \Theta(\rho,\sigma) = \arccos\bigg( \frac{d\:\tr[\rho\sigma] -1}{d\:\tr[\rho^2]-1}\bigg),
\end{equation}
using the expression for the overlap $\tr[\rho\sigma]$ in terms of GBVs. Note that the distance $\Theta(\rho,\sigma)$ does not depend on the basis chosen to represent the states, since the trace is basis-independent.
Our first result is a bound on the speed of unitary evolution of isospectral mixed states derived from the distance $\Theta$.
\begin{theorem}
\label{th:qsl_gba}
The minimal time $\tau$ required to evolve from state $\rho$ to state $\sigma$ by means of a unitary operation generated by the Hamiltonian $H_{t}$ is bounded from below by 
\begin{align}
    \label{eq:qsl_gba}
    & T_\Theta (\rho,\sigma) :=  \hbar\: \frac{\Theta(\rho,\sigma)}{ Q_\Theta},\;\textrm{where}\\
    \label{eq:gba_infinitesimal}
    &Q_\Theta := \frac{1}{\tau}\int_0^\tau dt \sqrt{\frac{2 \; \textrm{\emph{\tr}}[\rho_t^2 H_{t}^2-(\rho_t H_{t})^2]}{\textrm{\emph{\tr}}[\rho_t^2-\mathbb{1}/d^2]}}.
\end{align}
\vspace{5pt}
\end{theorem}
\noindent
To prove Thm.~\ref{th:qsl_gba} we first show that $\Theta$ is a distance, and then we use the geometric approach outlined in Sec.~\ref{s:traditional_unified_qsl} to bound the time required to unitarily evolve between any two states by evaluating the infinitesimal distance $\Theta(\rho_t,\rho_{t+dt})$ between two unitarily connected states. 
\vspace{5pt}

\begin{proofbox}{Proof of Theorem~\ref*{th:qsl_gba}}{proof:qsl_gb}
First, we prove that $\Theta$ is a distance. Let $\hat{\bm{r}}=\bm{r}/\lVert \bm{r} \rVert_2$, where $\lVert \bm{r} \rVert_2 = \sqrt{\bm{r} \cdot \bm{r}}$. Since $\hat{\bm{r}}\cdot\hat{\bm{s}} \in [-1,1] \Rightarrow \Theta(\rho,\sigma)\in[0,\pi]$, positivity holds. $\rho=\sigma \Rightarrow \hat{\bm{r}} = \hat{\bm{s}}$, thus $\Theta(\rho,\sigma) = 0$. $\Theta(\rho,\sigma)=0 \Rightarrow\bm{r}\cdot\bm{s}=\bm{r}\cdot\bm{r} =\bm{s}\cdot\bm{s}$, thus $\bm{r} =\bm{s}$ and so the identity of indiscernibility holds. Symmetry holds because $\hat{\bm{r}}\cdot\hat{\bm{s}} =\hat{\bm{s}}\cdot\hat{\bm{r}}$, thus $\Theta(\rho,\sigma) = \Theta(\sigma,\rho)$. Lastly, the triangle inequality holds, since $\hat{\bm{r}},\hat{\bm{s}}$ belong to a subset of a $d^2-1$-dimensional unit sphere~\cite{Byrd2003}. It holds for all elements of $S^{d^2-1}$ that the angle between $\hat{\bm{r}}$ and $\hat{\bm{s}}$ is smaller than the sum of the angles between $\hat{\bm{r}}$, $\hat{\bm{q}}$ and $\hat{\bm{s}}$, $\hat{\bm{q}}$. 

To prove Eq.~\eqref{eq:qsl_gba},
we consider the infinitesimal propagation between $t$ and $t+dt$
\begin{equation}
    \rho_{t+dt}=U_{t+dt,t} \ \rho_t \ U_{t+dt,t}^\dagger,
\end{equation}
where $U_{t_2,t_1}$ is the unitary that maps $\rho_{t_1}$ to $\rho_{t_2}$. We expand $U_{t+dt,t}$ up to the second order in $dt$
to obtain
\begin{equation}
\begin{split}
    \rho_{t+dt} = &\rho_t -i [H_t,\rho_t]dt -\frac{i}{2} [\partial_t H_t,\rho_t]dt^2- \{H_t^2,\rho_t\}\frac{dt^2}{2} +\\
    &+H_t\rho_t H_t dt^2 + \mathcal{O}(dt^3),
\end{split}
\end{equation}
where we set $\hbar \equiv 1$ for simplicity.
We then consider the inequality $\Theta(\rho,\sigma)\leq \int_0^\tau \Theta(\rho_t,\rho_{t+dt})$, that holds for any Hamiltonian $H_t$, where the equality may hold only when $H_t$ is the optimal Hamiltonian.
We calculate the infinitesimal distance
\begin{equation}
    \Theta(\rho_t,\rho_{t+dt}) = \arccos \bigg(1-\frac{\tr[\rho_t^2 H_t^2]-\tr[(\rho_t H_t)^2]}{\tr[\rho_t^2]-1/d}dt^2\bigg),
\end{equation}
and expand $\arccos(1-c) = \sqrt{2c}+\mathcal{O}(c)$ for small $c>0$, obtaining $\int_0^\tau\Theta(\rho_t,\rho_{t+dt}) = \tau Q_\Theta $, which leads to $\Theta(\rho,\sigma)\leq \tau Q_\Theta$, and thus to Eq.~\eqref{eq:qsl_gba}, after reintroducing $\hbar$. \hfill\qedsymbol
\end{proofbox}
\noindent
Before studying the performance and tightness of bound $T_\Theta$  in Sec.~\ref{s:performance_qsl_mixed}, let us make a few fundamental remarks.

\begin{remark}
\label{r:invariance_under_rescaling}
Bound~\eqref{eq:qsl_gba} is invariant under rescaling of the distance~\eqref{eq:gba} by a positive unitary invariant factor, thus, the arc length $\lVert\bm{r}\rVert_2 \:\Theta(\rho,\sigma)$ induces the same bound on the minimal time.
\end{remark}
\noindent
The proof of this remark is trivial, and follows directly from the fact that the purity is invariant under unitary transformations. Since the purity $\mathcal{P}(\rho)=\tr[\rho^2]$ is directly related with the length of the GBVs via,
\begin{equation}
    \label{eq:purity_gba}
    \lVert \bm{r} \rVert_2 = \sqrt{\frac{d\:\tr[\rho^2]-1}{d-1}}, 
\end{equation}
constant purity implies constant radius. 

For pure states we would like Eq.~\eqref{eq:qsl_gba} to reduce to the unified bound~\eqref{eq:qsl_pfifer}, obtained from the Fubini-Study metric. However, bound~\eqref{eq:qsl_gba} satisfies this requirement only for qubits:
\begin{remark}
\label{r:pure_limit}
Bound~\eqref{eq:qsl_gba} does not reduce to the QSL induced by the Fubini-Study metric for pure states, except for the case of a single two-level system ($d=2$).
\end{remark}

\begin{figure}[htbp]
    \centering
    \includegraphics[width=0.75\textwidth]{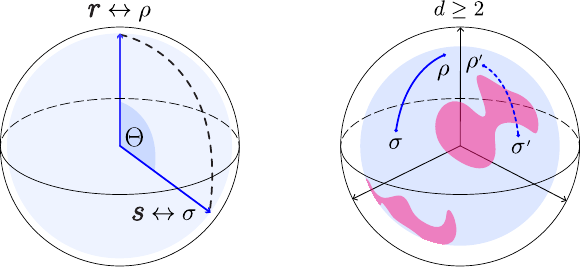}
    \caption{(\emph{Left}) The generalised Bloch angle $\Theta$ is a measure of distinguishability between two states $\rho \leftrightarrow \bm{r}$ and $\sigma \leftrightarrow \bm{s}$, that we can use as a distance to derive the QSL $T_\Theta$ for unitary evolution of density operators, given in Eq.~\eqref{eq:qsl_gba}. The arc length corresponding to $\Theta$ is also a suitable distance, and induces the same bound. (\emph{Right}) For Hilbert space dimension $d>2$, the group of unitary operations $SU(d)/\mathbb{Z}_d$ is only a subgroup of the rotations $SO(d^2-1)$. In this cartoon, the hypersphere embedding the generalised Bloch sphere is projected onto a 3-dimensional vector space. The magenta shaded areas, on the surface of a shell of fixed purity, represent regions of the hypersphere that are not associated with states. Unitary evolution avoids these \emph{forbidden regions}, while some rotations might go straight through them, underestimating the distance between the considered states.}
    \label{fig:gba_summary}
\end{figure}

\begin{proofbox}{Proof of Remark~\ref*{r:pure_limit}}{proof:pure_limit}
Since bound $T_\Theta$ expressed in Eq.~\eqref{eq:qsl_gba} is clearly different from the QSL for pure states, we show that they coincide when $d=2$. 
First, we show that $\Theta$ coincides with the Fubini-Study distance for the case of 2-dimensional systems.
Let $\rho = \ketbra{a}{a}$ and $\sigma=\ketbra{b}{b}$ for two pure qubit states $\ketbra{a}{b}$, with associated Bloch vectors $\bm{a}, \bm{b}$. Let us fix the basis such that $\ket{a}=\ket{(\varphi,\theta)}$ where $\ket{(\varphi,\theta)}$ relative to Bloch vector $\bm{a} = (\cos\varphi\sin\theta,\sin\varphi\sin\theta,\cos\theta)$ with $\varphi\in[0,2\pi], \; \theta\in[0,\pi]$, and where $\ket{b}=\ket{(0,0)}$ is aligned with the $\hat{z}$-axis. State $\ket{a} = \cos (\theta/2) \ket{b} + e^{i\varphi} \sin(\theta/2)\ket{\bar{b}}$, where $\ket{\bar{b}} = \ket{0,\pi}$ is orthogonal to $\ket{b}$. The Fubini-Study distance $d(\ket{a},\ket{b}) = \arccos{|\braket{a|b}|}=\theta/2$, while $\Theta(\rho,\sigma) = \theta$, thus the two distances are identical up to a factor of $1/2$. For pure qubit ($d=2$) states $Q_\Theta = 2 \Delta E$, since $\tr[\rho^2]=1$, therefore $T_\Theta =  \theta/\Delta E$. Hence, $T_\Theta$ and the Mandelstam-Tamm bound coincide for qubits. For dimension $d>2$, $Q_\Theta$ reduces neither to the standard deviation, nor to the average energy, while $\Theta$ does not become the Fubini-Study distance. \hfill\qedsymbol
\end{proofbox}
\noindent
The reason why $\Theta$ does not conform with the Fubini-Study distance for pure states of arbitrary dimension is due to the fact that, in general, the group of rotations on GBVs does not correspond to the unitary group on the subset of points corresponding to states\footnote{In the exceptional case of $d=2$, however, $\Theta$ does reduce to the Fubini-Study distance, since the set of all Bloch vectors forms a 2-sphere, i.e., $SO(3)\sim SU(2)/\mathbb{Z}_2$}. To be specific, the group of \emph{relevant} unitary operations $SU(d)/\mathbb{Z}_d$ is a subgroup of $SO(d^2-1)$ that does not correspond to generic rotations, but to those rotations that preserve $d-1$ independent functions $\tr[\rho^k]$ for $k=1$ to $d$ \cite{Bengtsson2008}. When going from initial to final state, unitary evolution avoids the \emph{forbidden regions} of this $d^2-1$-dimensional hypersphere that do not represent states, while some rotations might go straight through these regions, thus, underestimating the distance between the considered states, as shown in Fig.~\ref{fig:gba_summary}.

\section{An angle that reduces to the Fubini-Study distance}
In order to derive a speed limit that conforms with the QSL for pure states regardless of the dimension of the system, we introduce another distance on isospectral states,
\begin{gather}
\label{eq:Theta_tilde}
    \Phi(\rho,\sigma):= \arccos \Bigg(\sqrt{\frac{\mathrm{tr}[ \rho \sigma]}{\mathrm{tr} [\rho^2]}}\Bigg),
\end{gather}
which reduces to the Fubini-Study distance for the case of pure states.
As done with $\Theta$, we derive a bound on the speed of unitary evolution from this distance.
\begin{theorem}
\label{th:qsl_tilde}
The minimal time $\tau$ required to evolve from state $\rho$ to state $\sigma$ by means of a unitary operation generated by the Hamiltonian $H_t$ is bounded from below by 
\begin{gather}
    \label{eq:qsl_tilde}
    \begin{split}
    & T_{\Phi} (\rho,\sigma) = \hbar\:  \frac{\Phi(\rho,\sigma)}{ Q_\Phi},\; \textrm{where} \\
    & Q_\Phi = \frac{1}{\tau}\int_0^\tau dt \sqrt{\frac{\textrm{\emph{\tr}}[\rho_t^2H_t^2-(\rho_t H_t)^2]}{\textrm{\emph{\tr}}[\rho_t^2]}}.
    \end{split}
\end{gather}
\end{theorem}
\noindent 
The proof of Thm.~\ref{th:qsl_tilde} can be carried out using arguments similar to those for the proof of Thm.~\ref{th:qsl_gba}, and can be found in Sec.~\ref{s:proof_phi} of the Appendix.
Remarkably, the bound expressed in Eq.~\eqref{eq:qsl_tilde} reduces to the Mandelstam-Tamm bound for pure states, since $\Phi$ reduces to the Fubini-Study distance and $Q_\Phi$ reduces to $\Delta E$. To see this, note that the quantity {$\tr[(\rho_t H_t)^2]=\bra{a} H_t \ketbra{a}{a} H_t\ket{a} = |\tr(\rho_t H_t)|^2$} for {$\rho=\ketbra{a}{a}$}.

In contrast to bound~\eqref{eq:QSL_mixed}, the two QSLs derived here account for both the energetics of the dynamics and the purity of the driven state. The latter is accounted for by the denominators of $Q_\Theta$ and $Q_\Phi$, while the term  $\sqrt{\tr[\rho_t^2 H_t^2-(\rho_t H_t)^2]}$ in their numerators is a lower bound on the instantaneous standard deviation of the Hamiltonian $H_t$~\cite{Modi2016}. If states of different purity were considered, neither $\Theta$ nor $\Phi$ would be distances, since the symmetry and triangle inequality properties would be lost.

It is worth highlighting that the bounds derived from $\Theta$ and $\Phi$ are significantly easier to compute than the one expressed in Eq.~\eqref{eq:QSL_mixed} for the case of mixed states, since no square root of density operators needs to be calculated, and thus no eigenvalue problem needs to be solved.
More specifically, in order to compute the Bures angle one needs to perform two matrix multiplications and two matrix square roots\footnote{Optimal methods for matrix multiplication between two $d \times d$ matrices require $\mathcal{O}(d^{2.373})$ operations, whereas the evaluation of the square root of such a matrix requires $\mathcal{O}(d^3)$ operations~\cite{Frommer2010,Davie2013}.}, whereas only two matrix multiplications are needed to compute $\Theta$ or $\Phi$.
Accordingly, distances $\Theta$ and $\Phi$ can be experimentally estimated more efficiently than the Bures angle. The latter involves the evaluation of the root fidelity between the two considered states, harder to obtain than their overlap, which can be determined by means of a controlled-swap circuit\footnote{A similar experimental set up could be used to evaluate the bound derived by authors in Ref.~\cite{Mondal2016b}.}~\cite{Ekert2002,Keyl2001}. Finally, not only are our bounds simpler to compute and measure, they also outperform  Eq.~\eqref{eq:QSL_mixed}, as we will show next.

\section{Performance of the new bounds} 
\label{s:performance_qsl_mixed}
\begin{figure}[tb]
\centering
\includegraphics[width=0.8\textwidth]{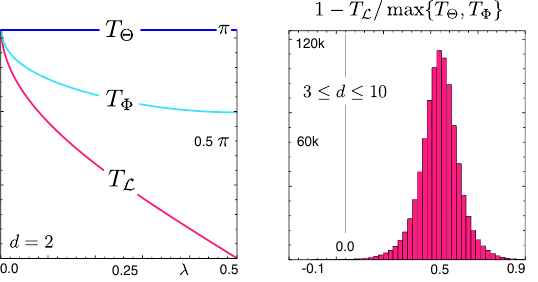}
\caption{(\emph{Left}) Bounds $T_{\mathcal{L}}$, $T_\Phi$, and $T_\Theta$, as a function of the eigenvalue $\lambda\in[0,1/2)$, for two mixed and antipodal ($\theta=\pi/2$) qubit states $\rho$ and $\sigma$, given in Eqs.~\eqref{eq:qubit_rho} and~\eqref{eq:qubit_sigma}. The unitary evolution is generated by the Hamiltonian of Eq.~\eqref{eq:optimal_qubit_hamiltonian}. The same hierarchy holds for non-antipodal mixed qubit states.
(\emph{Right}) Evaluation of $1-T_{\mathcal{L}}/\max[T_\Theta,T_\Phi]$ as a measure of the tightness of the new bounds, for $3\leq d\leq 10$, with a sample size of $10^6$ uniformly sampled random states and Hamiltonians. The traditional QSL $T_{\mathcal{L}}$ is found to be larger than $\max[T_\Theta,T_\Phi]$ for less than 0.1 \% of cases, and only with a difference of less than 1\% from the largest of the other bounds. 
}
\label{fig:qubit_uniform}
\end{figure}

We now study the bounds presented in Eqs.~\eqref{eq:qsl_gba} and~\eqref{eq:qsl_tilde}, and compare them to that in Eq.~\eqref{eq:QSL_mixed} for a given orbit defined by initial and final states $\rho$ and $\sigma$, and a Hamiltonian $H$ that connects them. For the case of mixed qubits we calculate all three bounds analytically. We consider 
\begin{align}
    \label{eq:qubit_rho}
    &\rho=\lambda \ketbra{r_1}{r_1} + (1-\lambda)\ketbra{r_2}{r_2}, \\
    \label{eq:optimal_qubit_hamiltonian}
    &H=e^{i\varphi} \ketbra{r_1}{r_2} + h.c.,
\end{align}
as the initial state and driving Hamiltonian, respectively, where $\varphi\in[0,2\pi]$ is a phase. This Hamiltonian generates the optimal unitary evolution for any choice of final state
\begin{equation}
    \label{eq:qubit_sigma}
    \sigma=\lambda \ketbra{s_1}{s_1} + (1-\lambda)\ketbra{s_2}{s_2},
\end{equation} 
for $\ket{s_1}=\cos\theta\ket{r_1} +e^{i\varphi}\sin\theta\ket{r_2}$.
Since $H$ is time-independent, the denominator of each of the considered bounds can be directly calculated without performing the time-average. For example, for bound $T_\mathcal{L}$, $\Delta E = \overline{\Delta E}$, since the unitary $U_t=\exp[-i H t]$ commutes with the Hamiltonian $H$.
Accordingly, in natural units ($\hbar \equiv 1$),
the bounds read
\begingroup
\allowdisplaybreaks
\begin{align}
\label{eq:analytic_Theta}
&T_\Theta(\rho,\sigma)=\theta, \\
\label{eq:analytic_Phi}
&T_\Phi(\rho,\sigma)=\arccos\bigg({\sqrt{\frac{1+k^2\cos2\theta}{1-k^2}}}\;\bigg)\sqrt{\frac{1-k^2}{2 k^2}}, \\
\label{eq:analytic_bures}
&T_{\mathcal{L}}(\rho,\sigma) = \arccos \bigg( F_+(\theta,\lambda)+F_-(\theta,\lambda) \bigg),
\end{align}
\endgroup
where $F_\pm(\theta,\lambda)=\frac{1}{2}\sqrt{1+k^2c_{2\theta} \pm 2kc_{\theta}\sqrt{1-k^2s_\theta^2}}$, with $c_x = \cos x$, $s_x = \sin x$, and $k=1-2\lambda$.
Note that these bounds are independent of the relative phase $\varphi$, as we expect, and only depend on the distance $\theta=d(\ket{r_1}, \ket{s_1})$ between the basis elements, and on the value of $\lambda$. Bound $T_\Theta$ is tight and attainable and does not depend on the spectrum. A simple plot of the bounds shows that $T_\Theta \geq T_\Phi \geq T_{\mathcal{L}}$ (see Fig.~\ref{fig:qubit_uniform}). The three bounds coincide for pure states $\lambda=0,1$ and for the trivial case of $\theta=0$. This results validates exactly the argument that we made in Sec.~\ref{s:attainable_qsl} with the aim of tightening the QSL for unitary evolution of mixed states. It also resolves the conceptual issue with the traditional bounds, which seem to suggest that noise increases the speed of evolution. On the contrary, Eq.~\eqref{eq:analytic_Theta} provides analytical evidence that mixing does not increase the speed of evolution when unitary driving is considered, in accordance with the results presented in Ref.~\cite{Marvian2016}.

\begin{figure}[hb]
    \centering
    \includegraphics[width=0.9\textwidth]{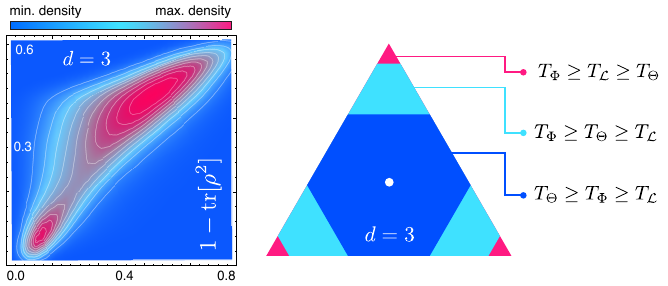}
    \caption{
    (\emph{Left})
    Density plot of the numerical estimation of $1-T_{\mathcal{L}}/\max[T_\Theta,T_\Phi]$ along the \emph{impurity} $1-\tr[\rho^2]$, for $d=3$ (qutrits), with sample size $10^5$. The sampled states have been forced to be distributed approximately uniformly along the purity: in this way it is possible to notice how bounds $T_\Phi$ and $T_{\mathcal{L}}$ coincide for pure states (bottom left), and differ for increasingly mixed states (top right).
    (\emph{Right}) For $d=3$, the hierarchy between the three bounds can be expressed with three regions of the polytope defined by the spectrum $\{\lambda_1, \lambda_2,\lambda_3\}$ of states $\rho$ and $\sigma$, as indicated in the legend. The corners of the triangle represent pure states, while its centre represent the maximally mixed state. The exact shape of the regions represented here reflects a specific choice of $H$, $\rho$ and $\sigma$, but similar features are common to those of any pair of states. For the case of qutrits, $T_{\mathcal{L}}$ is never larger than $\max[T_\Theta,T_\Phi]$ (see Sec.~\ref{s:qutrits} of the Appendix for more information).}
    \label{fig:uniform_qudits}
\end{figure}

In the general case of higher dimensions, we study the tightness of bounds $T_\Theta$ and $T_\Phi$ numerically (See details in {\fontfamily{phv}\selectfont\textbf{Box~\ref{tech:qsl_tightness}}}). To do so, we calculate and compare the three bounds along the same orbit, for a large number of choices of initial and final states $\rho$ and $\sigma$, and driving Hamiltonians $H$. As shown in Fig.~\ref{fig:uniform_qudits}, $T_{\mathcal{L}}$ is larger than $\max[T_\Theta,T_\Phi]$ for a negligible fraction of the sampled states, and only with a difference of less than 1\% from the largest of the new bounds, while the new bounds are found to strongly outperform the traditional one for the vast majority of the cases. However, there could be some exceptional regions where the latter is larger than the new bounds, such as along degenerate subspaces which form a subset of measure zero for the considered ensembles. In the absence of an analytic proof of strict hierarchy between these bounds, we cast our main result in the form of a unified bound 
\begin{gather}
\label{eq:unified}
\begin{split}
    T_{QSL}(\rho,\sigma) = \max \big\{
    T_{\mathcal{L}},T_\Theta,T_\Phi
\big\},
\end{split}
\end{gather}
where $T_{\mathcal{L}}$, $T_\Theta$ and $T_\Phi$ are given by Eqs.~\eqref{eq:QSL_mixed},~\eqref{eq:qsl_gba} and~\eqref{eq:qsl_tilde}, respectively.

\begin{techbox}{Numerical estimation QSL performance}{tech:qsl_tightness}
To test the performance of the considered QSLs numerically we generated uniformly sampled random states and Hamiltonians, from which we estimate the parameter $1-T_{\mathcal{L}}/\max[T_\Theta,T_\Phi]$, as a measure of the tightness of the new bounds. This quantity is positive if the new bounds are tighter than the traditional one. The states are sampled uniformly according to the Bures ensemble~\cite{Miszczak2012},
\begin{equation}
\label{eq:bures_sample}
    \rho = \frac{(\mathbb{1}+U)AA^\dagger(\mathbb{1}+U^\dagger)}{\tr[(\mathbb{1}+U)AA^\dagger(\mathbb{1}+U^\dagger)]},
\end{equation}
where $A$ is a random complex matrix from the Ginebre ensemble, and $U$ is a Haar-random unitary.
Hamiltonians are obtained from random unitaries
\begin{equation}
    \label{eq:haar_hamiltonians}
    H = i \log U,
\end{equation}
where $U$ is sampled uniformly according to the Haar measure. In particular, operators $A$ are complex square matrices, whose elements are given by random complex numbers $z=x+iy$, with $x$ and $y$ being random real number uniformly sample over the \textit{normal distribution} with $\mu=0$ and $\sigma=1$, while operators $U$ are obtained from the orthogonalisation of random matrices $A$.

We considered dimension $3\leq d\leq 10$, with a sample size of $10^6$ random states and Hamiltonians.
The bounds $T_\Phi$ and $T_{\mathcal{L}}$ coincide for pure states (as analytically shown above), and the difference between $\max[T_\Theta, T_\Phi]$ and $T_{\mathcal{L}}$ grows with decreasing purity, as shown in the left inset of Fig.~\ref{fig:uniform_qudits}.
\end{techbox}

\section{Chapter summary}
In this chapter, we have addressed the problem of attainability of quantum speed limits for the unitary evolution of mixed states. We first showed that the conventional bound given in Eq.~\eqref{eq:QSL_mixed} is not generally tight for mixed states, because the Bures distance, defined as the minimal Fubini-Study distance on a dilated space, corresponds to the length of a path that is, in general, not a unitary orbit in the reduced space. It is thus not a suitable choice of distance under the assumption of unitary evolution.

We have introduced and discussed two new distances between those elements of state space with the same spectrum, i.e., those that can be unitarily connected, and derived the corresponding QSLs. The first distance coincides with the angle between the generalised Bloch vectors of the states, and induces a tight and attainable speed limit for the case of mixed qubit states, but does not reduce to the unified bound in Eq.~\eqref{eq:qsl_pfifer} for pure states of arbitrary dimension. The second distance is designed to conform with the Fubini-Study distance for the case of pure states, while being as similar as possible to the generalised Bloch angle. These bounds arise from the properties of state space, when mixed states are represented as generalised Bloch vectors, thus providing a simple geometric interpretation. We have shown that the bounds obtained by these two distances are tighter than the conventional QSL given in Eq.~\eqref{eq:QSL_mixed} for the vast majority of states. Moreover, they are always easier to compute, as well as easier to measure experimentally. 

Beyond its fundamental relevance, our result provides a tighter, and hence more accurate bound on the rate of work extraction, information transfer and processing in the presence of classical uncertainty. For instance, in Ch.~\ref{ch:quantum_batteries} this improved bound will be used to tighten the results of Ref.~\cite{Campaioli2017} on the limits of charging power of cyclic unitary operations. As already mentioned in Sec.~\ref{s:performance_qsl_mixed}, this result solves the issue with traditional QSL in the presence of noise, which wrongly suggest that, in order to speed up the unitary evolution between states, one could simply add noise, reducing the purity of the considered states, with the effect of reducing the time required to evolve between them. This paradoxical situation is now systematically ruled out by our new bound, which demonstrates that even in the proximity of maximally mixed states the time required to perform any unitary evolution is finite, and comparable to the time required to perform the evolution between pure states. Finally, these QSLs constitute a novel and strong benchmark to assess the quality of unitary driving protocols, as discussed in Ch.~\ref{ch:fast_state_preparation} and Ref.~\cite{Campaioli2019}, where they are directly used to estimate the optimality of the numerical solutions to state-preparation problems.

There is a natural trade-off between the tightness of a QSL and its computational complexity. The ideas presented in this chapter represent a step towards finding a distance based on the explicit geometric structure of the space of states. Such a distance would allow for the derivation of a QSL that is guaranteed to be tight.
Another fundamental perspective is to apply the ideas developed here for the case of non-unitary dynamics. This generalisation requires modifying the proposed distances such that they accommodate changes in purity, and allow for considering arbitrary pairs of states. In the next chapter we present the results that we have obtained extending this geometric QSL to the case of arbitrary quantum processes.

\chapter[Quantum speed limits for general processes]{Quantum speed limits for general processes}
\label{ch:qsl_open} 

\begin{ch_abs}
\textsf{\textbf{Chapter abstract}}\vspace{5pt}\\
\small\textsf{By allowing for open dynamics, the minimal time required to evolve between two states can be drastically shorter than for unitary evolution. With the geometric approach used in the previous chapter we derive a quantum speed limit for arbitrary open quantum evolution, that serves as a fundamental bound on the time taken for the most general quantum process. 
This bound is provably robust under composition and mixing, features that largely improve the effectiveness of QSLs for open evolution. It is also easier to compute and measure than other bounds, while being tighter than them for almost all open processes. \vspace{10pt}\\
\emph{This chapter is based on publication \cite{Campaioli2018b}.}}
\end{ch_abs}\vspace{20pt}

\section[A brief review of QSLs for open quantum evolution]{A brief review of QSLs for open quantum evolution}
\label{s:speed_of_open_quantum_evolution}
In the previous chapter, we studied the bounds on the minimal time of evolution of isolated quantum systems, whose dynamics is prescribed by the Schr\"odinger (or equivalently, von Neumann) equation, and described in terms of unitary operators. In practice, however, systems are typically coupled to some uncontrollable environment, which might influence their dynamics in a non-negligible way. Even the very act of measuring a system's observables, necessary to test the predictions of a dynamical theory, requires coupling between system and measurement apparatus. For these reasons, quantum systems must, in general, be regarded as open~\cite{Breuer2002,Milz2017}. 

While system and environment can be described as closed composite system that evolves unitarily, the inaccessibility of the environment's degrees of freedom often requires the use of dynamical maps, whose evolution can be expressed by master equations~\cite{Milz2017}. These are used to model and characterise physical processes, such as the effect of a thermal bath on two-level systems~\cite{Silaev2014}, the evolution of an atom coupled to a cavity~\cite{Garraway1997,Wallquist2010}, the harvesting and transport of energy~\cite{Skochdopole2011}, as well as the energetic properties of molecules~\cite{Trugler2008}, and are particularly important for studying controlled quantum systems~\cite{Lloyd2001}. For example, the Gorini-Kossakowski-Sudarshan-Lindblad master equation, often known simply as Lindblad equation, provides a general dynamical model for the evolution of a quantum system in the presence of Markovian\footnote{In classical probability theory, a Markov process is one where the state of the system at any time depends conditionally on only the state of the system at the previous time step and not on the remaining history. Technically the Lindblad equation describes CP-divisible evolution~\cite{Taranto2019}. A rigorous definition of the Markov condition for quantum process has recently been introduced in Ref.~\cite{Pollock2018} and can be fully reconciled with its classic notion.} environment, and is widely used in all the area of quantum physics and quantum chemistry~\cite{Milz2017}.  

Accordingly, a satisfactory description of the bounds on the minimal time of evolution of a system must take into account the effect that the environment has on its dynamics, based solely on the information that can be practically accessed. Indeed, it is well known that, even for comparable energetic scales, coupling with an environment can accelerate the system's dynamics, driving the two parties through the space of entangled states~\cite{Giovannetti2004,Binder2015,Campaioli2018}, as well as \emph{freeze} its evolution, via the famous \emph{quantum Zeno effect}~\cite{Itano1990,Pascazio1994,Bernu2008,Maniscalco2008}. 

However, there are several hurdles in the derivation of an effective bound for the minimal time of evolution of open quantum systems. First of all, when a system evolves unitarily it is clear how to quantify the contribution of the Hamiltonian which drives the system, and that which does not. In Ch.~\ref{ch:qsl_mixed}, we saw how the commutator $[H_t,\rho_t]$ is intimately related to a notion of velocity in the space of states. Even if the norm of the Hamiltonian or the generator changes during the evolution, it is always possible to fairly compare the speed of two different evolutions by leveraging the notion of average speed. The latter, for instance, is used in Ch.~\ref{ch:quantum_batteries} to define uniform energetic constraints, and compare the power of classical and quantum work extraction protocols. In the case of open dynamics, instead, it is impossible to directly access information about the velocity of the underlying unitary evolution, and thus we must restrict ourselves to the marginal state of the system, tracing over the environmental degrees of freedom. In this case, care must be taken in comparing the speed of different evolutions, due to the coupling between system and environment, and the unknown environment's energetic structure. 

Another difficulty lies in the choice of a suitable measure of distinguishability, necessary to derive a QSL. In Secs.~\ref{s:attainable_pure_qsl} and~\ref{s:attainable_qsl}, we have seen how the Fubini-Study distance provides the perfect choice for a well-behaved distance for the unitary evolution of pure states, and how even for the unitary evolution of \textit{mixed states} there are many possible choices of distance, some of which induce extremely loose bounds. 
When considering the evolution of open quantum states, we would like our measure of distinguishability to reflect the fact that, under the effect of divisible \textit{noisy channels}, states must become increasingly less distinguishable. Indeed, it has been suggested that a \textit{bona fide} measure of distinguishability between states must be contractive under physical evolutions~\cite{Pires2016,Deffner2017}, given by \textit{completely positive and trace preserving} maps (CPTP)~\cite{Breuer2002,Milz2017}. 

Interestingly, it was thought for a long time that the only CPTP-contractive Riemannian metric on the space of states\footnote{The space of states, i.e., the space of positive, unit trace, Hermitian operators, is indeed a Riemannian manifold~\cite{Bengtsson2008,Pires2016}.} was given by the Fisher information metric~\cite{Cencov1982,Pires2016}. While this is not strictly true, as we will see soon, the first rigorous extensions of the Mandelstam-Tamm QSL to the case of open evolution has been indeed derived in Ref.~\cite{Taddei2013} by Taddei et al. using the quantum Fisher information (QFI),
\begin{align}
    \label{eq:QFI_metri}
    &F_Q(t)=\tr[\rho(t) L^2(t)], \\
    \label{eq:symm_log}
    &\:\dot{\rho}(t) = \frac{1}{2}\{\rho(t),L(t)\},
\end{align}
where $L(t)$ is the symmetric logarithmic derivative operator, implicitly defined by Eq.~\eqref{eq:symm_log}.
There, the authors obtain an implicit lower bound on the evolution time $\tau$ valid for arbitrary physical processes, for which a notion of speed is directly associated with the root of the QFI, averaged over the orbit. While the QFI and its metric assume a central role in quantum information, being deeply linked to many other fundamental figures such as the Bures distance and the quantum relative entropy~\cite{Bengtsson2008}, they don't necessarily have any dynamical relevance, and they are also often hard to evaluate, an obstacle that can severely undermine the efficacy of the QSL that it induces.

Another influential result on QSL for open evolution, which has been presented independently but almost synchronously with the work by Taddei, is given in the work by del Campo et al., in Ref.~\cite{DelCampo2013}. There, the authors derive a QSL for arbitrary quantum evolution from a much simpler measure of distinguishability, the so-called \emph{relative purity},
\begin{equation}
    \label{eq:relative_purity}
   \mathcal{P}(\rho_0|\rho_t) = \frac{\tr[\rho_0\rho_t]}{\tr[\rho_0^2]}.
\end{equation}
The bound derived from this quantity is considerably simpler to evaluate than that of Ref.~\cite{Taddei2013}, it predicts the inaccessibility of the Heisenberg limit under Markovian noise, and discerns different types of noise for general open evolution. However, the bound is also provably loose, another feature that severely diminishes the performance of QSLs. Since the work of Taddei and del Campo, several fundamental discoveries have advanced the knowledge and performance of speed limits for open evolution~\cite{Deffner2013b,Sun2015,Pires2016,Mondal2016}, with both foundational and practical relevance, and the field is still thriving due to the many open questions.

From the present state of the art, it appears evident that, for a QSL to be operationally useful, it needs to be feasible and tight.
The feasibility of a bound is quantified in terms of the computational or experimental resources required to evaluate or measure the bound~\cite{Deffner2017, Campaioli2018}. Bounds that require the evaluation of complicated functions of the states or the generators of the evolution are less feasible, and thus less useful, than otherwise equally performing bounds that are easier to compute or experimentally measure. The distance term in many QSLs requires the square root of the initial and final states, thus the knowledge of the spectrum of the considered states~\cite{Deffner2013, Deffner2013b, Pires2016, Mondal2016}. In contrast, the bounds that only involve the evaluation of the overlap $\tr[\rho\sigma]$~\cite{DelCampo2013, Sun2015, Campaioli2018}, including the one that we introduce in this chapter, are much easier to compute and measure~\cite{Keyl2001, Ekert2002, Mondal2016b}.
Aside from the distance, the other important feature of QSL bounds is the speed, directly linked to the infinitesimal distance and the generator of the evolution, that depends on the orbit of the evolution~\cite{Bengtsson2008, Russell2014a, Russell2017, Mondal2016,Mondal2016b, Pires2016, Campaioli2018}. A common criticism of QSLs is that calculating these bounds becomes as hard as solving the dynamical problem. In Sec.~\ref{s:feasibility}, we overcome this limitation by providing an operational procedure to experimentally evaluate the speed term for any type of process, and go on to discuss the purpose of QSLs in this context.

Finally, the tightness of QSLs represents how precisely they bound the actual minimal time of evolution. As we have seen in the previous chapter, the tightness becomes a problem as soon as we move away from the case of unitary evolution of pure states~\cite{Levitin2009}, which is, in practice, always an idealized description. All the available bounds for the general case of open evolution of mixed states are loose for certain types of dynamics, and in some cases, their performance gets worse as increasingly mixed states are considered.
This looseness is often a consequence of the choice of the distance used to derive the bounds. Different distances result in different speed limits, and a suitable choice that reflects the features of the considered evolution is the key to performance~\cite{Bengtsson2008, Pires2016, Campaioli2018}. 

In this chapter, we directly address these issues, starting from the definition of a suitable distance, and deriving a bound that is provably robust under composition and mixing, vastly improving the effectiveness of QSLs. In this sense, our results strongly complement the findings of Ref.~\cite{Pires2016}, where the authors used geometric arguments to obtain an infinite family of distances and their corresponding QSL bounds. While their result firmly and rigorously establishes the mathematical framework for a certain class of QSLs, it leaves open the task of \emph{choosing} a distance that leads to a QSL that is tight and feasible. We do exactly this by uncovering a distance measure on quantum states, which is based on the geometry of the space of density operators, leading to a QSL that is both tight and feasible for almost all states and processes.

\section{Euclidean metric as distance between states}
\label{s:euclidean_open}

Let us consider a $d$-dimensional system $S$, where $d=\dim\mathcal{H}_S$, coupled to its environment $E$, with total Hilbert space $\mathcal{H} = \mathcal{H}_S\otimes\mathcal{H}_E$, and denote its physical state space of positive, unit trace density operators by $\mathcal{S}(\mathcal{H}_S)$. Following the geometric approach used in the previous chapter, we represent states $\rho\in\mathcal{S}(\mathcal{H}_S)$ of the system using their GBVs $\bm{r}$, as prescribed by Eq.~\eqref{eq:rho_bloch}. We would like to measure the distance between two states $\rho \leftrightarrow \bm{r}$ and $\sigma \leftrightarrow \bm{s}$ using the length of the shortest path through $\mathcal{S}(\mathcal{H}_S)$ that connects $\rho$ and $\sigma$. However, solving this geodesic problem is, in general, a hard task, since the state space for $d>2$ is a \emph{complicated} subset of the $(d^2-1)$-ball~\cite{Byrd2003,Bengtsson2008}. Our approach will be to simplify this problem by lower bounding this distance by the length of the well-known geodesics of this ball.
\begin{figure}[tb]
    \centering
    \includegraphics[width=0.8\textwidth]{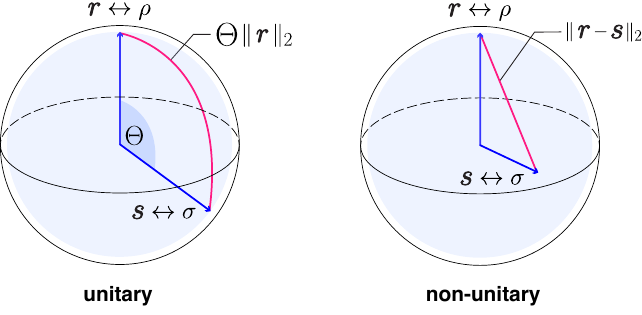}
    \caption{Mixed states $\rho$ and $\sigma$ are represented by their generalized Bloch vectors $\bm{r}$ and $\bm{s}$. In order to simplify the evaluation of the distance between states we approximate the state space with such $(d^2-1)$-dimensional ball. (\emph{Left}) For the case of unitary evolution we choose to measure the distance between $\rho$ and $\sigma$ as the length of the arc of great circle that connects $\bm{r}$ and $\bm{s}$, given by the product between the GBA $\Theta$ between the two vectors, and their length $\lVert \bm{r}\rVert_2$. As seen in Remark~\ref{r:invariance_under_rescaling}, this is equivalent to measuring the angle $\Theta$, since the length of the GBV is invariant under unitary operations. (\emph{Right}) For the case of arbitrary evolution, we use, instead, the norm of the displacement vector $\bm{r}-\bm{s}$, since the length of the GBV is allowed to change. The shortest path that connects the two states is given by the straight line between the two GBVs.}
    \label{fig:distances}
\end{figure}

With respect to this ball, the natural choice for the geodesic is given by arcs of great circles, if we are confined to surfaces of equal radius, like for the case of unitary evolution. Instead, for open dynamics, the natural choice for the geodesic is with respect to the Euclidean distance, which is just the straight line between $\bm{r}$ and $\bm{s}$, as depicted in Fig.~\ref{fig:distances}, whose length is given by 
\begin{gather}
    \label{eq:distance}
    D(\rho,\sigma)=\lVert \bm{r}-\bm{s}\rVert_2.
\end{gather}
From this distance we derive our speed limit, following the geometric approach outlined in Sec.~\ref{s:traditional_unified_qsl}, and used in Refs.~\cite{Pires2016,Campaioli2018}, and other QSL derivations. 

By definition, the distance is always smaller than or equal to the length of any path $\gamma$ associated with some dynamics $\rho_t$, that connects $\rho=\rho_0$ and $\sigma=\rho_\tau$.
We evaluate the infinitesimal distance $D(\rho_{t+dt},\rho_t)$ and rearrange to obtain 
\begin{equation}
    \tau \geq \frac{\lVert \bm{r}-\bm{s}\rVert_2}{\overline{\lVert \dot{\bm{r}}_t\rVert}_2},
\end{equation}
where $\overline{f(t)}=\int_0^\tau dt\: f(t)/\tau$, as usual, stands for the average of $f(t)$ along the orbit parameterized by $t\in[0,\tau]$. Expressing $\lVert \bm{r}-\bm{s}\rVert_2$ and $\lVert \dot{\bm{r}}_t\rVert_2$ in terms of $\rho$ and $\sigma$,
we obtain the bound $\tau\geq T_{D}$. Note that from here on we will consider natural units and set $\hbar\equiv 1$, unless specified otherwise.
\begin{theorem}
\label{th:euclidean_qsl}
The minimal time $\tau$ required to evolve from state $\rho$ to state $\sigma$ by means of general quantum evolution is bounded from below by 
\begin{equation}
    \label{eq:speed_limit_arbitrary}
    T_{D}(\rho,\sigma) := \frac{\lVert \rho - \sigma \rVert}{\overline{ \lVert \dot{\rho}_t\rVert}}.
\end{equation}
\end{theorem}
\noindent
Note how the Hilbert-Schmidt norm, $\lVert X \rVert = \sqrt{\tr[X^\dagger X]}$ for an operator $X$, arises in the form of bound $T_D$ as a consequence of equipping the space of GBVs with the standard Euclidean norm~\cite{Bengtsson2008}.

\begin{proofbox}{Proof of Theorem~\ref*{th:euclidean_qsl}}{proof:euclidean_qsl}
Given two states $\rho$, $\sigma\in\mathcal{S}(\mathcal{H}_S)$ of the system, with associated generalized Bloch vectors $\bm{r}$, $\bm{s}$, respectively, the function $D(\rho,\sigma)=\lVert \bm{r}-\bm{s}\rVert_2$ expressed in Eq.~\eqref{eq:distance} is clearly a distance, as it is the Euclidean norm of the displacement vector $\bm{r}-\bm{s}$~\cite{Bengtsson2008}. This distance can be expressed as a function of the dimension $d$ of the system and of the density matrices $\rho$ and $\sigma$, remembering that
\begingroup
\allowdisplaybreaks
\begin{align}
    \label{eq:euclidean_hilbertschmidt}
    \tr[(\rho-\sigma)^2] & = \tr\Big[\Big(\frac{c}{d} \sum_a (r_a-s_a)\Lambda_a\Big)^2\Big] \\
    & = \frac{d(d-1)}{2d^2}\sum_{a,b}(r_a -s_a)(r_b-s_b)\tr[\Lambda_a\Lambda_b], \\
    & =  \frac{d-1}{2d}\sum_{a,b}(r_a -s_a)(r_b-s_b) 2\delta_{ab}, \\
    & =  \frac{d-1}{d}\sum_{a}(r_a -s_a)^2, \\
    & =  \frac{d-1}{d}\lVert \bm{r} - \bm{s} \rVert_2^2.
\end{align}
\endgroup
Recalling that $\lVert \rho \rVert  = \sqrt{\tr[\rho^\dagger\rho]} = \sqrt{\tr[\rho^2]}$, we obtain
\begin{gather}
    D(\rho,\sigma) = \sqrt{\frac{d}{d-1}}\lVert \rho -\sigma \rVert .
\end{gather}
The proof for the QSL bound of Eq.~\eqref{eq:speed_limit_arbitrary} is carried out as follows: Consider a parametric curve $\gamma(s):[0,S]\in\mathbb{R}\to\mathbb{R}^N$ for some $N\geq1$, that connects two different points $A=\gamma(0)$ and $B=\gamma(S)$. Let $\lVert \cdot \rVert_\eta$ be some norm, specified by $\eta$, on $\mathbb{R}^N$, which induces the distance $D(A,B)=\lVert A-B\rVert_\eta$. The length of the path $\gamma$ is given by $L[\gamma_A^B]=\int_0^S ds \lVert \dot{\gamma}(s)\rVert_\eta$, where $\dot{\gamma}(s) = d\gamma(s)/ds$. Since $D(A,B)$ is the geodesic distance between $A$ and $B$, any other path between the two points can be either longer or equal, with respect to the chosen distance (associated by the chosen norm), thus $D(A,B)\leq L[\gamma]$. In particular, we choose $D(\rho,\sigma)=\lVert \bm{r}-\bm{s}\rVert_2$ and $\bm{r}(t)$ as the parametric curve generated by some arbitrary process, with $\bm{r}(0)=\bm{r}$, and $\bm{r}(\tau)=\bm{s}$. Accordingly, the length of the curve is given by $L[\gamma] = \int_0^\tau dt \lVert \dot{\bm{r}}(t)\rVert_2$. Finally, we express the length of the tangent vector in terms of the generator of the evolution, following the steps used in Eq.~\eqref{eq:euclidean_hilbertschmidt}, to obtain
\begin{equation}
    \label{eq:speed_euclidean_hilbertschmidt}
    \lVert \dot{\bm{r}}(t)\rVert_2 = \sqrt{\frac{d}{d-1}}\lVert \dot{\rho}_t\rVert,
\end{equation}
and obtain the bound. \hfill \qedsymbol
\end{proofbox}

Despite its surprisingly simple form, reminiscent of kinematic equations, the bound in Eq.~\eqref{eq:speed_limit_arbitrary} originates from a precise geometric approach and encompasses all the fundamental features of previous QSL bounds, including the orbit dependent term $\overline{ \lVert \dot{\rho}_t\rVert}$, which will be referred to as \emph{speed}, or \emph{strength of the generator}, that appears, under various guises, in the bounds of Refs.~\cite{Deffner2013, Deffner2013b, Taddei2013, DelCampo2013, Sun2015, Pires2016, Mondal2016, Mondal2016b}.
We now elaborate on several key properties of our geometric bound, directly addressing the issues with traditional QSLs that have haven been discussed in Sec.~\ref{s:speed_of_open_quantum_evolution}. 

\section{Robustness}
It has been suggested in the literature that a sensible measure of distinguishability should be contractive under physical noisy and divisible maps~\cite{Pires2016}, under which pairs of states are expected to become increasingly harder to discern. As discussed in Sec.~\ref{s:speed_of_open_quantum_evolution}, the QFI metric\footnote{Which is indeed the metric of the Bures metric on the space of mixed quantum states.} was believed to be the only Riemannian metric to be contractive under CPTP maps. In fact, there exists an infinite family of such metrics~\cite{Wootters1981,Braunstein1994} (bounded from above by the QFI metric itself), from which Pires et al. derived a familty of QSLs that can be adapted to the details of the considered dynamics~\cite{Pires2016}. On the contrary, it is well-known that the Hilbert-Schmidt norm, at the core of our bound $T_D$, is generally not contractive for CPTP dynamics~\cite{Perez-Garcia2006}. How does this impact our bound and its performance? We now show that the bound $T_D$ is, in fact, robust under composition with an ancillary system, and mixing with the fixed point of a dynamical map, which often undermine the performance of other bounds.

\subsection{Robustness under composition}
\label{s:robustness_composition}
\noindent
In order to represent physical processes, quantum maps have to be completely positive~\cite{Nielsen2000}. In a nutshell, a map $M$ is complete positive if the composite map with the same effect on the system and no-action on any ancillary system, $M\otimes I$, is also always a positive map, where $I$ here represents the identity map. For the same reason that physical maps must be completely positive, we would like a QSL bound to be invariant under any composition with an ancillary system that does not partake in the evolution. We now prove exactly this invariance.

A non-contractive distance such as $D$, based on the Hilbert-Schmidt norm, will in general change drastically when an ancillary system is introduced trivially. Let us consider the trivial composition of the system with an ancilla in state $\alpha$,
\begin{equation}
    \label{eq:separable_ancilla}
    \begin{split}
    & \rho\to\rho\otimes\alpha \\
    & \sigma\to\sigma\otimes\alpha.
    \end{split}
\end{equation}
The distance between the composite initial and final states becomes
\begin{equation}
    \label{eq:composite_distance}
    \begin{split}
     \| \rho\otimes \alpha-\sigma \otimes\alpha \|  &= \| (\rho-\sigma) \otimes\alpha \|\\
     & = \| \rho-\sigma \| \cdot \|\alpha\|.
     \end{split}
\end{equation}
The term $\|\alpha\|$ in Eq.~\eqref{eq:composite_distance} is the root of the purity of the ancillary system; therefore, the distance decreases by simply introducing an inert ancilla that is not pure~\cite{Piani2012}.
However, a desirable property for a QSL is that if the ancilla does not participate in the dynamics, the minimal time of evolution between $\rho\otimes\alpha$ and $\sigma\otimes\alpha$ should be bounded as that for the evolution between $\rho$ and $\sigma$. Indeed, for dynamics of the form $\rho\otimes\alpha\to\rho_t\otimes\alpha$ we have
\begin{equation}
    \| \partial_t (\rho_t \otimes \alpha) \| = \| \partial_t \rho_t \| \cdot \| \alpha \|,
\end{equation}
which means that numerator and denominator of Eq.~\eqref{eq:speed_limit_arbitrary} are affected by the same factor, and $T_D (\rho \otimes \alpha ,\sigma \otimes \alpha) =T_D (\rho,\sigma)$. This can be summarised by the following remark:
\begin{remark}
\label{r:invariance_under_composition}
Bound~\eqref{eq:speed_limit_arbitrary} is invariant under composition with a system that does not participate in the evolution. 
\end{remark}
\noindent
In general the initial state of the ancilla can be correlated with the system, and be part of the dynamics. It is well known that in such case the evolution between $\rho$ and $\sigma$ can be considerably faster than for trivial composition, thus affecting the QSL, as shown in Ref.~\cite{Giovannetti2004}.

Another type of composition under which distances and QSLs are desired to be robust is that of many copies that undergo the same evolution. When unitary evolution is considered, this scenario is described by a composite state $\rho_{(1)}\otimes\rho_{(2)}\otimes\cdots\otimes\rho_{(N)}$ that evolves under a composite unitary $U_{(1)}\otimes U_{(2)}\otimes\cdots\otimes U_{(N)}$. While the standard deviation of the Hamiltonian grows as $\sqrt{N}$, the geometric distances introduced in Ch.~\ref{ch:qsl_mixed} and in this Chapter do not suggest a similar behavior. The robustness of these geometric QSLs for many copies is still subject of investigation, and the bounds above could become looser as the number of copies increases. Nevertheless, there are bounds that can be used to circumvent this problem, such as the QSL induced by the Bures angle (see Eq.~\eqref{eq:QSL_mixed}), indeed used in Ch.~\ref{ch:quantum_batteries} for the powerful charging of an array of identical systems, and the log-based purity speed limit of Ref.~\cite{Uzdin2016}.

\subsection{Robustness under mixing}
In Sec.~\ref{s:attainable_pure_qsl} we saw how the traditional QSL, which is tight for unitary dynamics of pure states, becomes rather loose for mixed states. The reason for this, as we show in detail in Sec.~\ref{s:attainable_qsl}, is that the Bures distance $\mathcal{L}(\rho,\sigma)$, given in Eq.~\eqref{eq:bures_angle}, decreases rapidly under mixing.

\begin{figure}
    \centering
    \includegraphics[width=\textwidth]{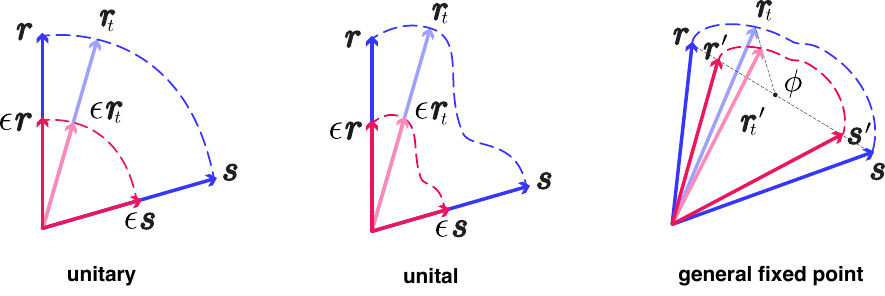}
    \caption{Bound $T_D$, expressed in Eq.~\eqref{eq:speed_limit_arbitrary}, is invariant under general mixing  with the fixed state of the considered dynamics. (\emph{Left}) Invariance under unitary evolution is satisfied for pure depolarisation. (\emph{Center}) More generally, all unital maps, i.e., maps that preserve the identity, also satisfy the condition for invariance under pure depolarisation. In terms of the GBV, the condition for unital invariance simply reduces to $\bm{r}_t'=\epsilon\bm{r}_t$. (\emph{Right}) The bound is also robust for general mixing $\mathcal{M}_\epsilon(\phi)$ for dynamics that preserves the fixed point $\phi$.
    }
    \label{fig:invariance}
\end{figure}

Let us consider the pure depolarisation map
\begin{gather}
\label{eq:depolarization}
    \rho\to\rho'=\mathcal{D}_\epsilon[\rho]:=\epsilon \rho + \frac{1-\epsilon}{d}\mathbb{1},
\end{gather}
with $\epsilon\in[0,1]$, which represents the special case of mixing with the maximally mixed state $\mathbb{1}/d$, i.e., the state with highest von Neumann entropy. When $\epsilon$ tends to 0, the Bures distance $\mathcal{L}(\rho',\sigma')$ vanishes faster than the denominator of the right-hand side of Eq.~\eqref{eq:QSL_mixed}, and so does the corresponding QSL. Now, note that the GBVs of $\rho$ ($\sigma$) and $\rho'$ ($\sigma'$) are $\bm{r}$ ($\bm{s}$) and $\epsilon\bm{r}$ ($\epsilon\bm{s}$) respectively. A unitary transformation that maps $\bm{r}$ to $\bm{s}$ will also map $\epsilon\bm{r}$ to $\epsilon\bm{s}$ in exactly the same time. That is, the value of $\epsilon$ is inconsequential. Based on this observation, we proposed the angle between the GBVs as distance because it is independent of $\epsilon$ and therefore robust under mixing, (see Fig.~\ref{fig:invariance}). This robustness is precisely the reason for the supremacy of the QSLs introduced in Ch.~\ref{ch:qsl_mixed} over that of Eq.~\eqref{eq:QSL_mixed}.

Even when open evolution is considered, it is of primary importance for a QSL to remain effective and tight for increasingly mixed initial and final states. We now show that the bound $T_D$, introduced in Eq.~\eqref{eq:speed_limit_arbitrary}, is robust under mixing not only for unitary dynamics, but also for any open evolution with a well-defined fixed point.
Let the dynamics from $\rho$ to $\sigma$ be due to a completely positive and trace preserving linear map $\mathcal{C}_t$ with fixed state $\phi$, that generates the orbit $\mathcal{C}_t[\rho]=\rho_t$ such that $\mathcal{C}_0(\rho)=\rho$ and $\mathcal{C}_\tau(\rho)=\sigma$.
A generalisation of the pure depolarisation map, of Eq.~\eqref{eq:depolarization}, to general mixing is given by
\begin{equation}
    \label{eq:generali_mixing}
    \mathcal{M}_\epsilon(\phi)[\rho]:=\epsilon\rho+(1-\epsilon)\phi.
\end{equation}
We apply $\mathcal{M}_\epsilon(\phi)$ to initial and final states $\rho$ and $\sigma$,
\begin{equation}
    \begin{split}
        & \rho'=\epsilon \rho + (1-\epsilon) \phi \\
        & \sigma'=\epsilon \sigma + (1-\epsilon) \phi.
    \end{split}
\end{equation}
This map shrinks the numerator of the Eq.~\eqref{eq:speed_limit_arbitrary} by $\epsilon \in [0,1]$,
\begin{equation}
    \|\rho'-\sigma'\| = \epsilon \|\rho-\sigma\|.
\end{equation}
Since the dynamics $\mathcal{C}_t$ preserves $\phi$, we have
\begin{equation}
    \mathcal{C}_t(\rho')=\epsilon\rho_t+(1-\epsilon)\phi,
\end{equation}
therefore, the denominator of Eq.~\eqref{eq:speed_limit_arbitrary} also shrinks by the same amount
\begin{equation}
    \frac{d}{dt}\mathcal{C}_t(\rho')=\epsilon\dot{\rho}_t,
\end{equation}
hence $T_D(\rho',\sigma')=T_D(\rho,\sigma)$, as summarised by the following remark.
\begin{remark}
\label{r:invariance_under_mixing}
Bound~\eqref{eq:speed_limit_arbitrary} is invariant under mixing with the fixed state of the dynamics.
\end{remark}
\noindent
Note how the above result contains the previous case of unitary dynamics, and all unital dynamics, as they preserve the maximally mixed state. In such cases the condition for robustness under mixing simply becomes a condition on the contraction factor for the length of the GBV $\bm{r}'_t=\epsilon \bm{r}_t$, as expressed in Fig.~\ref{fig:invariance}. Remarks~\ref{r:invariance_under_composition} and~\ref{r:invariance_under_mixing} finally answer the question posed in the preamble of this section, and prove that, not only the Hilbert-Schmidt norm can be used to define a \emph{bona fide} QSL for the evolution of open quantum system, but that the bound it naturally induces is also provably robust under composition and mixing. 
In the next section we study the form of the bound, with particular attention to the speed, for the fundamental types of quantum evolution.

\section{Form of the bound}
\label{s:forms}
The numerator of Eq.~\eqref{eq:speed_limit_arbitrary} is independent of the type of dynamics considered for the evolution between a pair of states $\rho$ and $\sigma$. The denominator, instead, depends on the orbit, and its form varies for different types of evolution. We thus study its form and interpret the different notions of speed that arise for different dynamics.
\subsection{Unitary evolution} 
When \emph{unitary evolution} is considered, the denominator of Eq.~\eqref{eq:speed_limit_arbitrary} is a simple function of the time-dependent Hamiltonian $H_t$
\begin{gather}
    \label{eq:unitary_denominator}
     \overline{\lVert \dot{\rho}_t \rVert}  = \overline{\sqrt{2\;\tr[H_t^2\rho_t^2-(H_t\rho_t)^2]}}.
\end{gather}
This term is proportional (up to a constant of motion) to the term in the denominator of the QSLs Eqs.~\eqref{eq:qsl_gba} and~\eqref{eq:qsl_tilde}, derived in Ch.~\ref{ch:qsl_mixed}. Moreover, for pure states, this quantity reduces to the time-averaged standard deviation of the Hamiltonian $H_t$, up to a factor of $\sqrt{2}$. However, the numerator of the QSL in Eq.~\eqref{eq:speed_limit_arbitrary} and the QSL in Eq.~\eqref{eq:qsl_gba} are different and the latter is always tighter. This is because, for this special case, the length of the  GBV $\bm{r}$ must be preserved along the evolution~\cite{Campaioli2018}, and the geodesics are arcs of great circles that connect $\bm{r}$ and $\bm{s}$ (see Fig.~\ref{fig:distances}). 
This observation should not be surprising, since the arc length $\|\bm{r}\|_2\Theta(\rho,\sigma)$ is always greater than the length of the displacement vector $D(\rho,\sigma)$, given in  Eq.~\eqref{eq:distance}. If we are promised that the evolution is unitary, then we are free to work with the tighter QSLs of Eqs.~\eqref{eq:qsl_gba} and~\eqref{eq:qsl_tilde}. However, if that information is not available, we must be conservative and work with the QSL in Eq.~\eqref{eq:speed_limit_arbitrary}.
We now consider the open evolution case, starting with Lindblad dynamics, before proceeding with more general non-Markovian evolutions.

\subsection{Lindblad dynamics} 
In the case of semigroup dynamics, $ \overline{\lVert \dot{\rho}_t \rVert} $ becomes a function of the Lindblad operators~\cite{DelCampo2013}. While this function is generally complicated, we can derive its form for some particular types of Lindblad dynamics, for which it substantially simplifies. Let us consider a general form of the Lindblad master equation 
\begin{equation}
    \label{eq:linblad_master_equation}
    \dot{\rho}_t=-i[H,\rho_t]+\sum_k\gamma_k\big(L_k\rho L_k^\dagger -\frac{1}{2}\{L_k^\dagger L_k,\rho_t\}\big),
\end{equation}
where typically the Lindblad operators are chosen to be orthonormal and traceless, i.e., $\tr[L_k L_l]=\delta_{kl}$, and $\tr[L_k]=0$. If the unitary part of the dynamics is irrelevant with respect to the dissipator, i.e., when $[H,\rho_t]$ is negligible when compared to the other terms, we obtain
\begin{align}
    \label{eq:lindblad_depolarization_denominator}
    \overline{\lVert \dot{\rho}_t \rVert} \leq 2 \sum_k \gamma_k^2 \overline{\lVert L_k \rVert^2},
\end{align}
where the inequality holds since
\begin{equation}\lVert\rho_t\rVert \lVert\dot{\rho}_t\rVert \leq 2\sum_k \gamma_k^2 \lVert L_k\rVert^2 \lVert\rho_t\rVert ^2,
\end{equation}
as shown in Ref.~\cite{Uzdin2016}, and $\lVert \rho_t\rVert \leq 1$. We can readily apply this result to three important cases: Pure dephasing dynamics, pure depolarization dynamics, and speed of purity change.  
\vspace{10pt}

\partitle{Pure dephasing dynamics} This type of dynamics models the idealized evolution of an open quantum system whose coherence decays over time due to the interaction with the environment. Under this kind of dynamics, a quantum system that evolves for a sufficiently long time is expected to lose its quantum mechanical features and exhibit a classical behavior.
Here, for the sake of clarity, we consider the case of \emph{pure dephasing of a single qubit}, described by the Lindblad equation 
\begin{equation}
\label{eq:pure_dephasing_qubit}
    \dot{\rho}_t = \gamma(\sigma_z\rho_t\sigma_z -\rho_t).
\end{equation}
The instantaneous speed can be written in terms of the components of the Bloch vector $\bm{r}_t = (r_1(t),r_2(t),r_3(t))$ associated to $\rho_t$,
\begin{equation}
    \label{eq:insta_speed_pure_dephasing_qubit}
    \lVert \dot{\rho}_t\rVert =\sqrt{2} \;\gamma \sqrt{r_1^2(t)+r_2^2(t)},
\end{equation}
and accordingly the time-averaged speed can be bounded as
\begin{gather}
    \label{eq:dephasing_denominator}
    \overline{\lVert \dot{\rho}_t \rVert}  \leq \sqrt{2}\gamma.
\end{gather}
Although considering the case of a single qubit might sound simplistic, this result can be applied to high-dimensional systems that effectively behave like qubits~\cite{Ilichev2003}, such as highly degenerate systems that can be approximately described as a two-level system.
\vspace{10pt}

\partitle{Pure depolarizing dynamics} Another interesting observation is that the bound $T_D$ is geometrically tight when purely depolarizing dynamics is considered:
\begin{equation}
    \label{eq:pure_depolatisation_dynamics}
    \rho_t = \mathcal{D}_{\epsilon(t)}[\rho_0] = \epsilon(t)\rho_0+\frac{1-\epsilon(t)}{d}\mathbb,
\end{equation}
which serves as an idealized model of noise for the evolution of an open quantum system that monotonically deteriorates towards the state of maximal entropy, i.e., the maximally mixed state. Geometrically, it corresponds to the re-scaling of the generalized Bloch vector $\bm{r}_t = \epsilon(t)\; \bm{r}_0$, where $\epsilon(0)=1$. Tightness is guaranteed by the fact that each vector $\bm{r}_t$ obtained in this way represents a state, along with the fact that the orbit of such evolution is given by the straight line that connects $\bm{r}_0$ to $\bm{r}_\tau$, whose length is exactly given by $D(\rho_0,\rho_\tau)$. In this case our bound reads
\begin{gather}
    T_D(\rho_0,\rho_\tau)=\frac{1-\epsilon(\tau)}{\overline{|\dot{\epsilon}(t)|}}.
\end{gather}
If we restrict ourselves to the case of strictly monotonic contraction (or expansion) of the GBV, the denominator becomes $(1-\epsilon(\tau))/\tau$, which further supports our argument for the tightness of our bound. In this case, it simply returns the condition for optimal evolution $T_D(\rho_0,\rho_\tau)=\tau$, i.e., the evolution time $\tau$ coincides with the bound, and thus with the minimal time.
\vspace{10pt}

\partitle{Speed of purity change} Since a contraction of the GBV corresponds to a decrease of the purity,
\begin{equation}
    \label{eq:purity}
    \mathcal{P}[\rho] =\tr[\rho^2],
\end{equation}
of the initial state, Eq.~\eqref{eq:speed_limit_arbitrary} provides a QSL for the variation in the purity $\Delta \mathcal{P}$, which is saturated when obtained by means of purely depolarizing dynamics with strictly monotonic contraction. In particular, the rescaling of the GBV $\bm{r}\to\epsilon\bm{r}$ implies a variation of the purity
\begin{equation}
    \tr[\rho_0^2]\to \epsilon^2\tr[\rho_0^2]+\frac{(1-\epsilon^2)}{d}\mathbb{1},
\end{equation}
which thus depends also on the dimension $d$ of the system. Similar QSLs have been derived by the authors of Ref.~\cite{Rodriguez-Rosario2011}, who express a bound on the instantaneous variation of the purity in terms of the strength of the interaction Hamiltonian and the properties of the total system-environment density operator, as well as by the authors of Ref.~\cite{Uzdin2016}, who provide a bound on the variation of the purity $\mathcal{P}[\rho_0]/\mathcal{P}[\rho_\tau]$ as a function of the non-unitary part of the evolution, both in Hilbert and Liouville space.

\subsection{Non-Markovian dynamics} 
For the most general non-Markovian dynamics, the denominator of bound~\eqref{eq:speed_limit_arbitrary} can be written in terms of a convolution with a memory kernel~\cite{Breuer2002}, e.g., in the form of the Nakajima-Zwanzig equation
\begin{equation}
    \label{eq:nz_master_equation}
    \dot{\rho}_t = \mathcal{L}_t\rho_t + \int_{t_0}^t ds\mathcal{K}_{t,s}\rho_s +\mathcal{J}_{t,t_0},
\end{equation}
where $\mathcal{L}_t$ is a time-local generator, like that of the Lindblad master equation. The memory kernel $\mathcal{K}_{t,s}$ accounts for the effect of memory, and $\mathcal{J}_{t, t_0}$ accounts for initial correlations between system and environment~\cite{Pollock2017}. The denominator of bound~\eqref{eq:speed_limit_arbitrary} can be simplified using the triangle inequality $\lVert A + B + C\rVert \leq \lVert A\rVert  +\lVert B\rVert +\lVert C\rVert $, at the cost of its tightness. Similarly, the memory kernel can be divided up into a finite sum of terms whenever it is possible to resort to a temporal discretization, in order to obtain the \textit{transfer tensor}, i.e., the exact discrete time memory kernel~\cite{Pollock2018}. In this case, one can express $\lVert\int_{t_0}^t ds\mathcal{K}_{t,s} \rho_s \rVert \sim\delta t\sum_k \lVert \mathcal{K}_{t_{k},t_{k-1}} \rho_{t_{k-1}} \rVert $, again, at the cost of reducing the tightness of the bound. 

Alternatively, the orbit dependent term can always be related to an underlying unitary evolution with a wider environment: $\dot{\rho}_t = -i \ \tr_E [H,\Pi_t]$, where $H$ and $\Pi_t$ are the Hamiltonian and the state of the joint system-environment, respectively. We can further break down the total Hamiltonian into $H = H_S+H_{\textrm{int}} +H_E $, where $H_S$ ($H_E$) is the Hamiltonian of the system (environment) and $H_{\textrm{int}}$ describes the interactions between the two. In this case the denominator of bound~\eqref{eq:speed_limit_arbitrary} reads
\begin{align}
    \label{eq:arbitrary_denominator}
    \overline{\lVert \dot{\rho}_t \rVert}  &= \overline{\lVert -i[H_S,\rho_t]-i\tr_E\{[H_{\textrm{int}}, \Pi_t]\}\rVert},
\end{align}
since $\tr_E\{[H_E, \Pi_t]\} = 0$. A less tight speed limit can be obtained by splitting the right hand side of Eq.~\eqref{eq:arbitrary_denominator}, using the triangle inequality and the linearity of the time average, to obtain $\overline{\lVert \dot{\rho}_t \rVert} \leq \overline{\lVert -i[H_S,\rho_t]\rVert}  + \overline{\lVert-i\tr_E\{[H_{\textrm{int}}, \Pi_t]\}\rVert}$, in order to isolate the contribution of $H_{\textrm{int}}$ from that of $H_S$, when convenient. 

Additionally, by considering the larger Hilbert space of system and environment combined, it is possible to appreciate the difference between the traditional QSL, $T_{\mathcal{L}}(\rho,\sigma)$, expressed in terms of the Bures angle (see Eq.~\eqref{eq:bures_angle}), and the bound $T_D$ of Eq.~\eqref{eq:speed_limit_arbitrary}. The Bures distance $\mathcal{L}(\rho,\sigma)$ corresponds to the minimal Fubini-Study distance between purifications of $\rho$ and $\sigma$ in a larger Hilbert space \cite{Bengtsson2008}, here denoted by $\ketbra{\psi}{\psi}$ and $\ketbra{\varphi}{\varphi}$, respectively. Such purified states must be entangled states of system and environment when $\rho$ and $\sigma$ are mixed. Moreover, unlike in Eq.~\eqref{eq:arbitrary_denominator}, these states may have nothing to do with the actual system-environment evolution. In general, in order to saturate the traditional QSL, one must have access to those (possibly fictional) entangled states, and be able to perform highly non-trivial operations over both system and environment, such as $\ketbra{\psi}{\varphi}+\ketbra{\varphi}{\psi}$, which can contain terms with high order of interaction \cite{Campaioli2017, Campaioli2018}. Since in practice one has little, if any control over the environment degrees of freedom, and nearly no access to the entangled state of the system and environment combined, the traditional QSL rapidly loses its efficacy.

In contrast, bound $T_D$, introduced in Eq.~\eqref{eq:speed_limit_arbitrary}, provides a conservative estimate of the minimal evolution time between two states $\rho$ and $\sigma$, under the assumption of no access to their purification. The speed of the evolution is assessed observing only the local part (the system) of a global evolution (the underlying unitary evolution of system and environment), as expressed by Eq.~\eqref{eq:arbitrary_denominator}, while still allowing for optimal driving of the purifications of $\rho$ and $\sigma$.
In addition to the ability of QSLs to represent an achievable bound for the minimal evolution time, their usefulness also depends on how easily they can be calculated and measured. We discuss this aspect in the next section, comparing the features of our bound $T_D$ to those of other QSLs.

\section{Feasibility}
\label{s:feasibility}
As discussed in Sec.~\ref{s:speed_of_open_quantum_evolution}, the usefulness of a QSL bound is directly linked to the feasibility of its evaluation, whether it be computational or experimental. There are two types of difficulties that one might encounter in the evaluation of a QSL. First, computing the distance, that in our case is given by the orbit-independent term in the numerator of Eq.~\eqref{eq:speed_limit_arbitrary}, and  second, evaluating the speed, that in our case is given by the orbit-dependent term in the denominator of Eq.~\eqref{eq:speed_limit_arbitrary}. While the latter is usually related to some norm of $\dot{\rho}_t$, the former changes remarkably from bound to bound. We address the distance first, before proceeding to a discussion of the speed.\vspace{10pt} 

\partitle{The distance} Among the majority of the QSL bounds known so far one can make a clear-cut distinction between the type of distances that have been used: Either they require evaluating the overlap $\tr[\rho\sigma]$ between the initial and the final states~\cite{Sun2015, DelCampo2013, Campaioli2018}, or they require to calculate $\sqrt{\rho}$ and $\sqrt{\sigma}$ (or similar functions)~\cite{Deffner2013b, Mondal2016, Pires2016}. The latter is much more complicated than the former, as it requires finding the eigenvalues and eigenvectors of $\rho$ and $\sigma$. Moreover, the overlap $\tr[\rho \sigma]$ between two density operators $\rho$ and $\sigma$ is easily measured experimentally using a \mono{CSWAP} (controlled-swap) and measurement on an ancillary system~\cite{Ekert2002}.

The same approach can be used to estimate the fidelity $\mathcal{F}$, given in Eq.~\eqref{eq:fidelity}, and the affinity $\mathcal{A}$ between $\rho$ and $\sigma$,
\begin{equation}
    \label{eq:affinity}
    \mathcal{A}(\rho,\sigma)=\tr[\sqrt{\rho}\sqrt{\sigma}],
\end{equation}
as well as any other measure of distinguishability based on the square roots of the density operators, as described in {\fontfamily{phv}\selectfont\textbf{Box~\ref{tech_measure_fidelity}}}. However, in this case, the number of measurements required to obtain a good estimate of this measure of distinguishability is much larger than for the case of the overlap, with respect to which our distance is conveniently defined. 

\begin{techbox}{Measuring the quantum affinity}{tech_measure_fidelity}
The quantum affinity (see Eq.~\eqref{eq:affinity}) is a measure of distinguishability between two density operators $\rho$ and $\sigma$ that requires the knowledge of the purifications $\sqrt{\rho}$ and $\sqrt{\sigma}$. In Ref.~\cite{Mondal2016} the authors provide a clever, yet experimentally expensive, way of measuring the quantum affinity based on a sequence of \mono{CSWAP} operations. To do so, one has to be able to prepare the states $\rho$ and $\sigma$, and to measure the overlap between any combination $\tr[\rho^k\sigma^l]$, with $k,l\in\{0,1,\cdots,d\}$, for $d$-dimensional systems. Each overlap measurement requires the use of a \mono{CSWAP} operation, as described in Ref.~\cite{Ekert2002}.

Measuring $\tr[\rho^k]$, for $k=2$ to $d$, allows to obtain the eigenvalues of $\rho$. This step requires exactly $d-2$ more \mono{CSWAP} measurement protocols than measuring just $\tr[\rho\sigma]$. 
Assuming to be able to prepare a state in the eigenbasis $\{\ket{r_k}\}_{k=1}^d$ of $\rho$, we can use its spectrum $\{\lambda_k\}_{k=1}^s$ in order to prepare
\begin{equation}
    \label{eq:sqrt_rho}
    \sqrt{\rho}=\frac{\sum_{k=1}^d\sqrt{\lambda_k}\ketbra{r_k}{r_k}}{\sum_{k=1}^d\sqrt{\lambda_k}}.
\end{equation} 
In case of unitary evolution, the final state $\sqrt{\sigma}$, which has the same spectrum as $\sqrt{\rho}$, is obtained from $\sqrt{\rho}\to\sqrt{\sigma} = U\sqrt{\rho}U^\dagger$, and the affinity is obtained directly measuring the overlap $\tr[\sqrt{\rho}\sqrt{\sigma}]$. For non-unitary evolution, instead, one has to directly measure $\tr[\sigma^k]$ for $k=2$ to $d$ in order to reconstruct its spectrum, to then prepare $\sqrt{\sigma}$ and measure its overlap with $\sqrt{\rho}$, for a total of $d-1$ additional measurements.

In conclusion, measuring the quantum affinity between $\rho$ and $\sigma$ requires between $d$ and $2d$ extra preparations and measurements with respect to the overlap $\tr[\rho\sigma]$. A detailed description of how to estimate linear and non-linear functionals of a quantum state can be found in Ref.~\cite{Ekert2002}.
\end{techbox}

\vspace{10pt} 

\partitle{The speed}
We now focus on to the orbit-dependent term  $\overline{\lVert \dot{\rho}_t\rVert}$, i.e., the denominator of Eq.~\eqref{eq:speed_limit_arbitrary}, which appears in different forms in virtually every QSL bound. This term is interpreted as the \textit{speed} of the evolution\footnote{Note that $\dot{\rho}_t$ is proportional to the tangent vector $\dot{\bm{r}}_t$, which can be regarded as the velocity. Accordingly, the norm of the latter is the speed, and it is proportional up to a constant of motion to the HS norm of $\dot{\rho}_t$.}, and it can be hard to compute, as it might require the knowledge of the solution $\rho_t$ to the dynamical problem $\dot{\rho}_t=L[\rho_t]$. For this reason, one might criticize QSLs as being impractical, or ineffective, if too hard to compute. Surely, when QSLs are easy to compute, they can be used to quickly estimate the evolution time $\tau$, required by some specific dynamics $\dot{\rho}_t=L[\rho_t]$ to evolve between $\rho$ and $\sigma$; however, their main purpose is rather to answer the question, \emph{can we evolve faster?} The evaluation of a QSL bound for the initial and final states $\rho$ and $\sigma$, along the orbit described by $\rho_t$, immediately tells us that it might be possible to evolve faster along another orbit that has the same speed, or confirms that we are already on a time-optimal orbit.

Besides, it is not always necessary to solve the dynamics of the system in order to evaluate the speed, which can be constant along the orbit. For example, the standard deviation of any time-independent Hamiltonian is a constant of motion, and can be directly obtained from the initial state of the system and the Hamiltonian, making the speed extremely easy to compute. 
In the more general case of an actually orbit-dependent speed, it is often possible to numerically and experimentally estimate  $\overline{\lVert \dot{\rho}_t\rVert}$ using the following approach. First, we can approximate $\dot{\rho}_t$ with the finite-time increment $\dot{\rho}_t \sim\rho_{t_2}-\rho_{t_1}/|t_2-t_1|$, where $t_{1,2} = t \pm \epsilon/2$, for small $\epsilon$.
We then proceed with the approximation 
\begin{gather}
    \label{eq:finite_time}
    \tr[\dot{\rho}_t^2]  \sim \frac{\tr[\rho_{t_2}^2]+\tr[\rho_{t_1}^2] - 2\tr[\rho_{t_2}\rho_{t_1}]}{|t_2-t_1|^2}.
\end{gather}
Each term on the numerator of the right-hand side of Eq.~\eqref{eq:finite_time} can be evaluated with a \mono{CSWAP} circuit, when disposing of an ensemble of identically prepared states, as one would do for $\tr[\rho\sigma]$, as described above and in Ref.~\cite{Ekert2002}. 
In this sense, the Euclidean metric considered here has an advantage over those featuring $\sqrt{\rho_t}$, such as those based on quantum fidelity and affinity, since in general it requires fewer measurements for each instantaneous sample of the speed of the evolution.
Nevertheless, obtaining a precise estimation of the average speed of the evolution is generally hard, requiring a number of samples that strongly depends on the distribution of the velocities of the considered process, independently of the notion of the considered metric. When such an estimation has to be approached, it is thus fundamental to reduce the number of measurements required to obtain each instantaneous sample of the speed of the evolution.
In the next section, we will show that, in addition to being more feasible, our bound also outperforms existing speed limits for the majority of processes.

\section{Tightness}
\label{s:tightness}
We are now going to study the performance of the bound
$T_D$, relative to other proposed QSLs. To this end, we must ensure that different bounds are fairly compared.
Since some QSLs originate from different metrics and depend on the orbit, the only meaningful way to compare them with each other is to evaluate them along a chosen evolution, for a given pair of initial and final states. We compare our bound to the most significant bounds for open quantum evolution appearing in the literature~\cite{Sun2015, DelCampo2013, Deffner2013b, Pires2016,Mondal2016} which either depend on the overlap $\tr[\rho\sigma]$ or require the evaluation of quantum fidelity $\mathcal{F}(\rho,\sigma) = \tr[\sqrt{\sqrt{\rho}\sigma\sqrt{\rho}}]$~\cite{Uhlmann1992b}, affinity $\mathcal{A}(\rho,\sigma)=\tr[\sqrt{\rho}\sqrt{\sigma}]$~\cite{Luo2004}, or Fisher information~\cite{Facchi2010}.

When comparing different bounds along different orbits, one might be led to assume that the hierarchy between the bounds depends on the process in question. However, the orbit-dependent term that appears at the denominator of  bounds from Refs.~\cite{DelCampo2013,Deffner2013b,Sun2015} is always given by $\overline{\lVert \dot{\rho}_t\rVert}$\footnote{In particular, we selected the Hilbert-Schmidt norm for analytical comparison, while we have evaluated operator norm and trace numerically, if required by the considered QSL.} (i.e., the \emph{strength} of the generator), or can be directly related to it, up to some orbit-independent factors. This fact allows us to reduce the hierarchy of some of these bounds to that of the distance terms that depend only on the initial and final states, regardless of the chosen process and orbit. When this direct comparison is not possible, such as for the case of the bound in Ref.~\cite{Mondal2016}, we need to resort to numerical comparison.

\begin{figure}[tb]
    \centering
    \includegraphics[width=0.8\textwidth]{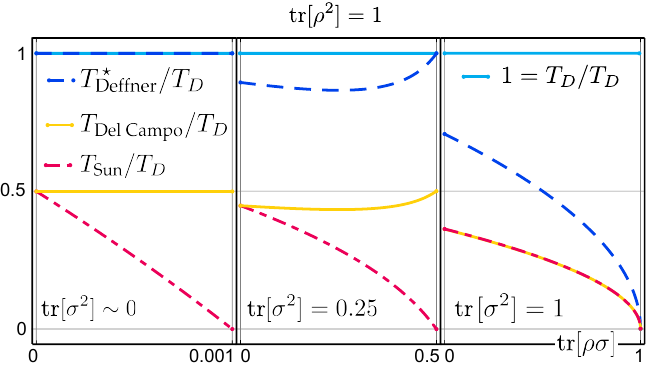}
    \caption{Analytic comparison of bounds from Refs.~\cite{Deffner2013b,DelCampo2013,Sun2015} with $T_D$, introduced in Eq.~\eqref{eq:speed_limit_arbitrary}, for the arbitrary evolution between $\rho=\rho^2$, and $\sigma$, expressed as the ratio between the considered QSL $T_{\textrm{Author}}$ and  $T_D$ as a function of $\tr[\rho\sigma]$ (which for $\rho^2=\rho$ also determines $\tr[(\rho\sigma)^2]$), as specified in the legend, and where $T_{\textrm{Deffner}}^\star\geq T_{\textrm{Deffner}}$ (see Eq.~\eqref{eq:over_deffner} for details). The three insets represent three different choices of purity $\tr[\sigma^2]$ for the final state, from left to right $\tr[\sigma^2]\sim 0$, $\tr[\sigma^2]=0.25$,  $\tr[\sigma^2]=1$. Note that, since the dimension $d$ of the system sets a bound for the minimal value $1/d$ of $\tr[\sigma^2]$, the central inset is meaningful from $d\geq 4$, while the left one is to be used in the limit of large $d$.}
    \label{fig:hierarchy}
\end{figure}

The orbit-independent term of our bound can be directly compared with those of Sun et al.~\cite{Sun2015} and del Campo et al.~\cite{DelCampo2013}, which depend on the overlap $\tr[\rho\sigma]$. In order to analytically compare our bound to that of Deffner et al.~\cite{Deffner2013b}, we over-estimate the orbit-independent term of the latter by replacing the fidelity with its lower-bound sub-fidelity, introduced in~\cite{Miszczak2008}, which  depends on the overlap $\tr[\rho\sigma]$, and on the additional quantity $\tr[(\rho\sigma)^2]$. For brevity, we will henceforth refer to previously introduced bounds by the corresponding first author's name.
As a result we find that, independently of the chosen process (i.e., for every choice of the generator $\dot{\rho}_t$), the bound $T_D$ expressed in Eq.~\eqref{eq:speed_limit_arbitrary} is tighter (i.e., greater) than Sun's, del Campo's, and Deffner's for every (allowed) choice of $\rho$ and $\sigma$
\begin{align}
    \label{eq:sun_delcampo}
   & T_D \geq \max\big\{T_{\textrm{Sun}},T_{\textrm{del Campo}}\big\}, \;\; \forall \rho,\sigma \; \in \mathcal{S}(\mathcal{H}_S), \\
   \label{eq:deffner_beat}
   & T_D \geq T_{\textrm{Deffner}} \;\; \forall \rho,\sigma, \;\; \text{for} \;\; \rho^2=\rho  \;\; \text{or} \;\; \sigma^2=\sigma,
\end{align}
as shown in Fig.~\ref{fig:hierarchy}. While Sun's and del Campo's QSLs are as easy to compute as our QSL given in Eq.~\eqref{eq:speed_limit_arbitrary}, they are also the loosest bounds. In contrast, Deffner bound's can be as tight as ours, but, since it requires the evaluation of $\sqrt{\rho}$ and $\sqrt{\sigma}$, it is less feasible.

\begin{figure}[htbp]
    \centering
    \includegraphics[width=0.8\textwidth]{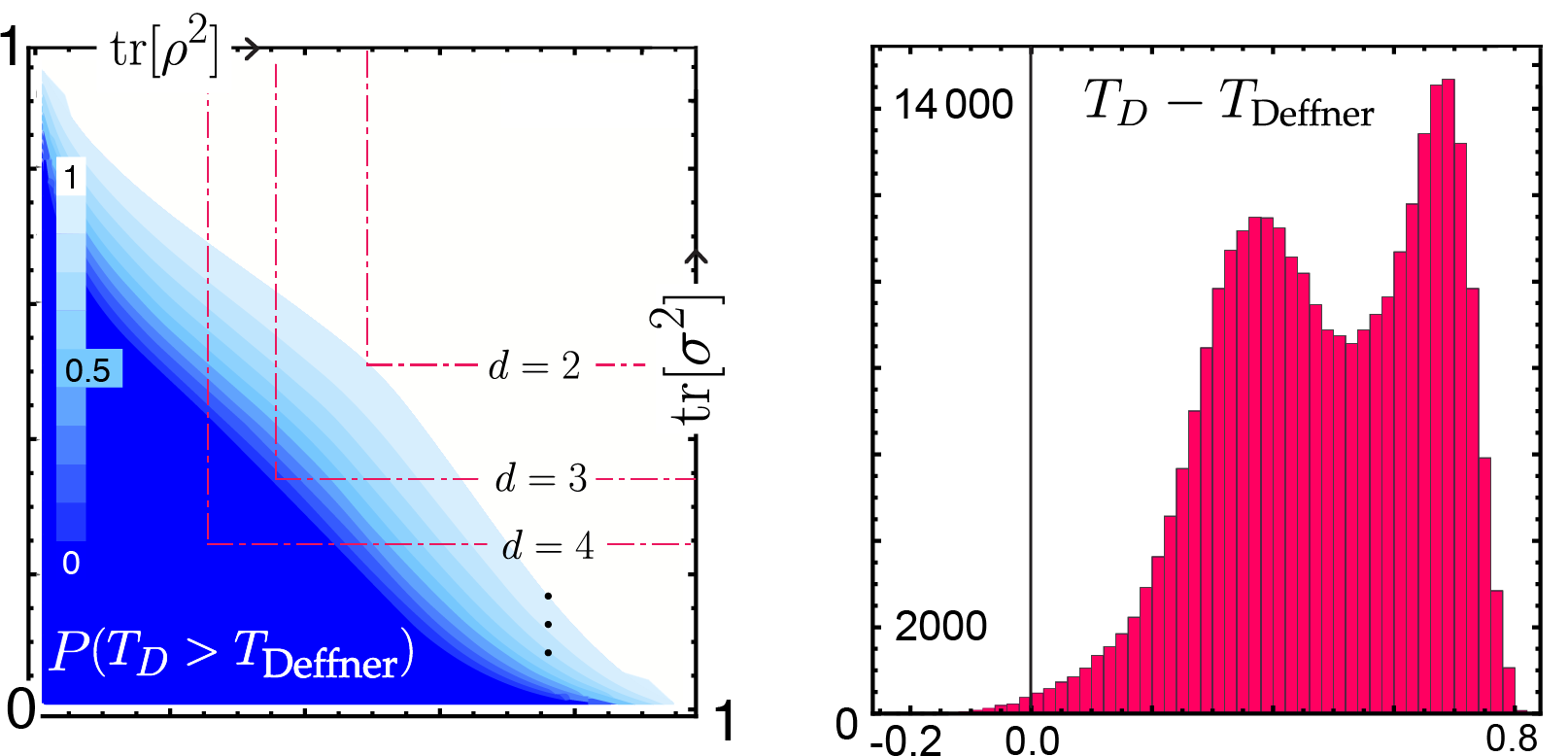}
    \caption{(\emph{Left}) The fraction of processes (quantified by their end points) for which $T_D$ outperforms $T_{\textrm{Deffner}}$, calculated as the ratio of the parameter space spanned by $\tr[\rho\sigma]$ and $\tr[\rho\sigma\rho\sigma]$ for which $T_D\geq T_{\textrm{Deffner}}$. A general \emph{rule of thumb} to evaluate the relative tightness between bound $T_D$ in Eq.~\eqref{eq:speed_limit_arbitrary} and Deffner's bound is given by $\tr[\sigma^2]\geq 1-\tr[\rho^2]\; \Rightarrow \;T_D\gtrsim T_{\textrm{Deffner}}$. Since $\tr[\rho^2],\tr[\sigma^2]\geq 1/d$, the region where $T_D$ outperforms Deffner's bound is always larger than that where the converse holds, as shown by the red dashed lines delimiting the physical region for $d=2,3,4$ (see Appending~\ref{a:deffner} for details).
    \emph{(Right}) Numerical estimation of relative tightness between $T_D$ and $T_{\textrm{Deffner}}$, obtained uniformly sampling $3 \cdot 10^6$ initial and final states from the Bures and the Ginibre ensembles. Bound $T_D$ is almost always tighter than $T_{\textrm{Deffner}}$.}
    \label{fig:hierarchy_deffner}
\end{figure}

In particular, Deffner's bound has been proven to be valid only when one of the two states is pure, i.e., for $\rho=\rho^2$ (or $\sigma=\sigma^2$)~\cite{Sun2015}. Under this condition our bound is always tighter than Deffner's. Additionally, we can analytically extend the validity of Deffner's bound to a larger class of cases by comparing it with our bound, and studying the region of the space of states for which $T_D$ is larger (see Fig.~\ref{fig:hierarchy_deffner}). All the details about the relative tightness of the considered bounds can be found in Appendix~\ref{a:tightness}.

Finally, we compare our bound to that of Ref.~\cite{Mondal2016} by Mondal et al., derived for the case of any general evolution, starting from the assumption that the initial state of the system $\rho_0$ is uncorrelated with that of the environment $\gamma_E$, i.e., $\Pi_0=\rho_0\otimes\gamma_E$. The orbit-independent term of their bound is a function of the affinity $\mathcal{A}(\rho_0,\rho_\tau)=\tr[\sqrt{\rho_0}\sqrt{\rho_\tau}]$ between initial and final states of the system, $\rho_0$ and $\rho_\tau$, respectively~\cite{Mondal2016}, which, as mentioned earlier, is hard to calculate and to measure as it requires the diagonalization of both density operators. The orbit-dependent term of their bound is a function of the root of $\rho_0$ and of an \emph{effective} Hamiltonian $\widetilde{H_S} = \tr_E[H\; \mathbb{1}\otimes\gamma_E]$, where $H$ is the total system-environment Hamiltonian. This function is not equivalent to $\overline{\lVert \dot{\rho}_t \rVert} $ (not even up to an orbit-independent factor), so we must calculate the two bounds for any given choice of dynamics, i.e., for any choice of total system-environment Hamiltonian $H$ and of initial state of the environment $\gamma_E$. 

\begin{figure}[tb]
    \centering
    \includegraphics[width=0.8\textwidth]{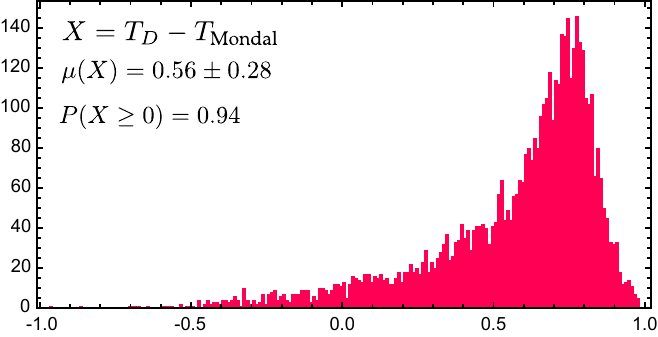}
    \caption{Numerical estimation of relative tightness between $T_D$ and $T_{\textrm{Mondal}}$, obtained sampling $5000$ orbits from random $H$, $\gamma_E$, and $\rho_0$, uniformly sampled as described in {\fontfamily{phv}\selectfont\textbf{Box~\ref{tech:qsl_tightness}}}. The dimension of the system's and environment's Hilbert spaces is uniformly sampled between $2$ and $10$. The tightness is measured using the parameter $X=T_D-T_{\textrm{Mondal}}$, which is bounded between $-1$ and $1$, given that the evolution is carried for a unit time $\tau=1$, and that both bounds have to be smaller than the evolution time $\tau$. The probability $P(X\geq 0)$ for our bound to be better than Mondal's is equal to $0.94$, with an average difference $\mu=0.56\pm0.28$.}
    \label{fig:hierarchy_mondal}
\end{figure}

As such, we proceed with a numerical comparison of the two bounds. We randomly generate total Hamiltonians $H$, initial states of the environment $\gamma_E$, and initial states of the system $\rho_0$, in order to choose the final state of the system $\rho_\tau = \tr_E [U_\tau \rho_0\otimes\gamma_E U_\tau^\dagger]$, where $U_\tau = \exp[-i H \tau]$, fixing $\tau=1$ for reference. We then compute both bounds for each instance of $H$, $\gamma_E$, and $\rho_0$ and compare their performance by measuring the difference $T_D - T_{\textrm{Mondal}}$. Remembering that $\tau=1$, and that both bounds must be smaller than $\tau$, the difference $T_D - T_{\textrm{Mondal}}$ must be bounded by $-1$ and $1$. Our numerical results provide convincing evidence of the performance of $T_D$ over $T_{\textrm{Mondal}}$, with the former being larger then the latter in 94\% of the cases for the considered sample, with an average difference of $0.57\pm0.28$ (see Fig.~\ref{fig:hierarchy_mondal} for the details about the numerical study). While Mondal's bound performs better than Deffner's, Sun's and del Campo's, it is arguably less feasible than all of them, as it involves the evaluation of $\sqrt{\rho}$ and $\sqrt{\sigma}$ for both distance and speed terms.

We have now shown that bound $T_D$ of Eq.~\eqref{eq:speed_limit_arbitrary} is tighter than the QSLs by del Campo et al.~\cite{DelCampo2013}, by Sun et al.~\cite{Sun2015}, and by Deffner et al.~\cite{Deffner2013b}, for all processes, while being just as easy to compute (if not easier). We have also provided numerical evidence of the superiority of our bound $T_D$ over the QSL by Mondal et al.~\cite{Mondal2016} for almost all processes, while being more feasible.

\section{Chapter summary}
\noindent
In this chapter we have presented a geometric quantum speed limit for arbitrary open quantum evolutions, which is based on the natural embedding of the space of quantum states in a high-dimensional ball, where states are represented by generalized Bloch vectors. The speed limit $T_D$ is induced by the Euclidean norm of the displacement vector $\bm{r}-\bm{s}$ between the two generalized Bloch vectors $\bm{r}$ and $\bm{s}$, associated with the initial and final states of the evolution. The measure of distinguishability that arises from this choice of distance corresponds to the Hilbert-Schmidt norm of the difference between initial and final states, $\rho$ and $\sigma$. The use of this norm has several benefits: It allows for the effective use of optimization techniques, such as convex optimization and semidefinite programming \cite{Abernethy2009}, it is easy to manipulate analytically and numerically, it has a straightforward geometric interpretation, and it is independent from the choice of the Lie algebra $\bm{\Lambda}$ of $SU(d)$ used to represent states as GBVs. The Hilbert-Schmidt norm is also widely used in experimental context, not only for quantum optimal control tasks, in order to impose finite energy bandwidth constraints on the control Hamiltonian \cite{Wang2015,Geng2016}.

We have considered the case of general open dynamics, in terms of a system-environment Hamiltonian and convolution with a memory kernel, as well as the special cases of unitary evolution and Lindblad dynamics. While the performance of many QSLs diminishes when increasingly mixed states are considered, our bound remains robust under composition, as well as under mixing with the fixed point of arbitrary CPTP maps. This result definitively confirms that the Hilbert-Schmidt norm can be used to derived a \textit{bona fide} QSL, despite being non-contractive under CPTP maps. 

We have discussed the form of our bound, with particular attention given to the speed of the evolution. We have highlighted the differences between our bound and the traditional QSL, induced by the Bures distance, and shed light on the reasons for the poor performance of the latter. Comparatively speaking, our bound outperforms several bounds derived so far in the literature~\cite{Sun2015, DelCampo2013, Deffner2013b, Mondal2016} for the majority of (if not all) processes. We have also addressed the physical interpretation of our bound, as well as that of similar QSLs, by providing a feasible experimental procedure that aims at the estimation of both the distance $D$ and the speed of the evolution $\overline{\lVert \dot{\rho}_t \rVert}$, while showing that our bound is easier to compute, as well as experimentally measure, than the other comparably tight bounds~\cite{Deffner2013b, Pires2016, Mondal2016}. These features indicate $T_D$ as the preferred choice of QSL. In particular, the versatility of this bound, as compared to that of Ref.~\cite{Campaioli2018}, allows it to be used for an much larger class of dynamics, which we have only just begun to approach with our examples in Section~\ref{s:forms}; a reflection that will hopefully be inspiring for further studies.

The efficacy of the QSL derived from this distance suggests that the use of a real vector space equipped with Euclidean metric to represent the space of operators, recently recognised also in Ref.~\cite{Brody2019}, could find application in the search for constructive approaches to time-optimal state preparation and gate design. This geometric picture might also offer solutions to some urgent outstanding problems, such as quantum optimal control in the presence of uncontrollable drift terms and constraints on tangent space, local quantum speed limits for multipartite evolution with restricted order of the interaction, and time-optimal unitary design for high-dimensional systems. The restrictions imposed by the constraints on the generators of evolution are known to dramatically change the geodesic that connects two states, and thus the bound on the minimal time of evolutions, as discussed in Refs.~\cite{Arenz_2014,Lee2018}. There, the authors introduced methods to bound the speed of evolution depending on the form of uncontrollable drift terms, control complexity and size of the system, obtaining accurate results for the case of single qubits in~Ref.~\cite{Arenz_2017}. Combining such considerations with the geometric approach used here could simplify the task of improving quantum speed limits and optimal driving of controlled quantum systems, by exploiting constants of motions that might be easier to represent in the generalised Bloch sphere picture.

Adapting this approach could find applications in other areas of quantum information, such as quantum metrology and quantum thermodynamics, where geodesic equations and geometric methods are routinely employed for the solution of optimization problems. While an attainable speed limit for arbitrary processes is yet to be found, its comprehension goes hand in hand with the understanding of the geometry of quantum states, as well as with the development of constructive techniques for time-optimal control. The latter is the central theme of the next chapter, where we shift the emphasis from bounding the minimal time of evolution, to finding the generator of the dynamics that saturates such bounds.

\chapter{Fast and efficient state preparation}
\label{ch:fast_state_preparation} 

\begin{ch_abs}
\textsf{\textbf{Chapter abstract}}\vspace{5pt}\\
\small\textsf{Fast driving of quantum systems is an important ingredient in many near future quantum technologies. However, finding the Hamiltonian that generates the fastest evolution is generally a hard task. In this chapter we introduce an iterative method to search for time-optimal Hamiltonians that drive a quantum system between two arbitrary states. The method is based on the idea of progressively improving the efficiency of an initial, randomly chosen, Hamiltonian that connects the two states, by reducing its components that do not actively contribute to driving the system. This iterative method converges rapidly even for large dimensional systems, and its solutions saturate any attainable bound for the minimal time of evolution. We provide a rigorous geometric interpretation of the iterative method by exploiting an isomorphism between the geometric phases acquired by the system along a path and the Hamiltonian that generates it. \vspace{10pt}\\
\emph{This chapter is based on publication \cite{Campaioli2019}.}}
\end{ch_abs}\vspace{20pt}

\section[The quantum brachistochrone problem]{The quantum brachistochrone problem}
In 1696 Johann Bernoulli challenged the brightest mathematicians of his times to find the curve of fastest descent between a point A and a lower point B on a vertical plane, for which he coined the name \textit{brachistochrone}, from the Greek words \textit{br\'akhistos khr\'onos}, meaning shortest time.
Bernoulli's deceptively simple problem succeeded in receiving the attention of some of the most illustrious mathematical minds, such as Leibniz and Newton, and became a milestone of infinitesimal calculus~\cite{Haws1995}. 

Inspired by this problem, the quantum brachistochrone problem (QBP) generally aims to find the fastest evolution between point A and point B in the space of quantum states, given some dynamical constraints. More specifically, in its first formulations, the QBP consisted in finding the time-independent Hamiltonian that generates the time-optimal evolution between two given quantum states~\cite{Bender2007, Assis2007, Mostafazadeh2007}. In this sense, it can be seen as the converse problem to that of deriving an attainable QSL for unitary evolution, formally addressed in Ch.~\ref{ch:qsl_mixed}. 

Since the original formulations, 
QBPs have been adapted and used to obtain accurate minimum-time protocols for the control of quantum systems~\cite{Wang2011,Russell2015, Geng2016}, clarify the role of entanglement and quantum correlations in time-optimal evolution~\cite{Borras2008, Basilewitsch2017}, study the speed of Hermitian and non-Hermitian Hamiltonians~\cite{Bender2007, Assis2007, Mostafazadeh2007}, and improve bounds on QSLs~\cite{Russell2017, Brody2015, Campaioli2018}. 

QBPs have a special role in quantum information theory and technology, where they are of fundamental importance to accurately performing tasks such as preparing a desired state of a system, or implementing a specific gate, while satisfying the strict physical and fault tolerance requirements imposed by the locality of interactions and short decoherence times~\cite{Caneva2009, Wang2015}. For this reason, their solutions have found applications in information processing~\cite{Werschnik2007,Reich2012, Jager2014,Dur2014,Huang2014,Schulte-Herbruggen2005}, quantum state preparation~\cite{Vandersypen2005, VanFrank2016, Islam2011, Senko2015, Jurcevic2014a, Zhou2016}, cooling~\cite{Wang2011,Machnes2012}, metrology~\cite{Kessler2014,Giovannetti2004c,Burkard1999}, and quantum thermodynamics~\cite{Rezakhani2009,DelCampo2013b,Binder2015,Campaioli2017}. Moreover, QBPs give a physical interpretation to the complexity of quantum algorithms, which emerges from the minimisation of the time required to obtain the unitary transformation that performs a desired gate~\cite{Carlini2006}.

Solving QBPs is generally hard, and accurate analytic and numerical solutions are only known for some special cases, such as the unconstrained unitary evolution of pure states and control problems of well structured quantum systems~\cite{Carlini2006, Levitin, Carlini2011, Carlini2012, Carlini2013, Brody2015, Wang2015}. There are methods that can be used to address a relatively large class of QBPs, but obtaining precise solutions becomes increasingly challenging as the dimension of the system grows, and the constraints on the  control Hamiltonian become more complex. For instance, the quantum brachistochrone equations proposed in Ref.~\cite{Carlini2006} involve the solution of ordinary differential equations (ODEs) with boundary conditions, for which there are no efficient numerical methods when high-dimensional systems are considered~\cite{krotov1993, Carlini2007, Rezakhani2009, Carlini2008, Wang2015}. One crucial open problem is the case of unconstrained unitary evolution between two mixed states. As opposed to the case of pure states, the solution to this QBP is not known, except for special cases with highly degenerate spectra. This is akin to the problem of finding tight bounds on the minimal time of evolution: As we have seen in Ch.~\ref{ch:qsl_mixed}, QSLs are attainable for pure states, but become often loose for mixed states due to the complicated structure of the space of density operators~\cite{Deffner2013, DelCampo2013, Sun2015}. Accordingly, the geometric methods used in the previous chapters to derive simple, efficient, and tight bounds for the time of evolution of mixed states~\cite{Campaioli2018, Campaioli2018b} have a crucial role in estimating the performance of the solutions to QBPs.

In this chapter we take a similar approach to solving the complementary problem, that of finding the optimal unconstrained unitary evolution between two mixed states. That is, we look for the generator $H$ of a unitary evolution $U_t = \exp(-i H t)$ that takes a mixed state $\rho$ to $\sigma = U_\tau \: \rho \: U_\tau^\dag$, such that the transition time $\tau$ is minimised. However, instead of solving a system of ODEs, we design an iterative method that progressively improves the efficiency of the Hamiltonian~\cite{Uzdin2012} that generates the evolution, to search for the optimal time-independent Hamiltonian, while respecting an energetic constraint.

We investigate the efficacy of our method using the bounds on the minimal time of evolution. By comparing the achievable upper bound, provided by our iterative method, with the inviolable lower bound offered by the QSLs introduced in Ch.~\ref{ch:qsl_mixed}, we demonstrate the strong synergy of the two results: When the two are found to be similar, we can be sure that both results are close to optimal. We study the performance of our method with respect to the size of the system, and its dependence on parameters such as the convergence threshold. We then discuss direct applications of the algorithm, its potential combination with time-optimal gate design and Monte-Carlo methods, and its geometric interpretation, juxtaposing it to that of Grover's quantum search algorithm.

\section{Time-optimal evolution and Hamiltonian efficiency} \label{s:hamiltonian_efficiency}
Let us begin by considering a simple QBP for a two-level system, defined by the initial state $\rho$ and the target state $\sigma$,
\begin{equation}
    \label{eq:QPB_example}
    \rho = \frac{\mathbb{1}+ p \, \Lambda_x}{2} \to \sigma = \frac{\mathbb{1}+ p \, \Lambda_y}{2},
\end{equation}
where $p \in (0,1]$, and where $\bm{\Lambda}=(\Lambda_x,\Lambda_y,\Lambda_z)$ is the vector of Pauli matrices, which form an orthonormal basis for the Lie Algebra $\mathfrak{su}(2)$. For this choice of $\bm{\Lambda}$, the Bloch vectors for the two states are given by $\bm{r} = (p,0,0)$ and $\bm{s} = (0,p,0)$, respectively. Any unitary $O(\bm{\varphi})$ of the form
\begin{equation}
    \label{eq:gate_example}
    O(\bm{\varphi}) = e^{i\varphi_1}\ketbra{s_1}{r_1}+e^{i\varphi_2} \ketbra{s_2}{r_2}
\end{equation}
can be used to map $\rho\to\sigma = O\rho O^\dagger$, where $\ket{r_k}$ and $\ket{s_k}$ are the eigenvectors of $\rho$ and $\sigma$, respectively, and where $\bm{\varphi}=(\varphi_1,\varphi_2)$ represents the geometric phase that the state gathers from $O(\bm{\varphi})$. Accordingly, the Hamiltonians
\begin{equation}
    \label{eq:hamiltonians_example}
    H(\bm{\varphi}) = i \log O(\bm{\varphi})
\end{equation}
define different evolutions, i.e., different orbits in the space of states, depending on the choice of $\bm{\varphi}$. Let us focus on two possible choices for these phases,
\begin{align}
    \label{eq:z_phase}
    &\bm{\varphi}_z = \Big(\frac{\pi}{4},\frac{\pi}{4}\Big), \\
    \label{eq:xy_phase}
    &\bm{\varphi}_{xy} = \Big(\frac{3}{4}\pi,-\frac{\pi}{4}\Big).
\end{align}
The associated Hamiltonians are given by
\begin{align}
    \label{eq:z_ham}
    &H(\bm{\varphi}_z) = \frac{\pi}{4}\Lambda_z, \\
    \label{eq:xy_ham}
    &H(\bm{\varphi}_{xy}) = \frac{\sqrt{2}}{4}\pi(\Lambda_x+\Lambda_y).
\end{align}
These two Hamiltonians generate different paths on the Bloch sphere. Under uniform energetic constraints, such as fixing the standard deviation of the Hamiltonians with respect to the initial state
\begin{equation}
    \label{eq:uniform_energetic_constraint_standard_dev}
    \Delta E = \sqrt{\tr[\rho H^2] - \tr[\rho H]^2} \equiv \omega,
\end{equation}
for some $\omega>0$, the lengths of these paths meaningfully correspond to evolution times\footnote{More generally, when the Hamiltonian is time-dependent, one can use either instantaneous energetic constraints, such as $\Delta E_t \equiv \omega$, or time-averaged energetic constraints $\overline{\Delta E} = \omega$, depending on the details of the problem (see Eqs.~\eqref{eq:standard_deviation} and~\eqref{eq:time_averaged_standard_deviation}).}. 
Such homogeneous energetic constraints induce a corresponding metric in the state space, which we use to measure the length of orbits and the evolution time, as seen in Sec.~\ref{s:traditional_unified_qsl}.
The evolution time is thus given by the total time required to obtain the gate $O(\bm{\varphi}) = U_\tau$ generated by $H$, which can be expressed as $U_t = \exp[-i H t]$ for time independent Hamiltonians. 
\begin{figure}
    \centering
    \includegraphics[width=0.8\textwidth]{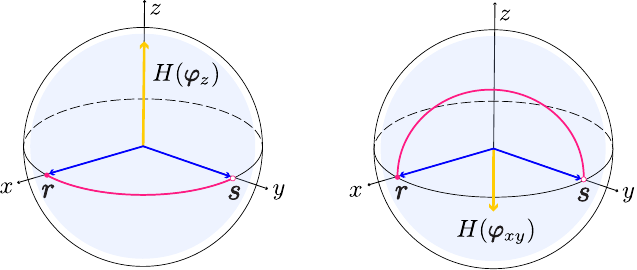}
    \caption{The pair of state $\rho\leftrightarrow\bm{r}$ and $\sigma\leftrightarrow\bm{s}$ defined in Eq.~\eqref{eq:QPB_example} is driven by the Hamiltonians $H(\bm{\varphi}_z)$ and $H(\bm{\varphi}_{xy})$ of Eqs.~\eqref{eq:z_ham} and~\eqref{eq:xy_ham}. The orbits generated by these Hamiltonians have different lengths in the space of states. Here lengths are calculated using the infinitesimal distance associated with the uniform energetic constraint of  Eq.~\eqref{eq:uniform_energetic_constraint_standard_dev}. Their Hamiltonian efficiency, given in Eq.~\eqref{eq:Hamiltonian_efficiency}, is indeed different, with $H(\bm{\varphi}_z)$ being twice as efficient as $H(\bm{\varphi}_{xy})$ for any purity of $\rho$ and $\sigma$.
    }
    \label{fig:efficiency}
\end{figure}

As represented in Fig.~\ref{fig:efficiency}, the path generated by $H(\bm{\varphi}_z)$ connects $\rho$ to $\sigma$ with an arc of great circle, i.e., the geodesic with respect to the Fubini-Study  metric, and thus it constitutes a solution to the considered QBP for the homogeneous energy constraint of Eq.~\eqref{eq:uniform_energetic_constraint_standard_dev}. In contrast, $H(\bm{\varphi}_{xy})$ draws a \emph{longer} path with respect to the same metric, which is thus not time-optimal\footnote{In fact, it does so for any uniform energetic constraint~\cite{Russell2017}.}.
A heuristic explanation for the variable performance of different Hamiltonians is that the vectors $\bm{h}$, associated with them via 
\begin{equation}
    \label{eq:Hamiltonians_vector}
    H = \bm{h}\cdot \bm{\Lambda},
\end{equation}
have a different orientation with respect to $\bm{r}$ and $\bm{s}$. In particular, when $\bm{h}$ is orthogonal to $\bm{r}$, it generates a rotation on a plane that passes through the origin of the Bloch sphere, while when $\bm{h}$ is not perpendicular to $\bm{r}$, it generates a rotation on a plane that does not. The slower evolution can be seen as due to a less efficient use of the resources encoded in constraints on the energy, which are \emph{wasted} on parts of the Hamiltonian that do not actively drive the system~\cite{Uzdin2012}. This notion of efficiency can be formalised for pure states as done in {\fontfamily{phv}\selectfont\textbf{Box~\ref{tech_Hamiltonian_efficiency}}}.
\begin{techbox}{Hamiltonian efficiency}{tech_Hamiltonian_efficiency}
The notion of Hamiltonian efficiency introduced in Ref.~\cite{Uzdin2012} for the unitary evolution of pure states is given by
\begin{equation}
    \label{eq:Hamiltonian_efficiency}
    \eta(H,\rho) := \frac{\Delta E}{\lVert H\rVert_{op}},
\end{equation}
where $\Delta E = \sqrt{\tr[\rho H^2]-\tr[\rho H]^2}$, with $\lVert \cdot\rVert_{op}$ being the operator norm. 
The efficiency measures how much of the energy available, quantified by $\lVert H\rVert_{op}$, is transferred to the system to 
drive the evolution, and thus \textit{converted} into speed $\Delta H$. 
Different notions of efficiency can be obtained for any choice of speed and total energy measure. For instance, the latter can be replaced with the Hilbert-Schmidt norm of the Hamiltonian.

We can assess the performance of the two Hamiltonians $H(\bm{\varphi}_z)$ and $H(\bm{\varphi}_{xy})$, of Eqs.~\eqref{eq:z_ham} and~\eqref{eq:xy_ham}, using $\eta$ (see Fig.~\ref{fig:efficiency}). 
We obtain
\begin{align}
    \label{eq:eta_z}
    & \eta(H(\bm{\varphi}_z),\rho)=p, \\
    \label{eq:eta_xy}
    & \eta(H(\bm{\varphi}_{xy}),\rho)=\frac{p}{2}.
\end{align}
The dependence on $p$, and thus on the purity, implies that this notion of efficiency is not saturated, in general, for mixed states. 
\end{techbox}
This intuitive geometric argument can be generalised for dimension $d>2$ by replacing the notion of orthogonality between vectors with commutation relations between states and Hamiltonians. Given an operator $\rho$, the space of Hamiltonians, i.e., the Lie algebra $\mathfrak{su}(d)$, splits into a maximal dimensional \textit{parallel} subspace, closed under multiplication, that commutes with $\rho$, and an orthogonal \textit{perpendicular} subspace, every element of which does not. This allows us to decompose the Hamiltonian $H$ into components $H_\|$ and $H_\perp$ which are elements, respectively, of these two subspaces, such that $[H_\|,\rho]=0$ and $[H,\rho]=[H_\perp,\rho]$. These correspond to elements of the vertical and horizontal tangent bundles in a fibre bundle representation where the space of states with a given spectrum is the base manifold and the phases $\mathbf{\varphi}$ are the fibres. This observation leads to the idea at the core of the iterative method that we will introduce in the next section, where efficient Hamiltonians for a QBP are achieved by requiring their parallel component to be vanishing at all times during the evolution.

\section{Iterative method for efficient Hamiltonians}\label{p:method} 
\begin{figure}
    \centering
    \includegraphics[width=0.95\textwidth]{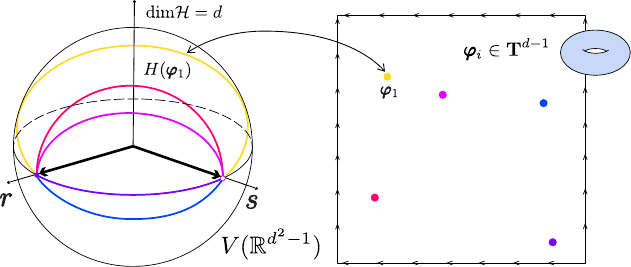}
    \caption{Let $\rho$, $\sigma$ be two isospectral $d$-dimensional states, here represented with their GBVs $\bm{r}$ and $\bm{s}$, respectively. In this cartoon, every unitary orbit connecting initial and final state $\rho$ and $\sigma$ generated by a time-independent Hamiltonian $H(\bm{\varphi})$ (\textit{left}) is associated with a $d$-dimensional phase vector $\bm{\varphi}$ via Eqs.~\eqref{eq:o_phi} and~\eqref{eq:hamiltonians_example}. By exploiting the periodicity of the phases and absorbing one component of those phases into a global phase, we obtain a one-to-one mapping between the points of a $(d-1)$-dimensional torus (\textit{right}), representing the space of measurable geometric phases, and the orbits of constant curvature connecting $\rho$ and $\sigma$.
    }
    \label{fig:isomorphism}
\end{figure}

\noindent
We can now consider the more general QBP defined by an arbitrary isospectral pair of initial and final states $\rho =\sum_k \lambda_k\ketbra{r_k}{r_k}$ and $\sigma =\sum_k \lambda_k\ketbra{s_k}{s_k}$ of finite dimension $d$. When non-degenerate states are considered, the operator
\begin{equation}
    \label{eq:o_phi}
    O(\bm{\varphi}) = \sum_k e^{i\varphi_k}\ketbra{s_k}{r_k},
\end{equation}
represents the most general unitary that connects initial and final states $\rho$ and $\sigma$, while $\varphi_k$ represent the geometric phases gained evolving along the path generated by $i\log O(\bm{\varphi})$. In other words, Eqs.~\eqref{eq:o_phi} and~\eqref{eq:hamiltonians_example} provide an isomorphism between the unitary orbits connecting $\rho$ and $\sigma$ and the points of the $(d-1)$-dimensional torus representing the space of relevant phases, as illustrated in Fig.~\ref{fig:isomorphism}. When degenerate states are considered, the phases $\varphi_k$  are replaced by unitary operators $U_{k} \in SU(m)$ on the subspace associated with eigenvalues $\lambda_{k}$ with multiplicity $m$.

Since any unitary $O(\bm{\varphi})$ maps $\rho\to\sigma$, we can choose an arbitrary initial phase $\bm{\varphi}^{(0)}$ to obtain the initial unitary
\begin{equation}
    \label{eq:initial_unitary}
    O^{(0)} = O(\bm{\varphi}^{(0)}).
\end{equation}
The Hamiltonian $H^{(0)} = i \log O^{(0)}$, canonically associated with $O(\bm{\varphi}^{(0)})$, is then split into its parallel and perpendicular components
\begin{align}
    \label{eq:parallel_split}
    &\mathcal{M}_\rho[H^{(0)}] = H^{(0)}_\|, \\
    \label{eq:perpendicular_split}
    &H^{(0)}_\perp = H^{(0)}-H^{(0)}_\|,
\end{align}
such that $[H^{(0)}_\|,\rho]=0$, via a map $\mathcal{M}_\rho$ that depends on $\rho$, and which will be referred to as \emph{mask}. We choose the mask to be
\begin{align}
\label{eq:mask}
    &\mathcal{M}_\rho[H]=D \big( M \circ D^\dagger H D) \big) D^\dagger,  \\
    \label{eq:components_of_the_mask}
    &M_{ij} = \delta_{\lambda_i\lambda_j},
\end{align}
where $D$ diagonalises $\rho$, i.e., its columns are given by the eigenstates $\{\ket{r_k}\}$ of $\rho$, and where $\circ$ is the Hadamard, i.e., element-wise, product~\cite{Horn2012}. This mask projects the Hamiltonian onto the maximal subalgebra $\mathfrak{c}\subset\mathfrak{su}(d)$ that commutes with $\rho$. When degenerate states are considered, the eigenvalues' multiplicity defines the structure of the mask via $\delta_{\lambda_i\lambda_j}$.
\begin{algorithm}[t]
\caption{Iterative method for efficient Hamiltonians}
\label{a:algorithm}
\begin{algorithmic}[1]
    \INPUT Initial state $\rho = \sum_{k=1}^d \lambda_k \ketbra{r_k}{r_k}$ and final state $\sigma = \sum_{k=1}^d \lambda_k \ketbra{s_k}{s_k}$. Initial phase $\bm{\varphi}^{(0)}$.
    Threshold for convergence $\varepsilon$.
    \OUTPUT Optimal Hamiltonian.
    \STATE Initialise $O^{(0)}=\sum _k e^{i \varphi_k^{(0)}} \ketbra{s_k}{r_k}$, $H^{(0)}=i \log O^{(0)}$, $H_\parallel ^{(0)}=\mathcal{M}_\rho[H^{(0)}]$; 
    \WHILE{$\|H_\parallel ^{(j)} \|_\mathrm{HS}>\varepsilon \| H^{(j)}\|_\mathrm{HS}$} \\
    Set $O^{(j)}=O^{(j-1)}e^{i H_\parallel ^{(j-1)}}$, $H^{(j)}=i \log O^{(j)}$, $H_\parallel ^{(j)}=\mathcal{M}_\rho[H^{(j)}]$;
    \ENDWHILE $\;\;$ 
    \STATE \textbf{return} the final Hamiltonian $H^{(n)}$ of the sequence $\{H^{(j)}\}_{j=0}^n$.
\end{algorithmic}
\end{algorithm}

We now notice that the unitary 
$U^{(0)} = \exp[i H_\|^{(0)}]$ can be composed with $O^{(0)}$ to obtain another unitary $O^{(1)} = O^{(0)} U^{(0)}$, formally equivalent to evolving with the time-dependent Hamiltonian $e^{-iH^{(0)}t}H^{(0)}_\perp e^{iH^{(0)}t}$, which also drives $\rho\to\sigma$ since $[\rho,H_\|]=0$. In general, the unitary $O^{(1)}$ is associated with a new geometric phase, and the Hamiltonian $H^{(1)} = i \log O^{(1)}$ draws a different path, which might be shorter (or longer) than that generated by $H^{(0)}$. In a second iteration, we apply the mask to the new Hamiltonian to obtain $H^{(1)}_\|=\mathcal{M}_\rho[H^{(1)}]$, and define another gate $O^{(2)}$ analogously. By iterating this method we obtain a sequence of Hamiltonians $\{H^{(j)}\}_{j=0}^n$ that drive $\rho\to\sigma$ via $\exp[-i H^{(j)}]$. The unitary at each step is related to the previous one by
\begin{align}
    \label{eq:gate_n}
    & O^{(j+1)} =O^{(j)} e^{-i \mathcal{M}_\rho[H^{(j)}]}, \\
    & H^{(j)} = i \log O^{(j)}.
\end{align}
A necessary and sufficient condition for the routine to reach a \emph{fixed point} $H^{(n)}$, such that $H^{(n+1)}=H^{(n)}$, is given by $\mathcal{M}_\rho[H^{(n)}]=0$, i.e., when the parallel part of the $n$-th Hamiltonian vanishes under the action of the mask $\mathcal{M}_\rho$, which follows trivially from the fact that in this case $e^{-i\mathcal{M}_\rho[H^{(n)}]}=\mathbb{1}$.

Note how the method introduced in this section involves removing the part of the Hamiltonian operator that commutes with the initial state $\rho$. A similar approach could be taken replacing Eq.~\eqref{eq:gate_n} with
\begin{equation}
    \label{eq:gate_n_backwards}
    O^{(j+1)}=e^{i\mathcal{M}_\sigma[H^{(j)}]}O^{(j)},
\end{equation}
where the part of the Hamiltonian that commutes with the final state $\sigma$ is removed, and the unitary joining the two states is acted on from the left in each iteration, instead of from the right. It is also possible to remove both components simultaneously, replacing Eq.~\eqref{eq:gate_n} with
\begin{equation}
\label{eq:gate_n_back_forth}
    O^{(j+1)}=e^{i\mathcal{M}_{\sigma}[H^{(j)}]}O^{(j)}e^{-i\mathcal{M}_{\rho}[H^{(j)}]}.
\end{equation} 
While these slightly different approaches can affect the number of iterations required for convergence for a given choice of initialisation phase $\bm{\varphi}^{(0)}$, they neither perform significantly better than the algorithm defined by Eq.~\eqref{eq:gate_n}, nor require less iterations to converge on average.

\begin{proofbox}{Derivation of Algorithm~\ref*{a:algorithm}}{tech_derivation_algorithm}
Let $O=\sum_k\ketbra{s_k}{r_k}$, and $i \log O=H=H_\|+H_\perp$, where $H_\|$ commutes with $\rho$. Consider the unitary $U_t$ generated by the time dependent Hamiltonian $O_t H_\perp O^\dagger_t$, where $O_t = \exp[-i H t]$, which satisfies the equation
\begin{equation}
    \label{eq:partial_derivative_U}
    \partial_t U_t = -i O_t H_\perp O^\dagger_t U_t.
\end{equation}
To leading order in time, this will rotate the perpendicular part of the Hamiltonian to follow the system's evolution. Now consider the unitary $\widetilde{U}_t=O^\dagger_t U_t$, whose equation of motion is given by
\begin{equation}
    \label{eq:eq_of_motion_U}
    \begin{split}
    \partial_t \widetilde{U}_t &= -i H_\perp \widetilde{U}_t
 + (\partial_t O^\dagger_t)U_t \\
 & = -i(H_\perp - H)\widetilde{U}_t \\
 & = i H_\| \widetilde{U}_t.
 \end{split}
\end{equation}
Eq.~\eqref{eq:eq_of_motion_U} implies that $\widetilde{U}_t = \exp[i H_\|t]$, thus that $\widetilde{U}_t$ commutes with $\rho$. Accordingly, $U=U_1=O_1\widetilde{U}_1$ transforms $\rho\to\sigma$, just as well as $O$ does. The procedure can be iterated until it reaches a fixed point, i.e., until $H = H_\perp$ and $H_\| = 0$.

The parallel part $H_\|$ of the Hamiltonian is not uniquely defined in terms of the commutator with the initial density matrix. In order to fix a precise definition one can consider the Hamiltonian's projection onto the maximal subalgebra $\mathfrak{c} \subset\mathfrak{su}(d)$ that commutes with $\rho$. This subalgebra is unique and given by $\mathfrak{c}=W^\dagger \bigotimes_j\mathfrak{su}(d_j)W$, where $d_j$ is the degeneracy of the $j$-th unique eigenvalue of $\rho$ and $W$ is the unitary that diagonalises it. The mask defined in Eq.~\eqref{eq:mask} achieves precisely this projection, in the sense that there is no component of $H_\perp$ that commutes with $\rho$. Let us consider a fully non-degenerate initial state $\rho=\sum_k \lambda_k \ketbra{k}{k}$, where for simplicity we relabelled the eigenvectors with $\ket{k}$ inspite of $\ket{r_k}$. Let us imagine that $G\in \mathfrak{su}(d)$ is also an component of $H_\perp$, i.e., $G=\sum_{i\neq j}g_{ij}\ketbra{i}{j}$ in the eigenbasis of $\rho$, as prescribed by the mask in Eq.~\eqref{eq:mask}. For $G$ to commute with $\rho$, we must have
\begingroup
\allowdisplaybreaks
\begin{align}
    [G,\rho] &= \sum_{i\neq j,k} g_{ij}\lambda_k (\ket{i}\!\braket{j|k}\!\bra{k}-\ket{k}\!\braket{k|i}\!\bra{j})  \\
    & = \sum_{i\neq j} g_{ij}(\lambda_j-\lambda_i)\ketbra{i}{j} = 0,
\end{align}
\endgroup
which vanishes for all the linearly independent components $\ketbra{i}{j}$ if and only if $g_{ij}\equiv 0$, since $\lambda_j \neq \lambda_i$ by hypothesis (non-degenerate spectrum). This means that $G$ can only commute with $\rho$ if it is trivial. Similar arguments can be made for the case of degenerate states, considering the form of the map $M_{ij} = \delta_{\lambda_i,\lambda_j}$. \hfill\qedsymbol
\end{proofbox}

This iterative method is summarised in its simplest form by Algorithm~\ref{a:algorithm}, available at Ref.~\cite{Campaioli2019b}, and can be interpreted as an optimisation of energy cost associated with the different geometric phases $\bm{\varphi}$, accomplished via the recursive suppression of ineffective components of the Hamiltonians. 

There is also a geometric interpretation of our method analogous to that of Grover's famous quantum search algorithm~\cite{Grover1997}. Grover's quantum search algorithm aims to find the unique input, encoded in a quantum \emph{target} state, of a function that produces a particular output value. Its action on an initialisation state $\ket{\psi^{(0)}}$ can be interpreted as rotation. Repeated iterations rotate the initialisation state closer to the target state, increasing the probability of finding the solution to the search problem.

The iterative method introduced here is also solving a searching problem, where the aim is to find the time-optimal Hamiltonian for a given QBP. Its actions is precisely interpreted as a sequence of rotations of some initialisation unit vector $\hat{\bm{h}}^{(0)}$, associated with the Hamiltonian via Eq.~\eqref{eq:Hamiltonians_vector}. However, unlike Grover's algorithm, the vectors of the sequence $\{\bm{h}^{(n)}\}$ do not span a two-dimensional real plane, but a high-dimensional subspace of $\mathbb{R}^{d^2-1}$.

\subsection{Numerical implementation and finite precision}
\label{ss:numerical_implementation_finite_precision}
When implementing the method numerically, one has to fix a convergence threshold to stop the routine as soon as the parallel component $H_\|^{(n)}$ becomes small enough with respect to the full Hamiltonian $H^{(n)}$.
We have chosen to quantify this threshold with a positive number $\varepsilon\ll 1$, such that the method stops when
\begin{equation}
    \label{eq:convergence_threshold}
    \lVert H_\|^{(n)}\rVert\leq\varepsilon \lVert H^{(n)}\rVert,
\end{equation}
where $\lVert \cdot\rVert$ is the Hilbert-Schmidt norm.
Accordingly, the number of iterations $n$ required for the method to converge implicitly depends on $\varepsilon$. Numerical evidence suggests that the sequence always converges towards a Hamiltonian $H^{(n)}$ that is fully \textit{perpendicular} with respect to $\rho$ along the whole evolution, in the sense that the parallel components eventually vanish within the precision defined by $\varepsilon$. 

When implementing this method experimentally, deviations from unit fidelity between target state $\sigma$ and prepared state $\rho(\tau^{(n)})$ can arise from a multitude of factors, such as finite precision on the duration of the driving $\tau\pm\Delta\tau$, as well as errors on the \textit{control parameters} $h_k\pm\Delta h_k$ that generate the Hamiltonians $\bm{h}\cdot\bm{\Lambda}$, but cannot be related to the convergence threshold. This is because each Hamiltonian in the sequence $\{H^{(j)}\}_{j=0}^n$ generates a unitary of the form of Eq.~\eqref{eq:o_phi}, driving $\rho$ exactly to $\sigma$,
\begin{equation}
    \mathcal{F}(\rho(\tau^{(j)}),\sigma) = 0 \; \forall\; H^{(j)}\in\{H^{(j)}\}_{j=0}^n.
\end{equation}
Thus, the finite numerical precision imposed by the convergence threshold $\varepsilon$ does not affect the distance between the target state $\sigma$ and the final state of the optimised evolution $\rho(\tau^{(n)})$.
This observation is particularly important for practical purposes,  such as for the case of quantum circuits, where small deviations from unity fidelity heavily affect the achievable circuit depth~\cite{Boixo2018}. Additionally, most optimisation method often require the maximisation of such fidelity as an additional task to the minimisation of the preparation time~\cite{Tibbetts2012}.

The effect of finite numerical precision does, however, affect the efficiency $\eta(H^{(n)},\rho)$ of the optimised solutions. To estimate this effect for the case of pure states, for which $H=H_\perp\Rightarrow\eta(H,\rho)=1$, we replace the Hilbert-Schmidt norm in Eq.~\eqref{eq:convergence_threshold} with the operator norm, and use the triangle inequality to split the contributions of parallel and perpendicular components of the Hamiltonian,
\begin{align}
\label{eq:effect_on_efficiency}
\frac{1}{\eta(H,\rho)}&=\frac{\lVert H\rVert_{op}}{\Delta E(H,\rho)}\\
                \label{eq:use_invariance_parallel}
                &=\frac{\lVert H\rVert_{op}}{\Delta E(H_\perp,\rho)}\\
                &\leq\frac{\lVert H_\perp\rVert_{op}+\lVert H_\|\rVert_{op}}{\Delta E(H,\rho)}\\
                \label{eq:inequality_precision}
                &\leq 1 + \varepsilon\frac{\lVert H\rVert_{op}}{\Delta E(H_\perp,\rho)}\\
                &= 1+\frac{\varepsilon}{\eta(H,\rho)},
\end{align}
where Eq.~\eqref{eq:use_invariance_parallel} holds since $\Delta E$ is invariant under $H\to H+P$ for any operator $P$ that commutes with $\rho$, while Eq.~\eqref{eq:inequality_precision} holds for Eq.~\eqref{eq:convergence_threshold}. Rearranging, and recalling that $\eta\leq1$, we obtain
\begin{equation}
    \label{eq:eta_finite_pericision}
    1-\varepsilon\leq \eta\leq 1,
\end{equation}
which means that, for finite numerical precision imposed by a convergence threshold $\varepsilon$, the solutions $H^{(n)}$ will at most suffer an $\varepsilon$-reduction of their efficiency from unity.
For the case of mixed states, the notion of efficiency given by Eq.~\eqref{eq:Hamiltonian_efficiency} cannot be saturated in general. A possible generalisation $\eta^\star$ to the case of mixed states is given in Eq.~\eqref{eq:efficiency_star}, for which we will discuss the effect of finite numerical precision in {\fontfamily{phv}\selectfont\textbf{Box~\ref{tech_eta_mixed}}}.

\section{Performance of the iterative method}
\begin{figure}[t]
    \centering
    \includegraphics[width=0.85\textwidth]{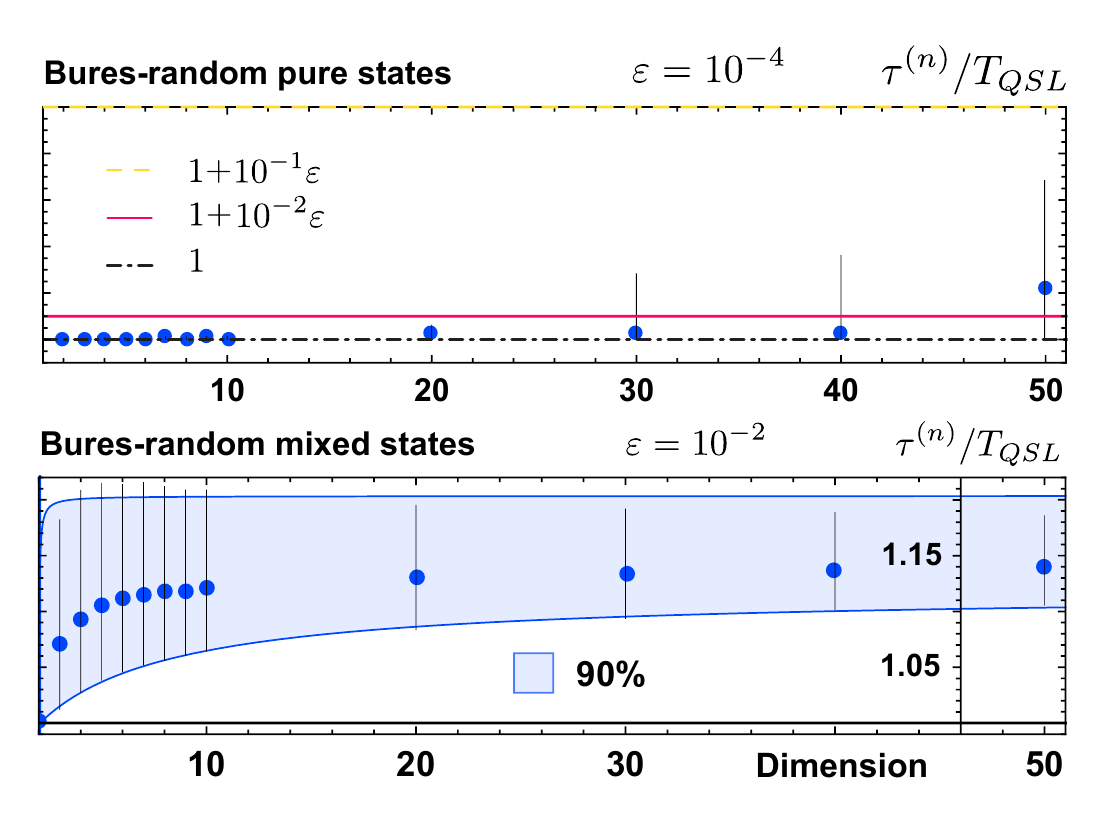}
    \caption{The performance of the method is here studied for Bures-random pure (\textit{top}) and mixed (\textit{bottom}) states, using the ratio $\tau^{(n)}/T_{QSL}\geq 1$ between the optimised time $\tau^{(n)}$ required to drive $\rho\to\sigma$ with the solution $H^{(n)}$, and the QSL $T_{QSL}$ of Eq.~\eqref{eq:unified}, evaluated for each pair of states $\rho$, $\sigma$, and optimised Hamiltonian $H^{(n)}$. The plotted points show the average performance for different Hilbert space dimensions $d$, with the \emph{error bars} representing 99\% (\textit{top}) and 90\% (\textit{bottom}) confidence intervals. The shaded area represents a fit of the 90\% confidence interval for the average $\tau^{(n)}$. The convergence threshold is $\varepsilon=10^{-4}$ for pure states (\textit{top}), and $\varepsilon=10^{-2}$ for mixed states (\textit{bottom}). The sample size is $10^4$ for each dimension. For pure states, the method returns solutions that converge on the QSL as the precision $\varepsilon^{-1}$ is increased, at the expense of requiring more iterations on average.}
    \label{fig:perf}
\end{figure}

\noindent
Our iterative method can be directly applied to solve QBPs defined by the unconstrained time-optimal unitary evolution of any isospectral pair of density operators. 
A sub-class of these problems with known solutions are those of unconstrained unitary evolution between pure states or between mixed states whose eigenvalues are all degenerate except one.
To demonstrate the performance of our method, we have tested it on random pairs of pure states of dimension $d=2,\dots,100$, uniformly distributed with respect to the Bures ensemble, successfully obtaining Hamiltonians that are time-optimal and fully efficient with respect to the notion of efficiency introduced in Eq.~\eqref{eq:Hamiltonian_efficiency}. We can be confident that the solutions we obtain are globally optimal by exploiting the fact that the QSL for unitary evolution of pure states is attainable~\cite{Levitin}. 
Comparing the evolution time $\tau^{(n)}$ of the optimized Hamiltonian $H^{(n)}$ with the minimal evolution time $T_{QSL}$~\cite{Deffner2017} we find $\tau^{(n)} \approx T_{QSL}$ within the precision imposed by $\varepsilon$, for all cases, as shown in Fig.~\ref{fig:perf}.

A more challenging test was run on random\footnote{Also uniformly distributed with respect to the Bures ensemble.} pairs of mixed states of dimension $d=3,\dots,100$, considering initial states with both degenerate and non-degenerate spectra, for which a general solution to the unconstrained unitary QBP is not known. Since the QSL for the unitary evolution of mixed states is in general not tight~\cite{Campaioli2018}, it is harder to benchmark the quality of the solutions provided by our method in this case. On the other hand, the natural synergy between this iterative method and the QSLs introduced in Ref.~\cite{Campaioli2018} can be used to gauge the performance of the former and the tightness of the latter. The \emph{actual} minimal time of evolution $\tau_{\text{min}}$ for a given choice of $\rho$ and $\sigma$ is bounded as
\begin{gather}
    \label{eq:bound_on_t_actual}
    T_{QSL} \leq \tau_{\text{min}} \leq \tau^{(n)},
\end{gather}
where $T_{QSL}$ corresponds to the inviolable lower bound offered by QSLs of Ref.~\cite{Campaioli2018}, and $\tau^{(n)}$ to an achievable upper bound, provided by the solutions obtained with our iterative method. Note how the inequality on the left-hand side of Eq.~\eqref{eq:bound_on_t_actual} can be saturated for pure states and for states with $d-1$ degenerate eigenvalues. In this case, the second inequality can be saturated up to the precision imposed by $\varepsilon$, which can be seen as an implicit trade-off between $n$ and $\varepsilon$.

By looking at the difference between $\tau^{(n)}$ and $T_{QSL}$ for the given solution $H^{(n)}$, we can assess the quality of both results, which we find to coincide within the precision set by $\varepsilon$ for some choices of $\rho$ and $\sigma$, even when their degeneracy structure differs from that of pure states.
We now discuss the rate at which the algorithm converges.
\begin{figure}[htbp]
    \centering
    \includegraphics[width=0.8\textwidth]{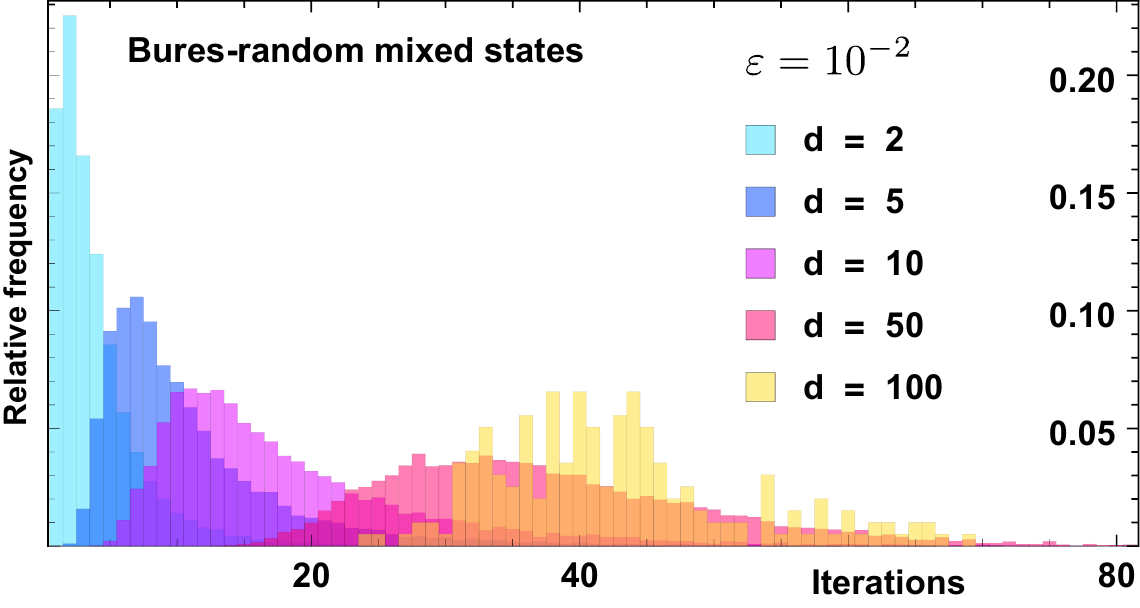}
    \caption{Relative frequency of the number of iterations required for convergence for Bures-random mixed states, with the considered system dimensions $d=2,\dots,100$, as shown in the legend. The sample size is $10^4$ for each dimension $d\leq50$, and $200$ for $d>50$, while the convergence threshold is chosen to be $\varepsilon=10^{-2}$.}
    \label{fig:iter_histo}
\end{figure}

\section{Convergence of the iterative method} 
\begin{figure}[htbp]
    \centering
    \includegraphics[width=0.8\textwidth]{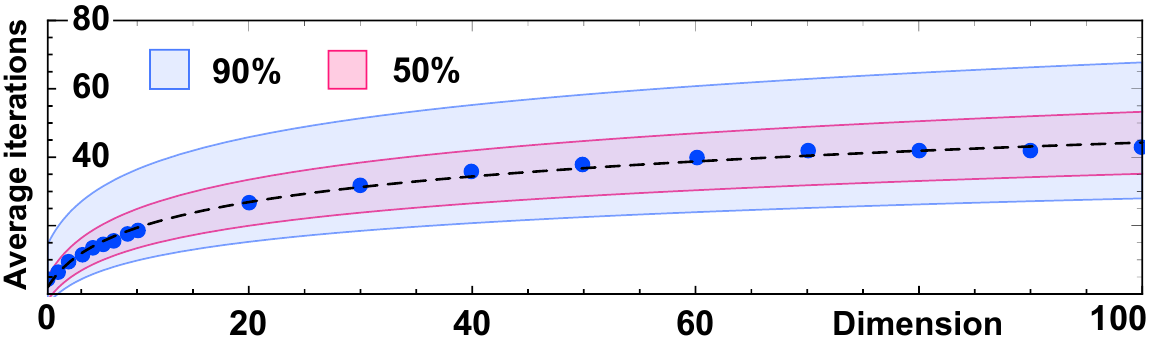}
    \caption{Average number of iterations required for convergence for Bures-random mixed states of the considered system dimensions $d=2,\dots,100$, with their 50\% and 90\% confidence regions. The black dashed line represents a logarithmic fit. The sample size is $10^4$ for each dimension $d\leq50$, and $200$ for $d>50$, while the convergence threshold is chosen to be $\varepsilon=10^{-2}$.}
    \label{fig:iter}
\end{figure}

The number $n$ of iterations required for convergence generally grows with the dimension $d$ of the system, and the strictness of the precision set by $\varepsilon$. Recalling that the number of elements of the matrices associated with non-commuting density operators $\rho$ and $\sigma$ grows quadratically with $d$, one might expect conservatively that the average number of iterations $\bar{n}$ would grow in the same way. However, $\bar{n}$ grows logarithmically with $d$, as shown in Figs.~\ref{fig:iter} and~\ref{fig:iter_histo}, with a slower growth for the case of pure states and highly degenerate mixed states. Equivalently, for composite systems $\bar{n}$ grows linearly with the number of constituent subsystems.

When dealing with degenerate states, $\overline{n}$ becomes smaller, as one might expect. In general, we could classify the types degeneracy for a $d$ dimensional state $\rho$ with the list of the multiplicities  $\bm{m}_\rho = (m_1,\dots,m_s)$ of the $s$ different eigenvalues of $\rho$, where $\sum_l m_l =d$. Numerical evidence suggests that $\overline{n}$ grows with $(\sqrt{\sum_l m_l^2})^{-1}$, i.e., the inverse of the length of the \textit{multiplicity vector}. For this reason, we expect the logarithmic growth of $\overline{n}$ with $d$ to persist in general, with fully non-degenerate states being the worst case scenario in terms of effective \textit{run-time} of the algorithm.

\begin{figure}[htbp]
    \centering
    \makebox[\textwidth][c]{\includegraphics[width=1.05\textwidth]{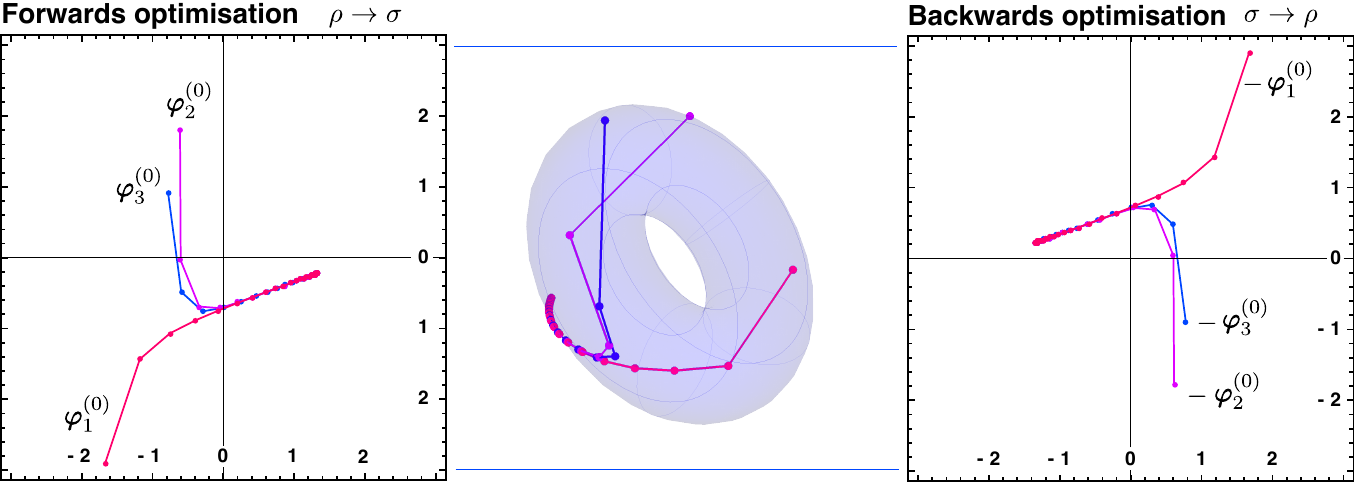}}
    \caption{Due to the symmetry of the problem, running the iterative method forwards (\emph{left}) from $\rho$ to $\sigma$ (Eq.~\eqref{eq:gate_n}) with initialisation phase $\bm{\varphi}^{(0)}_i$ is equivalent to running it backwards (\emph{right}) from $\sigma\to\rho$ (Eq.~\eqref{eq:gate_n_backwards}) with initialisation phase $-\bm{\varphi}^{(0)}_i$, up to a trivial rotation, and generate the same sequence of Hamiltonians, here represented by the sequence of geometric phases on the torus (\emph{center}), for a $3$-level system.}
    \label{fig:back_forth}
\end{figure}

Naturally, the choice of initial phase vector $\bm{\varphi}^{(0)}$ affects the number of iterations required to converge, since the corresponding Hamiltonian $H(\bm{\varphi}^{(0)})$ can be arbitrarily close to a fixed point of the iterative method. Remarkably, numerical evidence shows that the choice of initial geometric phase (and thus of initial Hamiltonian) does not noticeably affect the performance of the end point $H^{(n)}$. Interestingly,
running the algorithm backwards from $\sigma$ to $\rho$ (see Eq.~\eqref{eq:gate_n_backwards}) is equivalent to fixing a particular choice of initial phase vectors, simply given by $\bm{\varphi}^{(0)}\to-\bm{\varphi}^{(0)}$, returning not only the same optimised Hamiltonian, but also the same sequence of geometric phases $\{\bm{\varphi}^{(0)}\}$ up to rotations, as one might expect from the symmetry of the problem (See Fig.~\ref{fig:back_forth}). 

While the average number of iterations required for convergence $\bar{n}$ grows with $d$ and $\varepsilon^{-1}$, a good choice of initial geometric phase $\bm{\varphi}^{(0)}$ can still lead to rapid convergence, with runs that can take less then 10 iterations even for $d=100$ and $\varepsilon = 10^{-4}$. This observation allows us to take advantage of the strong sensitivity of the convergence rate to the choice of the initialisation phase in order reduce the effective run-time of the algorithm. The strategy is to deploy $L$ simultaneous runs of the iterative method, for the same pair of initial and final states $\rho$ and $\sigma$, with each run $i$ initialised with a different, randomly chosen geometric phase $\bm{\varphi}^{(0)}_i$. These tasks can be trivially parallelised, so that the whole computation is stopped as soon as one of these runs converges to a solution. In this way, the number $n$ of iterations required for convergence is guaranteed to be smaller than $\overline{n}$.

\begin{figure}[htbp]
    \centering
    \makebox[\textwidth][c]{\includegraphics[width=\textwidth]{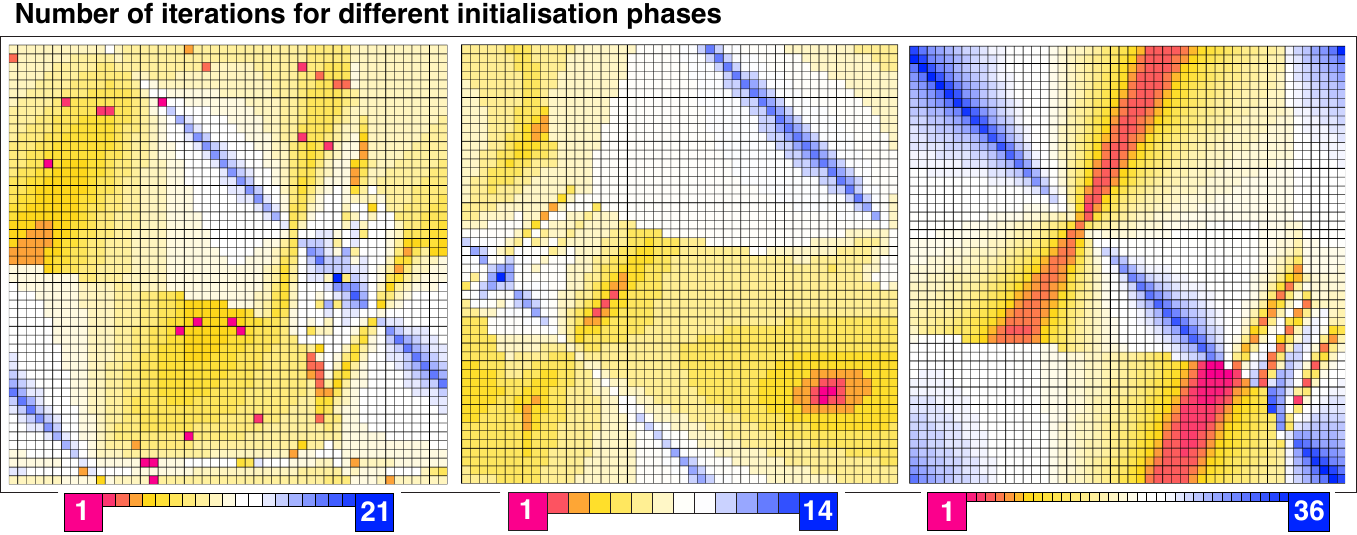}}
    \caption{Number of iterations required for convergence along initialisation phases $(0,\bm{\varphi}^{(0)}_1,\bm{\varphi}^{(0)}_2)$ for a $3$-level system, and three different choices of $\rho$, $\sigma$. The best choices of $\bm{\varphi}^{(0)}$, for which $n<4$, can occur in different patterns, such as scattered (\textit{left}), isolated (\textit{centre}), and aligned (\textit{right}). Note that, in general, Different initial phases can lead different optimised Hamiltonians with the same performance, and that the number of iterations strongly depends on the convergence threshold $\varepsilon$.}
    \label{fig:iter_phases}
\end{figure}

An interesting task is to identify convenient low-dimension subspaces of the full space of relevant initialisation phases, i.e.,  $\bm{T}^{d-1}$, from which to sample $\bm{\varphi}^{(0)}$. In Fig.~\ref{fig:iter_phases}, the number of iterations required for convergence is plotted over different choices of initialization phases for three different pairs of $\rho$ and $\sigma$, with Hilbert space dimension $d=3$. One can see how the best choices of $\bm{\varphi}^{(0)}$, for which $n<4$, can occur in different patterns, such as \textit{scattered}, \emph{isolated}, and \textit{aligned}. 

\begin{figure}[h]
    \centering
    \includegraphics[width=0.8\textwidth]{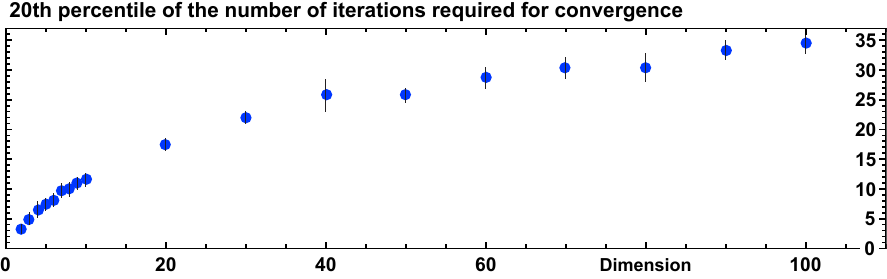}
    \caption{The average 20th percentile of the number of iterations required for convergence plotted along the Hilbert space dimension $d$. The \emph{error bars} represent the standard deviation for each dimension, with a sample of $10^2$ pairs of initial and final states $\rho$ and $\sigma$, and $10^2$ random initial phases for each pair.}
    \label{fig:monte_carlo_worthy}
\end{figure}

Even when sampling the phases over their full space $\bm{T}^{d-1}$, it is possible to take advantage of rapidly converging initialisation phases to improve the runtime. We study the dependence of $n$ on the choice of initial geometric phases by running several instances of the method for the same pair of states $\rho$ and $\sigma$, and sampling each phase $\varphi^{(0)}_j$ uniformly in the interval $[0,2\pi]$. The slow growth of the 20th percentile of the number of iterations required for convergence, shown in Fig.~\ref{fig:monte_carlo_worthy}, supports the potential of combining Algorithm~\ref{a:algorithm} with Monte-Carlo sampling methods~\cite{Hastings1970,Vanderbilt1984,Kalos2008}.

\section{Robustness under perturbations}
\label{s:robustness_method}
The iterative method defined by Algorithm~\ref{a:algorithm}, as well as its variants in Eqs.~\eqref{eq:gate_n_backwards} and~\eqref{eq:gate_n_back_forth}, is stable under perturbations of the initial and final states $\rho$ and $\sigma$. First, let us consider the case of convex mixtures
\begin{equation}
    \label{eq:convex_mixture}
    \rho\to\rho' = (1-\delta) \rho + \delta \; \chi,
\end{equation}
where $\chi\in\mathcal{S}(\mathcal{H})$ is a random state of the same dimension as $\rho$ and $\sigma$, while $\delta\in(0,1)$ is the perturbation strength. To make sure that the perturbed final state $\sigma'$ is isospectral with $\rho'$, we choose it as
\begin{equation}
    \sigma' = O\rho'O^\dagger,
\end{equation}
where $O$ can be any unitary that maps $\{\ket{r_k}\}_{k=1}^d\to\{\ket{s_k}\}_{k=1}^d$.
This type of perturbation is expected to change the spectral properties of the considered states, with $\rho$ having in general different spectrum than $\rho'$. In this case, both the optimised Hamiltonian and its performance may vary discontinuously in the perturbation strength with respect to the unperturbed case. Numerically, the states $\chi$ have been sampled uniformly according to the Bures ensemble (see {\fontfamily{phv}\selectfont\textbf{Box~\ref{tech:qsl_tightness}}}).

We also considered unitary perturbations, here defined by 
\begin{align}
    \label{eq:unitary_transform}
    \rho\to\rho' = e^{i \widetilde{H} \delta} \rho \; e^{-i \widetilde{H} \delta}, \\
    \sigma\to\sigma' = e^{i \widetilde{H} \delta} \sigma \; e^{-i \widetilde{H} \delta},
\end{align}
where $\widetilde{H}$ is a random Hamiltonian of unit Hilbert-Schmidt norm. Numerically, the Hamiltonians $\widetilde{H}$ have been obtained from uniformly sampled Haar-random unitaries, as explained in {\fontfamily{phv}\selectfont\textbf{Box~\ref{tech:qsl_tightness}}}.

\begin{figure}[htpb]
    \centering
    \includegraphics[width=0.8\textwidth]{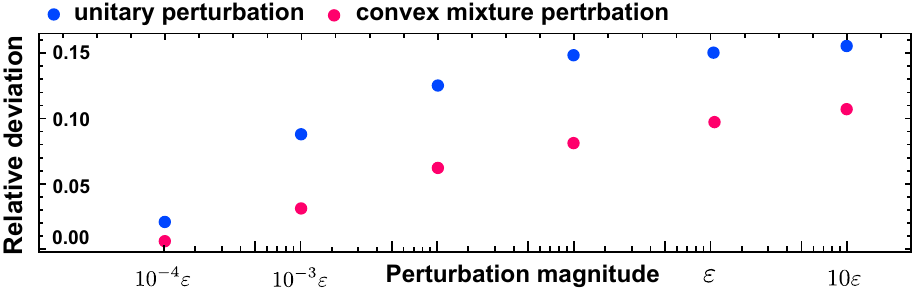}
    \caption{The relative deviation $\bm{\Delta}(\delta)$ from the unperturbed solutions, given in Eq.~\eqref{eq:relative_perturbation}, is plotted against the perturbation magnitude $\delta$ expressed in units of the numerical precision $\varepsilon$, with $\varepsilon=10^{-3}$, and plotted in logarithmic scale.}
    \label{fig:perturbation}
\end{figure}

To study the effect of perturbations we use the \textit{relative deviation} $\bm{\Delta}(\delta)$ from the unperturbed solutions, defined as
\begin{equation}
    \label{eq:relative_perturbation}
    \bm{\Delta}(\delta):=\frac{\lVert H^{(n)} - {H'}^{(n')}\rVert}{\lVert H^{(n)} \rVert},
\end{equation}
where ${H'}^{(n')}$ is the solutions to the perturbed problem, which in general requires a number $n'$ of iterations different from $n$ for the same chosen convergence threshold $\varepsilon$, and where $\lVert\cdot\rVert$ is the Hilbert-Schmidt norm. Numerical evidence suggests that the relative deviations grow slowly as a function of the perturbation strength $\delta$, and are negligible for $\delta \ll \varepsilon$, as shown in Fig.~\ref{fig:perturbation}.

\section{Signatures of non-monotonicity}
\label{s:non-mono}
In Sec.~\ref{s:hamiltonian_efficiency} we considered a notion of Hamiltonian efficiency, introduced in Ref.~\cite{Uzdin2015} for the case of pure states, and we showed it cannot be saturated $(\eta = 1)$ when mixed states are considered (see Eqs.~\eqref{eq:eta_xy} and~\eqref{eq:eta_z}). Nevertheless, when it comes to estimate how much energy of the driving Hamiltonian is wasted along components of the Lie algebra that do not actively contribute to driving the system, it is possible to adapt such notion of efficiency to the case of density operators. A possible generalisation is given by $\eta^\star(H,\rho)$, introduced in Eq.~\eqref{eq:efficiency_star} (See {\fontfamily{phv}\selectfont\textbf{Box.~\ref{tech_eta_mixed}}}). 

\begin{figure}[htbp]
    \centering
    \includegraphics[width=0.9\textwidth]{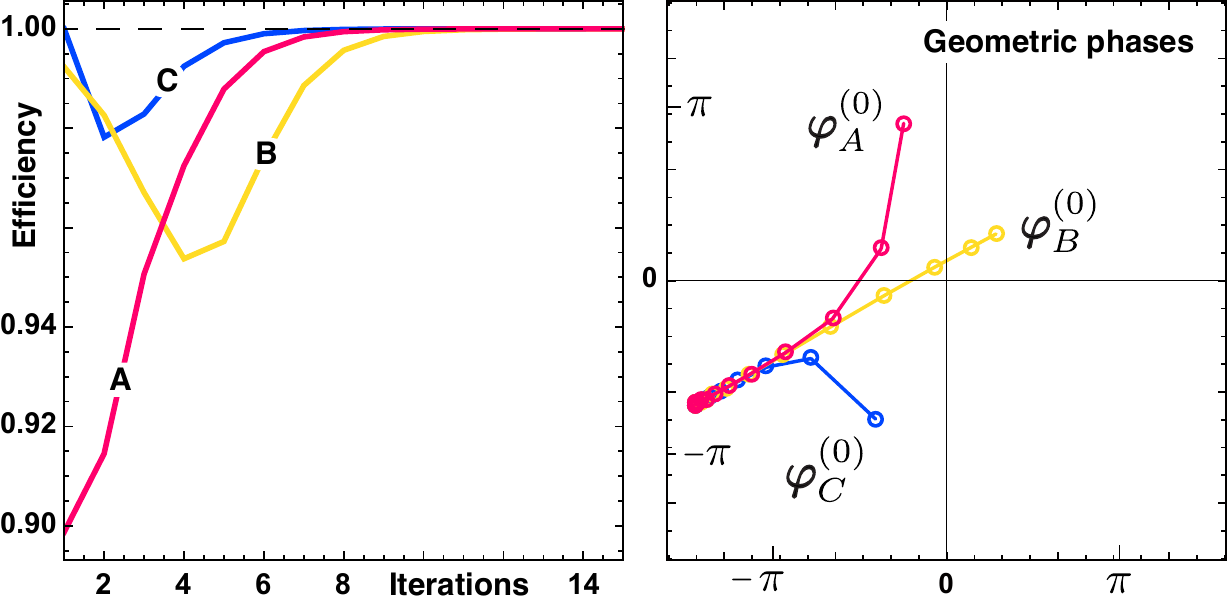}
    \caption{Algorithm~\ref{a:algorithm} is here applied on the same choice of $\rho$ and $\sigma$ for three different choices of initial geometric phases $\bm{\varphi}^{(0)}_A$, $\bm{\varphi}^{(0)}_B$, and $\bm{\varphi}^{(0)}_C$. The Hamiltonian efficiency $\eta^\star(H^{(j)},\rho)$ of the Hamiltonians in the three different sequences is plotted along the iterations $i$ for each run (\textit{left}). Since each Hamiltonian (and unitary) has an associated geometric phase as prescribed by Eq.~\eqref{eq:o_phi}, we can plot the trajectory of the geometric phase for each run, such as as $\bm{\varphi}^{(j)}_A=(\varphi^{(j)}_{A,1},\varphi^{(j)}_{A,2},\varphi^{(j)}_{A,3})$ for run $A$. By neglecting the first phase, which can be absorbed into a global phase, we plot the remaining components on a plane (\textit{right}). All runs converge to the same solutions, however, while the efficiency of run $A$ converges monotonically, the others converge non-monotonically. The trajectories of their geometric phases (\textit{right}) do not carry any obvious signature of non-monotonicity, and straight paths can correspond to the slowest non-monotonic descent towards the fixed point. Moreover, such trajectories can show richness and variety for different choices of $\rho$ and $\sigma$, and of initial geometric phases.} 
    \label{fig:non_mono}
\end{figure}

When applying this notion of efficiency to the sequence of Hamiltonians $\{H^{(j)}\}_{j=0}^{n}$ obtained with this iterative method, it is possible to witness a transient non-monotonic behavior for $\eta^\star$, followed by a rapid monotonic ascent towards unit efficiency, as shown in Fig.~\ref{fig:non_mono}. While we cannot rule out the existence of functionals that are strictly-monotonic over the sequence $\{H^{(j)}\}_{j=0}^{n}$, this signature of non-monotonicity can be seen in other, much simpler, figures of merit, such as the energy $\tr[\rho H^{(j)}]$ of the initial state $\rho$ with respect to the Hamiltonians $H^{(j)}$ of the sequence.

\begin{techbox}{Hamiltonian efficiency for mixed states}{tech_eta_mixed}
A possible notion of Hamiltonian efficiency that can be saturated for the case of density operators is given by
\begin{gather}
    \label{eq:efficiency_star}
    \eta^\star(H,\rho) = \frac{\sqrt{\tr[\rho^2H^2]-\tr[(\rho H)^2]}}{\sqrt{\tr[\rho^2 H^2]-(\tr[(\rho H)^2]-\tr[\rho H]^2)}},
\end{gather}
which is a positive function smaller than 1, that is saturated for $H_\| = 0$. The numerator of $\eta^\star$ is proportional to the speed $\lVert \dot{\rho}\rVert_{HS}$ of the generator $\dot{\rho} = -i[H,\rho]$~\cite{DelCampo2013,Deffner2013,Campaioli2018}, and reduces to $\Delta H$ when pure states are considered, for which $\eta^\star =1 \leftrightarrow \eta =1$. 

Like the numerator of $\eta$, that of $\eta^\star$ is also invariant under the addition of a parallel component $H_\|$ with respect to $\rho$. Accordingly, the Hamiltonians generated by the iterative method are of unit efficiency for all the considered QBPs, which reflect the ability to converge to a fully-perpendicular Hamiltonian. However, the convergence of this iterative method is, in general, non-monotonic with respect to this notion of efficiency, which can decrease before reaching $\eta^\star =1$ after several iterations, as shown in Fig.~\ref{fig:non_mono}.
\end{techbox}

\section{Chapter summary}
\noindent
In this Chapter, we have introduced an iterative algorithm to obtain efficient time-independent Hamiltonians that generate fast evolution between two $d$-dimensional isospectral states $\rho$ and $\sigma$. Such a method could be used to address fast driving, state preparation, and gate design problems.
Recently, such problems have been experimentally tackled in Refs.~\cite{Schaff2011, An2016,Du2016} by means of \textit{shortcuts to adiabadicity} and \textit{transitionless quantum driving}~\cite{Berry2009, DelCampo2012a,Santos2016a, Campbell2017b} Incidentally, these methods are closely linked to ours, with a key common element being the optimisation of geometric phases.
Other possible applications of the iterative method introduced here could be found in its combination with quantum optimal control methods. Algorithm~\ref{a:algorithm} can be used to identify a unitary $\widetilde{O}$, associated with an efficient Hamiltonian $\widetilde{H}$, to be implemented by means of gate design methods, such as those experimentally realised in Ref.~\cite{Geng2016}, or that introduced in Ref.~\cite{Innocenti2018} for the case of restricted generators. More generally, our method has broad application in further elucidating the geometric structure of quantum control, since it has been shown that quantum brachistochrone problems can be recast as those of finding geodesics in the space of unitary operators~\cite{Wang2015}.

A challenging outlook is to extend and adapt this method to more general control problems. However, it's not clear that, for open quantum evolutions, simply replacing the Hamiltonian with a Liouvillian as the generator would preserve the complete positivity of the dynamics. Nevertheless, it is possible to adopt the current method for open dynamics by applying it to purification of $\rho$ and $\sigma$ or dilation of the dynamics. Another challenge that remains is to build in constraints on the form of the generators, naturally imposed by physical restrictions on  the order and range of interactions. These can dramatically change the time required to perform a given evolution, as we will show in detail in the next chapter, where we apply QSLs to bound the power of work extraction and deposition for many-body quantum systems. 
\chapter[Bounds on work extraction and power of quantum systems]{Bounds on work extraction and power \\ of quantum systems}
\label{ch:quantum_batteries} 

\begin{ch_abs}
\textsf{\textbf{Chapter abstract}}\vspace{5pt}\\
\small\textsf{In this chapter, time-energy uncertainty relations are applied to derive a bound on the achievable power of work extraction and energy deposition for isolated many-body quantum system. First, we review some fundamental concepts and results of quantum thermodynamics, such as ergotropy and passive states. Then, we introduce the notion of quantum battery in order to study the bounds on power, by applying some of the results and methods described in the previous chapters of this thesis. This allows us to demonstrate the power advantage of using global \emph{entangling} operations over local ones, and show that it can grow at most extensively with the number of subsystems, while being strongly limited by the order of the interactions available for controlling them. \vspace{10pt}\\
\emph{This chapter is based on publications \cite{Campaioli2017} and \cite{Campaioli2018a}.}}
\end{ch_abs}\vspace{20pt}

\section{Quantum thermodynamics and quantum batteries}
\label{s:quantum_thermo}
The current rate of technology miniaturisation requires us to carefully consider the fundamental laws of physics of the microscopic domain, where are increasingly expected to function. This regime is replete with thermal and quantum fluctuations, which must be accounted for in any complete physical description. When  dealing with technologies working in the quantum regime, familiar thermodynamic concepts like work, heat, and entropy need to be applied with great care and consideration. For this reason, there has been an intense effort to understand how the laws of thermodynamics generalize to arbitrary quantum systems away from equilibrium. This effort is known as quantum thermodynamics and, given current interest in the development of quantum technologies, it is receiving attention across a wide range of scientific communities~\cite{Goold:16, Millen:16, Vinjanampathy:15}. 

Despite current momentum in the field of quantum thermodynamics, the precise role of genuinely quantum features in the operation of thermal machines is not fully understood. A common issue raised is that the universal applicability of thermodynamics is rooted in the theory's lack of respect for microscopic details and is bound to overlook some fundamental features of quantum mechanics~\cite{Gardas}. Nevertheless, if one relaxes the assumptions of large system size and quasi-static conditions, it is absolutely reasonable to get corrections based on the fine details of the working medium~\cite{Bender2000, Scully2003, Brunner2014, Uzdin2015}. Can such quantum features be harnessed to improve other meaningful figures of merit, such as power?

Collective quantum phenomena are known to offer advantages in areas such as computation, secure communication, and metrology. These advantages have received a great deal of attention in the context of work extraction and deposition~\cite{Alicki:13, Hovhannisyan:13, Giorgi2015, Bruschi:15, Friis:15, Huber:15, Binder15b, Perarnau-Llobet:15,  BinderThesis}. In particular, Alicki and Fannes suggested that entangling operations lead to increased work extraction from an energy storage device which they coined a \emph{quantum battery}~\cite{Alicki:13}. Nonetheless, while entangling operations are necessary for optimal work extraction, it has been shown that protocols exist for which no entanglement is actually created during optimal work extraction~\cite{Hovhannisyan:13, Giorgi2015}. Furthermore, considering a regime where entangling operations do not increase the extractable energy, entangling operations can still improve the charging power of arrays of quantum batteries~\cite{GelbwaserKlimovsky2015329}, as shown in Ref.~\cite{Binder15b} for the case of two-level systems.

Since their introduction in the literature, quantum batteries have been studied using the methods of quantum information and quantum thermodynamics \cite{Hovhannisyan2013}, with interesting conclusions and insights. While the first works focused on work extraction, and on the limits on the amount of work that can be extracted from such devices by means of cyclic unitary operations~\cite{Alicki2013,Hovhannisyan2013}, a more recent line of research has aimed to understand the role of entanglement, and entangling operations, in tasks like work extraction and charging \cite{Hovhannisyan2013,Binder2015,Campaioli2017,Andolina2019,Andolina2018a}. It has been shown that global, entangling operations can outperform local ones when charging an array of batteries, and that this advantage can grow extensively with the number of units involved~\cite{Binder2015}. 
It is no surprise that the advantage that entangling operations have over local ones is directly connected with time-energy uncertainty relations, interpreted as bounds on the minimal time required to evolve a system between different energetic levels in order to deposit or extract work, as we show in Sec.~\ref{s:enhancing}. 

\section{Work extraction}
\label{s:work_extraction}

\noindent
The study of work extraction from small quantum systems via reversible cyclic operations starts with the aim of defining the thermodynamical bounds and principles that are valid at those scales where a quantum mechanical description becomes necessary \cite{Alicki2013}. The intention is to look at the limits of extractable work allowed by quantum mechanics and compare them to their classical counterpart, while looking for possible advantages.

\begin{figure}[b!]
    \centering
    \includegraphics[width = 0.65\textwidth]{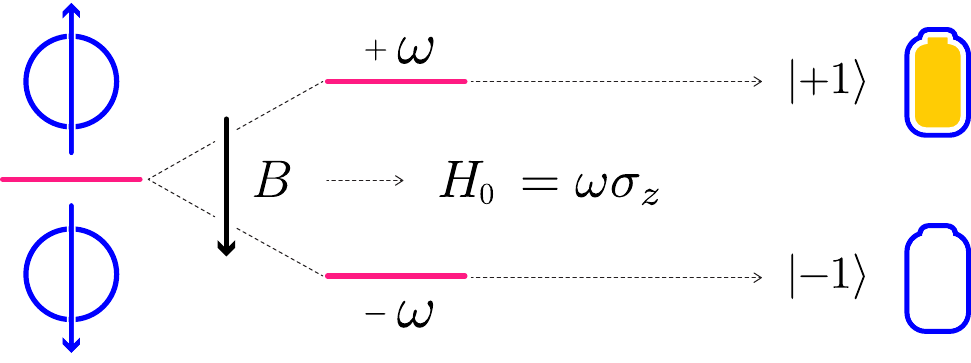}
    \caption{A simple quantum battery could consist of a non-degenerate two-level system, such as the spin of an electron immersed in a uniform magnetic field $B$, whose internal Hamiltonian $H_0= \omega \sigma_z$
has energy levels $\pm\omega$, associated with the eigenstates $\ket{\pm 1}$, respectively. The pure state $\ket{+1}$ ($\ket{-1}$) is considered a charged battery (discharged battery), since no work can be deposited onto (extracted from) it, with respect to the internal Hamiltonian $H_0$.}
    \label{fig:2-levle-battery}
\end{figure}

Let us start from a single \emph{closed}\footnote{In thermodynamics a \emph{closed} system is only allowed to exchange either work or heat, in contrast with an \emph{isolated} system which is not allowed to exchange either of them. However, here \emph{closed} stands for an isolated quantum system undergoing Schr\"odinger evolution, whose initial state can nevertheless be mixed.} quantum battery, the fundamental unit of this discussion. It consists of a $d$-dimensional system with associated internal Hamiltonian $H_0$
\begin{equation}
    \label{eq:internal_hamiltonian}
    H_0=\sum_{j=1}^d\omega_j\ketbra{j}{j}, 
\end{equation}
with non-degenerate energy levels $\omega_j<\omega_{j+1}$. From this definition, a simple quantum battery could consist of a non-degenerate two-level system, such as the spin of an electron immersed in a uniform magnetic field, as shown in Fig. \ref{fig:2-levle-battery}.

A time-dependent field $V(t)$ is used to reversibly extract energy from a battery via unitary evolution generated by $H(t) = H_0 +V(t)$, where with a slight change of notation with respect to the previous chapter, we expressed the dependence on time with round brackets. Given that the battery is found in some initial state described by the density operator $\rho$, the time evolution of the system is obtained from the von Neumann equation
\begin{equation}
    \label{eq:unitary_charging}
    \dot{\rho}(t) = -i[H(t),\rho(t)],
\end{equation}
with $\rho(0)=\rho$, and where the left-hand side represents the time derivative of $\rho(t)$. A solution of Eq.~\eqref{eq:unitary_charging} is given by $\rho(t)=U(t)\:\rho\: U^\dagger(t)$, where the unitary operator $U(t)$ is obtained as the time-ordered exponential of the generator $H(t)$, given in Eq.~\eqref{eq:dyson_series}, which can correspond to any unitary transformation on the battery's Hilbert space $\mathcal{H}$.

In practice, the control field $V(t)$ may be strongly limited by the operations that one is able to perform on the considered system. For example, one might be able to implement a \mono{NOT} gate on a single qubit, generated by some Pauli operator $\sigma_i$, while being unable to perform a \mono{CNOT} (controlled-not) gate acting on two qubits, because of lack of control over some two-body generators, such as $\sigma_i\otimes\sigma_j$ \cite{Nielsen2000}. 
For now we will assume the ability to perform any operation, unless specified otherwise, for the sake of simplicity. We will discuss the case of limited control later in Sec.~\ref{s:enhancing}.

\subsection{Ergotropy \& passive states}
\label{ss:ergotropy}
\begin{figure}[b!]
    \centering
    \includegraphics[width = 0.7\textwidth]{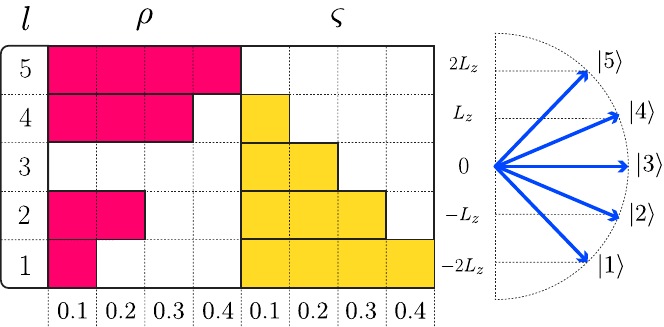}
    \caption{A 5-level system with internal Hamiltonian $H_0=L_z\sum_{l=1}^5 (l-3) \ketbra{l}{l}$ in state $\rho=0.1 \ketbra{1}{1}+0.2\ketbra{2}{2}+0.3\ketbra{4}{4}+0.4\ketbra{5}{5}$ has an associated passive state $\varsigma=0.4 \ketbra{1}{1}+0.3\ketbra{2}{2}+0.2\ketbra{3}{3}+0.1\ketbra{4}{4}$. From $\rho$ one can extract the ergotropy $W_{\max}=1.8 L_z$ by means of a unitary operation, such as $U=\ketbra{1}{5}+\ketbra{2}{4}+\ketbra{5}{3}+\ketbra{3}{2}+\ketbra{4}{1}$. This unitary operation is not unique, since an arbitrary relative phase can be introduced for each term $\ketbra{i}{j}$. Here, the eigenvalues of $\rho$ and $\varsigma$ are represented by the length of the coloured rows associated with the energy levels $l$ that they occupy. To obtain the passive state from $\rho$ one is only allowed to permute the rows. This constraint represents the conservation of the spectrum imposed by unitary evolution.}
    \label{fig:passive-5}
\end{figure}

\noindent
Let us go back to the battery that we have introduced above in Eq~\eqref{eq:internal_hamiltonian}, which evolves according to the dynamics described by Eq.~\eqref{eq:unitary_charging}.
The work extracted after some time $\tau$ by this unitary cycle is given by
\begin{align}
    \label{eq:work}
    W(\tau) &=\tr[\rho H_0]-\tr[\rho(\tau) H_0]
    \\
    &=\tr[\rho H_0]-\tr[U(\tau)\rho U^\dagger(\tau) H_0],
\end{align}
which corresponds to a measurement of the decrease of energy of the system in the time interval $\tau$, with respect to the interal Hamiltonian $H_0$.
Since we are interested in reversible work extraction, we look for the maximal amount of extractable work, known as \emph{ergotropy} \cite{Allahverdyan2004,Francica2017},  optimizing $W$ over all unitary operations,
\begin{equation}
    \label{eq:ergotropy}
    W_{\max}:=\tr[\rho H_0]-\min_{U\in SU(d)}\big\{\tr[U\rho U^\dagger H_0]\big\}.
\end{equation}
When no work can be extracted from a state $\varsigma$, it is said to be \emph{passive}, i.e., when $\tr[\varsigma H_0]\leq\tr[U\varsigma U^\dagger H_0]$ for all unitaries $U$ \cite{Pusz1978,Lenard1978,Alicki2013}.
Accordingly, a state is passive if and only if it commutes with the internal Hamiltonian $H_0$ and has non-increasing eigenvalues, i.e., if
\begin{equation}
    \label{eq:passive}
    \varsigma=\sum_{j=1}^d \omega_j \ketbra{j}{j}, \; \omega_{j+1}\leq \omega_j,
\end{equation}
as shown in Ref. \cite{Alicki2013}. From this definition it is possible to see that for any state $\rho$ there is a unique passive state $\varsigma = \bm{\mathsf{p}}(\rho)$,
\begin{equation}
\label{eq:passive_of_state}
    \bm{\mathsf{p}}(\rho) :=V\rho V^\dagger,
\end{equation}
that maximizes the extractable work
\begin{equation}
    \label{eq:ergotropy_passive}
    W_{\max}=\tr[\rho H_0]-\tr[\varsigma H_0],
\end{equation}
obtained via some unitary operation $V$ that rearranges the eigenvalues of $\rho$ in non-increasing order, as illustrated in Fig. \ref{fig:passive-5}.

\subsection{Bounds on extractable work}
\label{ss:bounds_on_extractable_work}

\noindent
A practical way to bound the work extractable from $\rho$, given in Eq. \eqref{eq:ergotropy}, is to consider a thermal state with the same von Neumann entropy as the considered state $\rho$, which also minimizes the energy with respect to $H_0$. 
It has been shown that a lower bound to the ergotropy for some state $\rho$ is given by
\begin{equation}
    \label{eq:bound_ergotropy}
    W_{\max}\leq \tr[\rho H_0]-\tr[\mathcal{G}_{\bar{\beta}}H_0],
\end{equation}
where $\mathcal{G}_{\bar{\beta}}$ is the Gibbs state whose von Neumann entropy is equal to that of $\rho$~\cite{Alicki2013}, i.e., $S(\mathcal{G}_{\bar{\beta}}) = S(\rho)$, as prescribed by,
\begin{equation}
\label{eq:von_neumann_entropy}
S(\rho) = -\tr[\rho\log\rho],
\end{equation}
and
\begin{align}
\label{eq:gibbs_state}
\mathcal{G}_\beta = \frac{\exp[-\beta H]}{\mathcal{Z}}, \;\;\;
 \mathcal{Z} = \tr[\exp[-\beta H]].
\end{align}
\vspace{15pt}

\begin{figure}[b!]
    \centering
    \includegraphics[width = 0.9 \textwidth]{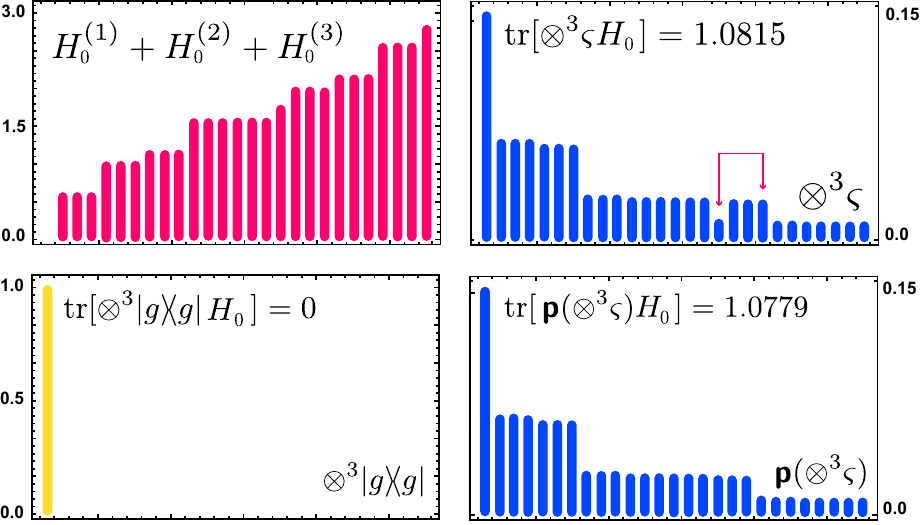}
    \caption{Here, a three-level battery is defined by an internal Hamiltonian $H_0^{(l)}=\diag(0,0.579,1)$. State $\varsigma=\diag(0.538,0.237,0.224)$ is passives, as it is diagonal in the basis of $H_0^{(l)}$, and has non-increasing eigenvalues. The 27 eigenvalues of state $\otimes^3\varsigma$ are represented as a bar chart (\textit{top right}), in a basis for which the local internal Hamiltonian $H_0^{(1)}+H_0^{(2)}+H_0^{(3)}$ has non-decreasing eigenvalues (\textit{top left}). It is easy to see that the eigenvalues of $\otimes^3\varsigma$ are not non-increasing as indicated by the arrows (\textit{top right}), thus $\otimes^3\varsigma$ is not a completely passive state. The passive state of the copies of $\varsigma$ is $\bm{\mathsf{p}}(\otimes^3\varsigma)$ (bottom right), and it is obtained by swapping the eigenvalues indicated by the arrows. On the contrary, the pure state $\ketbra{g}{g}=\diag(1,0,0)$ is completely passive, since the state of any of its copies, such as $\otimes^3\ketbra{g}{g}$ (\textit{bottom left}), is still passive with respect to the local internal Hamiltonian.}
    \label{fig:passive_example}
\end{figure}

A key observation is that all thermal states are passive, while not all passive states are thermal. A notable exception, as usual\footnote{Quantum two-level systems have unique features that cannot always be straightforwardly generalized to higher dimensional cases.}, is given by the case of two-level systems, for which all passive states are thermal, as discussed in {\fontfamily{phv}\selectfont\textbf{Box~\ref{tech:two_level_thermal}}}. Even more interestingly, the product of two or more copies of a passive state $\varsigma$ is not necessarily the passive state of the copies of $\rho$. Using the notation $\otimes^N A = A\otimes A\otimes\cdots\otimes A$ to represent the tensor product of $N$ copies of some operator $A$, we can express the previous statement as
\begin{equation}
    \label{eq:passive_state_copies}
    \otimes^N \bm{\mathsf{p}}(\rho) \neq \bm{\mathsf{p}}(\otimes^N \rho).
\end{equation}
This observation, illustrated and discussed with an example in Fig.~\ref{fig:passive_example}, leads to the definition of \emph{completely passive states}, as those whose $N$-copy ensembles are still passive for any $N$ \cite{Alicki2013}. It has been shown that a state is completely passive if and only if it is a thermal state \cite{Pusz1978,Lenard1978}, an observation that can be used to beat the bound given in Eq. \eqref{eq:ergotropy_passive} for many copies of the same battery
by means of entangling operations \cite{Alicki2013}. Let us see how this works in the next subsection. 

\begin{techbox}{Temperature of two-level systems}{tech:two_level_thermal}
Let us see how it is always possible to define a (positive or negative) temperature for a qubit. In its energy eigenbasis, a qubit can be written as
\begin{equation}
    \label{eq:qubit_energy_eigenbasis}
    \rho = p_0 \ketbra{0}{0} + (1-p_0)\ketbra{1}{1},
\end{equation}
where $p_0$ and $1-p_0$ are the probabilities of finding the system in the states $\ketbra{0}{0}$ and $\ketbra{1}{1}$, with energy $\omega_0$ or $\omega_1$, respectively. If we assume the qubit to be in thermal equilibrium, we can represent it as the Gibbs state $\rho = \exp[-\beta (\omega_0 \ketbra{0}{0}+\omega_1\ketbra{1}{1})]/\mathcal{Z}$, thus
\begin{equation}
    \label{eq:qubit_temperature_probabilities}
p_0 = \frac{\exp[-\beta \omega_0]}{\mathcal{Z}}, \;\;\;
1 - p_0 = \frac{\exp[-\beta \omega_1]}{\mathcal{Z}},
\end{equation}
from which we can calculate the inverse temperature
\begin{equation}
    \label{eq:qubit_temperature}
    \beta = \frac{\log\frac{p_0}{1-p_0}}{(\omega_1-\omega_0)}.
\end{equation}
However, for systems with dimension greater than two, the Gibbs state $\mathcal{G}_{\bar{\beta}}$ is in general different from the passive state $\varsigma$ associated with $\rho$. 
\end{techbox}

\subsection{Optimal work extraction}

\noindent
Consider a battery given by an ensemble of $N$ copies of the same $d$-dimensional unit cell defined by Eq. \eqref{eq:internal_hamiltonian}. This new battery has an associated internal  Hamiltonian $H_0$ given by the sum of the local internal Hamiltonians $H_0^{(l)}\otimes_{l'\neq l}\mathbbm{1}^{(l')}$ of the subsystems that form the global system
\begin{equation}
    \label{eq:interal_array_hamiltonian}
    H_0=\sum_{l=1}^N H_0^{(l)},
\end{equation}
where we omit the identities to simplify the notation. Since this Hamiltonian would likely have eigenvalues with high multiplicity, we will no longer assume they are non-degenerate, as long as there are at least two different energy levels, associated with distinguishable eigenstates.

\begin{figure}[h]
    \centering
    \includegraphics[width = 0.8\textwidth]{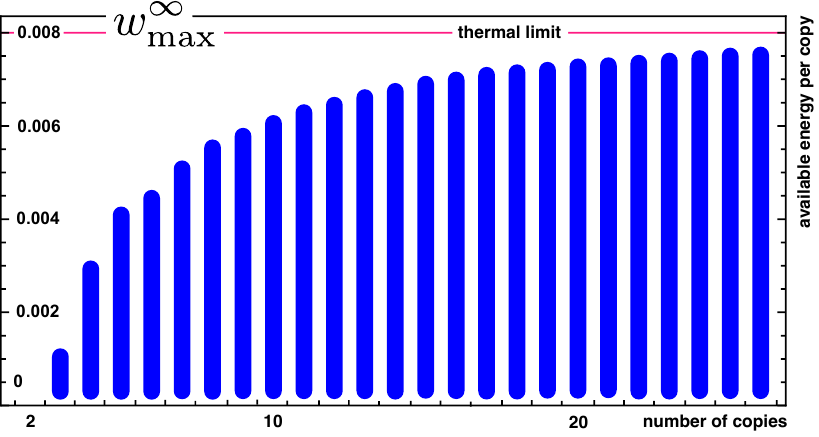}
    \caption{We consider an ensemble of $N$ copies of the same passive state $\varsigma$. As discussed in this section, there is a non-trivial amount of work per copy $w_{\max}(N)$ that we can extract from such an ensemble by means of entangling operations (at least 2-body operations), given by $\frac{1}{N}\{\tr[(\otimes^N\varsigma-\bm{\mathsf{p}}(\otimes^N\varsigma)\}) H^{(N)}]\}$, where $\bm{\mathsf{p}}(\otimes^N\varsigma)$ is the passive state for $\otimes^N \varsigma$. This figure represents this additional available energy per copy of passive state, for the three-level system considered in Fig.~\ref{fig:passive_example}, with energy levels $\{0,0.579,1\}$ and passive state $\varsigma$ with eigenvalues $\{0.538,0.237,0.224\}$ \cite{Alicki2013}. In particular, the maximal amount of extractable energy $ w^\infty_{\max}$, as in Eq.~\eqref{eq:limit_work_available}, is obtained in the limit of $N\to\infty$.}
    \label{fig:energy-copy}
\end{figure}

Recalling that the composite state of $N$ copies of a passive state $\varsigma$ might not be passive, our goal is to extract some additional work from $\otimes^N \varsigma$ until a completely passive state is reached. 
How much work can be extracted in this way? 
In the limit of large $N$, the maximal amount of available work \emph{per copy} of a battery $w_{\max}(N)$ in state $\rho$ is tightly bounded as in Eq. \eqref{eq:bound_ergotropy}, 
\begin{equation}
\label{eq:limit_work_available}
    \lim_{N\to\infty}w_{\max}(N)=\tr[\rho H_0^{(l)}]-\tr[\mathcal{G}_{\bar{\beta}} H_0^{(l)}],
\end{equation} 
where $\bar{\beta}$ represent the inverse temperature of a Gibbs state with von Neumann entropy equal to that of $\rho$, and where the maximal work per copy is written in terms of the passive state of $N$ copies of $\rho$,
\begin{equation}
    \label{eq:ergotropy_per_copy}
    w_{\max}(N):=\frac{1}{N}\bigg(\tr\Big[\Big(\otimes^N\rho-\bm{\mathsf{p}}(\otimes^N\rho)\Big) H_0\Big]\bigg).
\end{equation}
The proof of Eq.~\eqref{eq:limit_work_available} relies on the idea that for a large ensemble, the energy of the passive state $\bm{\mathsf{p}}(\otimes^N \rho)$ differs from that of $\otimes^N \mathcal{G}_{\bar{\beta}}$ only by a small amount that tends to vanish as $N$ increases  \cite{Alicki2013}. A numerical evaluation of the maximal work per copy is shown in Fig.~\ref{fig:energy-copy}.

The passive state $\bm{\mathsf{p}}(\otimes^N \rho)$ associated with any $\otimes^N\rho$ is diagonal in the eigenbasis of the local Hamiltonian of Eq.~\eqref{eq:interal_array_hamiltonian}, thus it is separable. However, as we will see later in this section, in order to unitarily connect $\otimes^N\rho$ to its passive state, at least 2-body operations are required. This remark led Alicki and Fannes to hypothesize that, in order to reach optimal work extraction, the unit cells of such an $N$-fold battery have to be dynamically entangled. However, as proven in Ref. \cite{Hovhannisyan2013}, optimal work extraction can be achieved while keeping the composite system in a separable state at all times. Even if non-local operations (i.e., at least two-body operations) are required to beat the classical limit of Eq. \eqref{eq:ergotropy_passive}, it is always possible to reach optimal work extraction without creating any entanglement, at the expense of requiring more operations, and thus additional time, as described  in {\fontfamily{phv}\selectfont\textbf{Box~\ref{tech_optima_work_extraction}}}.\\

\begin{techbox}{Separable optimal work extraction for arbitrary dimension}{tech_optima_work_extraction}
Let us consider a simple example that illustrates how to perform optimal work extraction from multiple copies of the same state without generating entanglement: A three-level system with (increasing) energy levels, associated with eigenvalues $\{\omega_1,\omega_2,\omega_3\}$, and two copies of some initial state $\rho=p\ketbra{2}{2}+(1-p)\ketbra{3}{3}$, with $p\in(0,1/2)$. The objective is to transform the initial compisite state
\begin{equation}
    \label{eq:initial_composite}
    \begin{split}
    \otimes^2\rho=\;\;&p^2\ketbra{22}{22}+p(1-p)\Big(\ketbra{23}{23}+\ketbra{32}{32}\Big)+\\
    &+(1-p)^2\ketbra{33}{33},
    \end{split}
\end{equation}
into a passive state 
\begin{equation}
    \label{eq:passive_composite}
    \begin{split}
    \bm{\mathsf{p}}(\otimes^2\rho)=\; &(1-p)^2\ketbra{11}{11}+p(1-p)\Big(\ketbra{12}{12}+0.21\ketbra{21}{21}\Big)+\\
    &+p^2\ketbra{13}{13},
    \end{split}
\end{equation}
by means of the permutation that maps $\ket{33}\to\ket{11}$, $\dots$, and $\ket{22}\to\ket{13}$. In order to avoid entanglement one can perform each swap in several steps, such as $\ket{33}\to\ket{13}$ followed $\ket{13}\to\ket{11}$, each of which keeps the state in a separable form at all times, if performed by means of controlled permutations and unitaries. 

This idea can be generalized to the case of arbitrary dimension $d$ and for any number $N$ of copies of the initial state: An $N$-body battery in an initial non-passive state $\otimes^N\rho=\diag(p_1,\cdots,p_{d^N})$, where $p_a\geq0$ and $\sum_a p_a=1$. To perform optimal work extraction we need to evolve $\otimes^N\rho$ to the passive state $\bm{\mathsf{p}}(\otimes^N\rho)=\diag (s_1,\cdots,s_{d^N})$, where $s_{a+1}\leq s_a$ and $\bra{a}\bm{\mathsf{p}}(\otimes^N\rho)\ket{a} = s_a=\Pi_{ab}p_b$, for some permutation $\Pi_{ab}$. To do so, each transposition $a\leftrightarrow b$ that swaps $p_a$ with $p_b$ is addressed separately by transforming $\ket{a}$ to $\ket{b}$ (and vice versa) with a sequence of steps: First, $\ket{a}=\ket{i_1^a i_2^a \cdots i_N^a}$ is mapped to $\ket{a'}=\ket{i_1^b i_2^a \cdots i_N^a}$, then to $\ket{a''}=\ket{i_1^b i_2^b \cdots i_N^a}$, and so on until it reaches $\ket{b}=\ket{i_1^b i_2^b\cdots i_N^b}$, after $N$ steps. Each of these steps is obtained by a unitary operator 
\begin{equation}
\label{eq:step_unitary}
    U_{aa'}(t)=\sum_{\mu\neq aa'}\ketbra{\mu}{\mu}+u_{aa'}(t),
\end{equation}
generated by some 2-body control interaction $H_{aa'}(t)$, that has, in principle, the power to generate bipartite entanglement. The state $\rho(t)$ of the system at time $t$ obtained via such unitary is
\begin{equation}
    \label{eq:state_step}
    \begin{split}
    \rho(t)=\;\;&U_{aa'}(t)(\otimes^n\rho) U_{aa'}^\dagger \\
    =\;\;&(p_a +p_{a'})\rho_1(t)\otimes\ketbra{i_2^a\cdots i_N^a}{i_2^a\cdots i_N^a}\\
    &+\sum_{\mu\neq aa'}p_\mu\ketbra{\mu}{\mu},
    \end{split}
\end{equation}
with $\rho_1$ being itself a state. The overall state $\rho(t)$ is thus separable at every step of the procedure, and after $2N-1$ total steps the target final state $\bm{\mathsf{p}}(\otimes^N\rho)$ is reached.
\end{techbox}

Of all the possible unitary cycles that connect the state $\otimes^N\rho$ to its passive state $\bm{\mathsf{p}}(\otimes^N\rho)$, those that preserve the system in a separable state are inevitably slower than those that generate entanglement. Accordingly, the authors of Ref. \cite{Hovhannisyan2013} indicate a relation between the rate of entanglement generation and the power of work extraction -- defined as the ratio between extracted work and time required for the extraction -- leaving the open problem of quantifying such a relation to successive work. This question paves the way for the study of charging and extracting power, as described in the next section. 

\section{Powerful charging}
\label{s:powerful_charging}
\noindent
We now consider the task of \emph{charging} quantum batteries via unitary operations. The deposited energy is the opposite of the work extracted, and as long as we consider closed systems, the two tasks are essentially equivalent. In this section we discuss the relation between charging (and extraction) power and entangling operations.

\subsection{Average and instantaneous power}
\label{ss:average_instantaneous_power}
\noindent
For our discussion, we can evaluate the average power of a unitary cycle that charges $\rho\to\rho(\tau)=U(\tau)\rho U^\dagger(\tau)$ as the ratio between the energy $W(\tau)$ deposited on the battery during the procedure, with respect to its internal Hamiltonian $H_0$, and the time required to perform the unitary operation, 
\begin{equation}
    \label{eq:average_power}
    \langle P\rangle=\frac{W(\tau)}{\tau},
\end{equation}
remembering that from now on $W(\tau)$ has opposite sign with respect to that of Eq. \eqref{eq:work}.
Similarly, the instantaneous power $P(t)$ at some time $t$ is given by the time derivative of the energy deposited at time $t$ along the unitary charging,
\begin{equation}
    \label{eq:istantaneous_power}
    P(t) =  \frac{d}{dt}W(t) = \frac{d}{dt} \bigg\{ \tr[\rho(t)H_0]-\tr[\rho H_0] \bigg\},
\end{equation}
which becomes $P(t)=-i\:\tr\{[H(t),\rho(t)]H_0\}$, using the von Neumann equation. 
In the next section we show that entangling operations are more powerful than local ones, when fairly compared, and  that they can yield a power advantage that grows with the number $N$ of units that compose the battery \cite{Binder2015,Campaioli2017}.

\subsection{Charging with global operations}
\label{ss:charging_global_operations}
Let us consider a battery given by $N$ copies of a $d$-dimensional unit cell, defined by a local internal Hamiltonian $H_0$, given by Eq. \eqref{eq:interal_array_hamiltonian}. Assuming that the energetic structure of the individual Hamiltonians $H_0^{(l)}$ is the same for each copy,
\begin{equation}
    \label{eq:internal_hamiltonian_copy}
    H_0^{(l)}=\sum_{j=1}^d\omega_j\ketbra{j}{j}
\end{equation}
the highest and lowest energy states are $\ket{G}:=\otimes^N\ket{1}$ and $\ket{E}:=\otimes^N\ket{d}$, respectively. The energy deposited onto the battery after an evolution from $\ket{G}$ to $\ket{E}$ is equal to $W=N(\omega_d-\omega_1)$.
We are going to compare the charging power of an optimal local Hamiltonian with that of an optimal global one, to illustrate the power of entangling operations.
From Ch.~\ref{ch:qsl_mixed}, we know that the Hamiltonian that generates the shortest unitary orbit between two orthogonal states $\ket{\psi}$, $\ket{\phi}$ is given by $H\propto\ketbra{\psi}{\phi}+h.c.$, as discussed in Fig~\ref{fig:saturate_qsl_pure}. 
Thus, when we restrict ourselves to local Hamiltonians, we drive each subsystem with a local Hamiltonian $H^{(l)}\propto\ketbra{1}{d}+h.c.$, in order to obtain the total parallel charging Hamiltonian
\begin{equation}
    \label{eq:parallel}
    H_\|=\alpha_\| \sum_{l=1}^N (\ketbra{1}{d}+h.c.)\otimes_{l'\neq l}\mathbbm{1}^{(l')}.
\end{equation}
Instead, when we allow ourselves to use global operations, we can drive the whole $N$-body system be means of a collective charging Hamiltonian,
\begin{align}
    \label{eq:collective}
    H_\sharp=\alpha_\sharp (\ketbra{E}{G}+h.c.).
\end{align}
Indeed, these two Hamiltonians generate the shortest path between the considered states, with respect to the Fubini-Study metric, for the given constraints on the space of generators, given that parallel driving can only contain local terms, while global driving is not constrained. 

\begin{figure}[htbp]
    \centering
    \includegraphics[width = 0.8\textwidth]{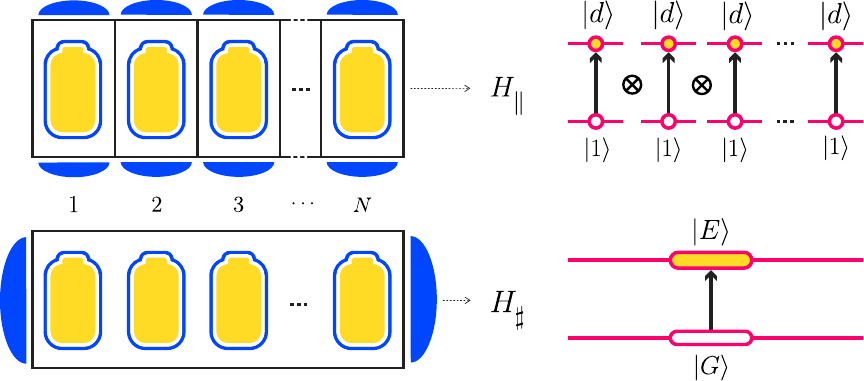}
    \caption{Parallel (\textit{top}) and collective (\textit{bottom}) charging procedures are here schematically represented. The optimal local driving $H_\|$ of Eq.~\eqref{eq:parallel} couples each individual ground state $\ket{1}$ to its respective excited state $\ket{d}$, while the optimal global driving $H_\sharp$ of Eq.~\eqref{eq:collective} couples the collective ground state $\ket{G}=\otimes^N\ket{1}$ to the collective excited state $\ket{E}=\otimes^N\ket{d}$. When fairly compered, collective charging can be up to $N$ times more powerful than parallel driving.}
    \label{fig:parallel_collective}
\end{figure}

To fairly compare the power of the two Hamiltonians we require them to satisfy the energy constraint
\begin{equation}
\label{eq:operator_norm_uniform_constraint}
    \lVert H \rVert_{\textrm{op}} = E_{\max},
\end{equation}
for some energy scale $E_{\max}>0$~\cite{Binder2015}. Accordingly, we obtain
\begin{equation}
    \label{eq:constraint_applied}
    \begin{split}
    &\alpha_\|=\frac{E_{\max}}{N}, \\
    &\alpha_\sharp=E_{\max}.
    \end{split}
\end{equation}
The time required to drive the initial to the final state by the two different Hamiltonians can be calculated analytically,
\begin{equation}
\label{eq:extensive_advantage}
     \tau_\| = N\frac{\pi}{2}\frac{1}{E_{\max}}, \;\; \tau_\sharp = \frac{\pi}{2}\frac{1}{E_{\max}}.
\end{equation}
We now evaluate the average power using Eq. \eqref{eq:average_power} to obtain $P_\sharp = N P_\|$. The power of the entangling operation is $N$ times larger than that of local ones. These two charging procedures are schematically represented in Fig.~\ref{fig:parallel_collective}.

\section[Enhancing the charging power of quantum batteries]{Enhancing the charging power of quantum batteries}
\label{s:enhancing}
\noindent
The advantage of using global operations has a profound geometric interpretation: While the collective Hamiltonian of Eq. \eqref{eq:collective} drives the initial state along the shortest path, through the space of entangled states, the local Hamiltonian generates a longer orbit that, in return, keeps the state separable for all times \cite{Binder2015}. In this section we discuss the relation between entangling operations and charging power, showing how it is possible to obtain an extensive advantage even without generating entanglement during the evolution, and addressing the order of the interaction as fundamental resource to obtain a speed-up with respect to local charging procedures.

\subsection{Quantum advantage}
\label{ss:quantum_advantage}
\noindent
We are now in a position to define the \textit{quantum advantage} for collective charging
\begin{gather}\label{qadvantage}
\Gamma:=\frac{P_\sharp}{P_{\|}} = \frac{\tau_{\|}}{\tau_\sharp}, 
\end{gather}
where $P_\|$  is the power of some optimal local driving, while $P_\sharp$  is the power of some driving with at least two-body interactions. Accordingly, $\tau_\|$ and $\tau_\sharp$ are the charging times required for the driving methods, respectively.
The right-hand side of the equality in Eq.~\eqref{qadvantage} is a consequence of our requirement that the work done evolving between states $\otimes^N\rho$ and $\otimes^N\sigma$ is independent of the charging method. Here, \emph{quantum} refers specifically to an enhancement over charging with the best local (i.e., non-entangling) operations. That is, to compute the quantum advantage we must take the optimal values for $P_{\|}$ for given $\rho$ and $\sigma$.
For reference, the form of the local Hamiltonian is
\begin{align}
     \label{eq:charging_parallel_hamitlonian}
    H_\| = \sum_{l=1}^N h_l(t)\otimes_{l'\neq l }\mathbb{1}^{(l')},
\end{align}
where $h_l(t)$ acts on the $l$-th subsystem.
This Hamiltonian generates the unitary evolution $U_\|(t) = \otimes_{l=1}^N U_l(t)$, where $U_l(t)$ is the unitary generated by $h_l(t)$.  The collective unitary generated by the Hamiltonian $H_\sharp(t)$ will be referred to as $U_\sharp(t)$, while its action on the initial state $\otimes^N\rho$ will be denoted by $\rho_{\sharp}(t)=U_\sharp(t)\otimes^N\!\!\rho\; U_\sharp^\dagger(t)$.

In the case of local charging, every battery evolves independently, and the time taken to charge $N$ batteries is equal to the single-battery charging time, $\tau_{\|} = \tau$. The deposited work scales extensively, $W_{\|}=N W$, where $W$ is the work per battery, leading to a charging power $P_{\|} = N W/\tau$ that grows linearly with the number of batteries. As we have seen in the previous section, the power of using global interactions can be $N$ times larger than $P_\|$, yielding an advantage $\Gamma = N$, as long as both charging Hamiltonians have the same operator norm, however different uniform energetic constraints can alter the achievable advantage. We now show how to obtain an upper bound to the quantum advantage $\Gamma$ depending on the chosen energetic constraint on the Hamiltonians.

In choosing an energetic constraint, the aim is to isolate the advantage due to collective quantum effects, without worrying about other consequences of introducing interactions between batteries. To do so, we require any charging Hamiltonian $H_\sharp(t)$ to have a similar \emph{energy scale} to some optimal local charging Hamiltonian $H_\|$
for the considered problem $\otimes^N \rho \to \otimes^N\sigma$. Without constraints, we could freely increase the total energy of the collective charging Hamiltonian to achieve faster driving, making the advantage arbitrarily large. 
Noting that the variance and mean energy are extensive quantities for non-interacting systems, we consider the following constraints on $H_\sharp$:

\vspace{5pt}
\noindent\textbf{C1} --- The time-averaged standard deviation in energy during the collective evolution for time $\tau_\sharp$ should not exceed $\sqrt{N}$ times the time-averaged standard deviation in energy of a single battery,
\begin{equation}
\label{con:var}
    \overline{\Delta E}_\sharp \le \sqrt{N} \; \overline{\Delta E},
\end{equation}
with $\overline{\Delta E}_\sharp$ and $\overline{\Delta E}$ being the time-averaged standard deviation of the charging Hamiltonians $H_\sharp(t)$ and $h_l(t)$, respectively (see Eq.~\eqref{eq:time_averaged_standard_deviation}). Note that $H_\sharp$ drives the initial state of $N$ copies, $\otimes^N\rho$, while $h_l$ drives the initial state of a single battery $\rho$.

\noindent\textbf{C2} --- The time-averaged energy during the collective evolution for time $\tau_\sharp$ should not exceed $N$ times the time-averaged energy of a single battery,
\begin{equation}
    \label{con:en}
    \overline{E}_\sharp \le N \;\overline{E},
\end{equation}
with $\overline{E}_\sharp$ and $\overline{E}$ being the time-averaged energy of the charging Hamiltonians $H_\sharp(t)$ and $h_l(t)$, respectively (see Eq.~\eqref{eq:time_averaged_energy}). Again, note that $H_\sharp$ drives the initial state of $N$ copies, $\otimes^N\rho$, while $h_l$ drives the initial state of a single battery $\rho$.

For example, applying constraint \textbf{C1} to the time-independent Hamiltonians of Eqs.~\eqref{eq:parallel} and~\eqref{eq:collective} yields $\Delta E_\sharp = \Delta E_\|$, and $\Gamma = \sqrt{N}$. However, constraints \textbf{C1} and \textbf{C2} are weaker than imposing a uniform energetic constraint, such as $\Delta H = \omega$ for every charging Hamiltonian $H$.
These choices of constraints are additionally motivated by the form of QSL bounds, as discussed in Ch.~\ref{ch:qsl_mixed}. While, as will show, \textbf{C1} leads to a stricter upper bound on the quantum advantage, there is no reason \emph{a priori} to choose one over the other. 

\subsection{Bound on the quantum advantage}
\label{ss:bound_quantum_advantage}
Since the quantum advantage defined in Eq.~\eqref{qadvantage} amounts to a ratio of transition times, we can use the quantum speed limit to obtain an upper bound for it, given a constraint. The traditional QSL states that the time required to transform $\otimes^N \rho$ into $\otimes^N \sigma$ is lower bounded as
\begin{gather}\label{eq:QSL}
\tau_\sharp\geq T^{(N)}_{\rm QSL} = \frac{\mathcal{L}(\otimes^N \rho,\otimes^N \sigma)}{ \min \big\{ \overline{E}_\sharp, \overline{\Delta E}_\sharp\big\}},
\end{gather}
where $\mathcal{L}(\rho,\sigma)$ is the Bures angle, given in Eq.~\eqref{eq:bures_angle}. The two constraints, \textbf{C1} and \textbf{C2}, are clearly related to the QSL, as $\overline{\Delta E}_\sharp$ and $\overline{E}_\sharp$ can be computed using Eqs.~\eqref{con:var} and \eqref{con:en} respectively.
If we could change the Hamiltonian at will to include arbitrary interaction terms, and only concern ourselves with the state transformation, we could choose $H_\sharp$ to be the optimal time-independent Hamiltonian connecting $\otimes^N \rho$ and $\otimes^N \sigma$ in the $N$-partite state space~\cite{PhysRevLett.110.157207, PhysRevLett.111.260501, PhysRevA.90.032110}. In this case, the QSL reduces to the usual inequalities due to Mandelstam-Tamm~\cite{Mandelstam1945} and Margolus-Levitin~\cite{Margolus1998}, where $\overline{E}_\sharp$ is replaced by the average initial energy $E_\sharp$ and $\overline{\Delta E}_\sharp$ is replaced by the average initial standard deviation $\Delta E_\sharp$.

To derive the upper bound, we first confine ourselves to constraint \textbf{C1}. We proceed by using the QSL to bound the minimal time required by the collectively charging $N$ quantum batteries,
\begin{equation}
    \label{eq:qsl_collective}
    \tau_\sharp\geq \frac{\mathcal{L}(\otimes^N \rho,\otimes^N \sigma)}{\Delta E_\sharp}.
\end{equation}
In the case of parallel charging each battery evolves independently under the effect of a local term with standard deviation equal to $\Delta E$, therefore the optimal time $\tau_{\|}$ required for the parallel charging is given by
\begin{equation}
    \label{eq:qsl_parallel}
    \tau_\| = \bm{\mathsf{s}}\; T_{QSL}^{(1)} = \bm{\mathsf{s}}\; \frac{\mathcal{L}(\rho,\sigma)}{\Delta E},
\end{equation}
where $\bm{\mathsf{s}} = \tau_\|T_{QSL}^{(1)}$ quantifies the potential inability to saturate the QSL in the parallel case. 
We can now combine Eqs.~\eqref{qadvantage},~\eqref{eq:qsl_collective},~\eqref{eq:qsl_parallel} and use the constraint $\Delta E_\sharp\leq\sqrt{N}\Delta E$ to obtain
\begin{gather}\label{conservativebounds}
\Gamma_{\textrm{\textbf{C1}}} \le \bm{\mathsf{s}} \sqrt{N} \frac{\mathcal{L}(\rho,\sigma)}{\mathcal{L}(\otimes^N \rho,\otimes^N \sigma)} 
\quad \mbox{and} \quad
\Gamma_{\textrm{\textbf{C2}}} \le \bm{\mathsf{s}} N \frac{\mathcal{L}(\rho,\sigma)}{\mathcal{L}(\otimes^N \rho,\otimes^N \sigma)},
\end{gather}
for constraints \textbf{C1} and \textbf{C2} respectively, since
 a similar argument can be made with constraint \textbf{C2}.

Two remarks are in order: Firstly, for orthogonal pure initial and final states, the QSL can be saturated (see Sec.~\ref{s:traditional_unified_qsl}) and $\bm{\mathsf{s}}=1$.  Though the quantum advantage for power could be larger in other cases, including where the battery states are mixed~\cite{Pires2016}, the improvement cannot grow with the number of batteries; i.e., $\bm{\mathsf{s}}$ is a constant function of $N$. Secondly, we have excluded cases where $\rho$ and $\sigma$ do not lie on the same unitary orbit, as there is no way of transforming the former into the latter using the scheme outlined above; the two states will therefore necessarily have the same spectrum~\cite{Binder15a}.
The two bounds in Eq.~\eqref{conservativebounds} are independent from each other, and constraint \textbf{C1} is stronger than \textbf{C2}, as it leads to a stricter bound on the quantum advantage. Many other bounds can be derived by considering other extensive constraints. The quantum advantage is tight for orthogonal initial and final states, due to the example given in Ref.~\cite{Binder15b}, though the Hamiltonian used to saturate the bound involves $N$-body interactions.
\begin{figure}[htbp]
    \centering
    \includegraphics[width = 0.9\textwidth]{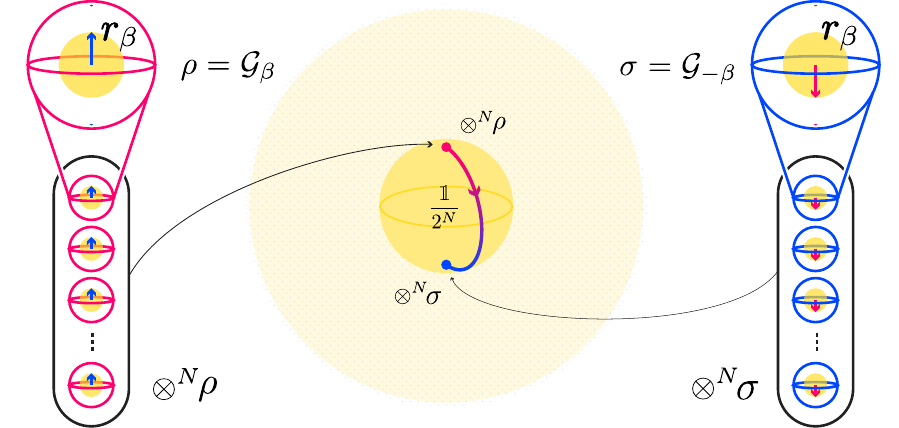}
    \caption{The product state $\otimes^N\rho$ of $N$ copies of a two-level system in the thermal state $\rho = \mathcal{G}_\beta$ with inverse temperature $\beta$, here represented by its GBV $\bm{r}_\beta$, is unitarily evolved into $\otimes^N\sigma$, where $\sigma = \mathcal{G}_{-\beta}$. For large $N$, if the inverse temperature $\beta$ is chosen to be small enough, the state of the copies lies the \textit{separable ball}, a region centered on the maximally mixed state $\mathbb{1}/2^N$ that contains only separable states. Unitary evolution preserves the distance from the maximally mixed state, keeping the system in a separable state at all times. The global Hamiltonian $H_\sharp = \alpha_\sharp (\ketbra{\omega_0}{\omega_1}+h.c.)^{\otimes N}$ has quantum advantage $\Gamma_{\textrm{\textbf{C1}}} =\sqrt{N}$, and $\Gamma_{\textrm{\textbf{C2}}} = N$~\cite{Campaioli2017}.}
    \label{fig:sep_ball}
\end{figure}

The significance of entanglement for quantum enhancement has previously been studied in the context of quantum speed limits for pure states: it was shown that, for non-interacting systems, initial entanglement is required for an enhancement in the speed of evolution \cite{Giovannetti2003,Zander2007}, while for
interacting systems a speedup may be achieved for initially
separable states, since intermediate entangled states are
accessible \cite{Giovannetti2003b,Xu2016}. In the more general case of mixed states,
the necessity of entanglement for an enhancement may not be
directly inferred, though it has been argued that, in general,
larger quantum Fisher information of the state with respect to
the generator of evolution leads to enhanced speed \cite{Frowis2012,Toth2014}.
In fact, as we now show, entanglement does not appear to
be necessary for a nontrivial quantum advantage.
\begin{proposition}
\label{thm:sepball}
An extensive quantum advantage can be attained even for highly mixed states, including those confined to the separable ball throughout the charging procedure.
\end{proposition}

\begin{proofbox}{Proof of Proposition~\ref*{thm:sepball}}{proof_example_separableball}
Consider $N$ two-level batteries with the same internal Hamiltonian characterised by non-degenerate energy levels $\omega_0=0$ and $\omega_1=1$. Let the initial state be the thermal state with inverse temperature $\beta$ with respect to the internal Hamiltonian, $\rho = \mathcal{G}_\beta$ (see Eq.~\eqref{eq:gibbs_state}), and the final state be $\sigma = \mathcal{G}_{-\beta}$. The optimal local charging scheme is achieved in time $\tau_\|=\pi/2$ by applying the Hamiltonian $\ketbra{\omega_0}{\omega_1}+h.c.$ to each battery independently. In contrast, the joint charging of $N$ batteries is achieved in $\tau_\sharp = \tau_\|/\alpha_\sharp$ using the global Hamiltonian $H_\sharp = \alpha_\sharp (\ketbra{\omega_0}{\omega_1}+h.c.)^{\otimes N}$, where the positive constant $\alpha_\sharp$ is introduced to satisfy the chosen constraint.
In both cases (local and global) the deposited work is identical, thus, the quantum advantage is simply the ratio of $\tau_\|$ to $\tau_\sharp$: $\Gamma = \alpha_\sharp$, which can be evaluated for the choice of constraint. We find $\Gamma_{\textrm{\textbf{C1}}} =\sqrt{N}$, while $\Gamma_{\textrm{\textbf{C2}}} = N$.

For $N$ quantum systems of dimension $d$, there exists a ball centered on the maximally mixed state and containing only separable states, known as \emph{separable ball}. Though an exact form for the radius of the separable ball is not known, it has been bounded from below and above \cite{Gurvits2002,GurvitsBarnum2005,AubrunSzarek2006}. Since the distance from the maximally mixed state cannot change under unitary evolution, for a small enough choice of $\beta$, the joint state of $N$ batteries will lie within this ball throughout the evolution. Yet, the quantum advantage remains extensive (See Fig.~\ref{fig:sep_ball}). \hfill \qedsymbol

Remarkably, neither $\tau_\|$ nor $\tau_\sharp$ depend on $\beta$, while the total work done does. In other words, no matter how mixed the battery is, a quantum advantage that scales with the number of batteries involved is always achievable, granted that one can implement the considered global Hamiltonian $H_\sharp$. The trade-off of using highly mixed states is that the charging power suffers, in absolute terms, as $\beta$ becomes smaller and smaller. Proposition~\ref{thm:sepball} implies that, while a quantum advantage requires entangling operations, the joint state of $N$ batteries does not have to be entangled during the charging process. However, other types of quantum correlations can survive under the depolarisation described in the example above, such as the quantum \textit{discord}~\cite{Giorgi2015}.
\end{proofbox}

The Hamiltonian used in the example above, and in
Ref. \cite{Binder2015}, to saturate the bound for quantum advantage involves $N$-body interactions. Such interactions are notoriously difficult to engineer. In the next section, we consider physically realizable interactions, and study the dependence of the enhancement on the order of interaction, i.e., the number of batteries that take part in a single interaction term.

\subsection{K-Local charging}
\label{ss:local_charging}
We now discuss the achievability of a significant quantum advantage in a regime where arbitrary multipartite entanglement generation is possible during the charging process. In particular, we demonstrate that, although a nontrivial quantum advantage is achievable in physical systems characterised by at most $k$-body interactions, this advantage -- upper bounded by a quantity that depends at most quadratically on $k$ -- cannot scale with the number $N$ of batteries that compose the system.

\begin{figure}
\centering
\includegraphics[width=0.9\textwidth]{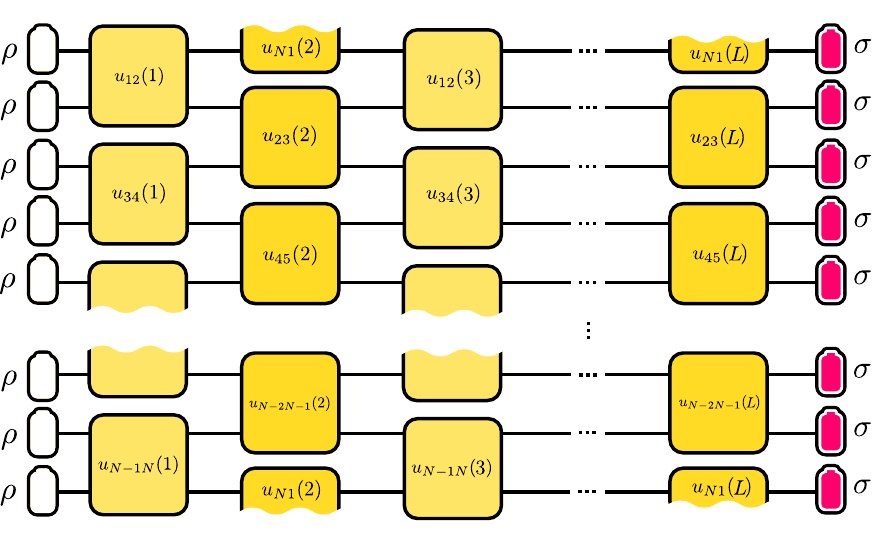}
\caption{A system with an even number $N$ of parties is charged with a unitary circuit, obtained from a piecewise time-independent Hamiltonian with at most $2$-body interactions.}
\label{fig:pairwise_circuit}
\end{figure}

First, we consider the situation where work is deposited onto the battery by means of a unitary circuit generated by a piecewise time-independent Hamiltonian. An example of such a circuit for $k=2$ is depicted in Fig.~\ref{fig:pairwise_circuit}. 
In the figure, a system with an even number $N$ of parties is charged with a piecewise time-independent Hamiltonian with at most $2$-body interactions. At each time step $t$, the driving Hamiltonian consists of a set of $s=N/2$ terms, each of which involving a different pair of batteries $i,j$. The result is $s$ independent unitary operations $u_{ii+1}(t)$ acting on pairs $i,i+1$. Note that at each step, there is no overlap between different pairs of batteries, while at every successive step the pairs are changed in order to allow the formation of highly entangled states. 

A circuit of this type can be used to approximate any time-dependent unitary evolution $U(t)$~\cite{Poulin2011}, with precision that increases with the number of steps $L$. The implementation of $U$ requires an extra amount of time that depends on the number of non-commuting terms in the Hamiltonian.
This model is reminiscent of the circuit model of universal quantum computation, which is known to outperform its classical counterpart. In this case, the collective state of $N$-batteries will, in general, be highly entangled. This scheme allows us to study how the quantum advantage is related to the number of batteries that are simultaneously interacting.

We consider an $N$-body system composed of $d$-level subsystems, with local internal Hamiltonian $H_0$, as in Eqs.~\eqref{eq:interal_array_hamiltonian} and~\eqref{eq:internal_hamiltonian_copy}. More explicitly, and without loss of generality, we assume that
\begin{equation}
    \label{eq:energy_gap}
    \omega_d-\omega_1 = 2\omega_d >0,
\end{equation}
with all the eigenvalues arranged in increasing order. The time interval $[0,\tau_\sharp]$ is divided up into $L$ steps. At each step the charging Hamiltonian $H_\sharp(t)$ is the sum of $s=\lceil N/k \rceil$ terms, each acting on a different set of $k$ batteries. In order to allow the formation of highly entangled states, these $s$ partitions could be different at each step, as illustrated in Fig.~\ref{fig:pairwise_circuit}. More specifically, at any time $t$, the $k$-local Hamiltonian can be written as
\begin{equation}
    \label{eq:k_local_hamiltonian}
    H_\sharp(t)=\sum_{\mu=1}^s h_{\mu}(t)\otimes\mathbb{1}_{\bar{\mu}},
\end{equation}
where each term $h_{\mu}(t)$ acts on a different $k$-partition of the Hilbert space, identified by the set $\mu=(\mu_1,\dots,\mu_k)$ of $k$ indices\footnote{If $N/k$ is not an integer these partitions can have \emph{at most} $k$ indices.}, while $\mathbb{1}_{\bar{\mu}}$ indicates the identity over all the other indices non included in $\mu$. At different time steps, different partitions are allowed.
In order to make a meaningful statement in this scenario, we need to introduce a constraint on the operator norm of the driving Hamitlonians, similar to that of Eq.~\eqref{eq:operator_norm_uniform_constraint}, but less restrictive, in the spirit of constraints \textbf{C1} and \textbf{C2}:
\vspace{5pt}

\noindent\textbf{\textbf{C0}} -- The time-averaged operator norm of the driving Hamiltonian $H_\sharp(t)$ during the collective evolution for time $\tau_\sharp$ should not exceed $N$ times that of the of a single battery driving Hamiltonian,
\begin{equation}
    \label{con:opnorm}
    \overline{\mathcal{E}}_\sharp \leq N \overline{\mathcal{E}},
\end{equation}
where $\overline{\mathcal{E}}_\sharp$ and $\overline{\mathcal{E}}$ represent the time-averaged operator norm of the driving Hamiltonians $H_\sharp(t)$ and $h_l(t)$, respectively, 
\begin{gather}
 \overline{\mathcal{E}}_\sharp  := \overline{\lVert H_\sharp(t) \rVert}_{op}\;\, \mbox{and} \;\,
 \overline{\mathcal{E}} := \overline{\lVert h_l(t) \rVert}_{op}.
\end{gather}
Note that $h_l(t)$ is still the optimal local Hamiltonian acting only on susbystem $l$.
Constraint \textbf{C0} guarantees that both the time-averaged standard deviation and the time-averaged energy are bounded from above, as shown in Sec.~\ref{s:bounds} of the Appendix. There, we show that $\overline{\mathcal{E}}_\sharp$ upper bounds both $E_\sharp/2$ and $\Delta E_\sharp$. In this sense, it is a stricter constraint than \textbf{C1} or \textbf{C2}.
We now show that, with this constraint, the upper bound on the quantum advantage depends on the interaction order $k$:

\begin{theorem}\label{thm:circuit}
For a circuit based charging procedure with interaction order of at most $k$, under constraint \emph{\textbf{C0}}, the achievable quantum advantage is upper bounded as $\Gamma_{\emph{\textrm{\textbf{C0}}}} < \gamma  k$, where $\gamma$ does not scale with the number $N$ of batteries.
\end{theorem}

The proof of Theorem~\ref{thm:circuit} can be found in Sec.~\ref{s:th2} of the Appendix. In the important case where $\rho$ and $\sigma$ are the ground and maximally excited states respectively, $\gamma=\pi/2$. By construction, this bound on the quantum advantage is not tight. For comparison, $\Gamma_{\textrm{\textbf{C1}}} = \sqrt{k}$ and $\Gamma_{\textrm{\textbf{C2}}} = k$ are  achievable if the total number of batteries $N$ can be divided by $k$, i.e., if $N/k=s\in \mathbb{N}$, as discussed in Refs.~\cite{Giovannetti2004,Binder15b}. In this particular case, such a speed-up can be obtained for pure states using the time-independent Hamiltonian
\begin{align}
    &H = \sqrt{s} \sum_{\mu=1}^s h_\mu, \\
    &h_\mu = |1\rangle^{\otimes k} \langle d|^{\otimes k} + h.c.,
\end{align} 
assuming that each $h_\mu$ acts on a completely different set of $k$ batteries, i.e., $[h_\mu,h_{\mu'}]=0$ for all $\mu, \mu'$. In the same situation, using constraint \textbf{C0} we obtain $\Gamma_{\textrm{\textbf{C0}}}= k$, suggesting that the strict inequality in Theorem~\ref{thm:circuit} is only different by a constant factor from an achievable bound.

The result of Theorem~\ref{thm:circuit} can be extended to a more general class of Hamiltonians, where $k$-body time-dependent interactions can occur between overlapping sets of batteries, with the restriction that each battery is simultaneously interacting with at most $m$ others. This restriction is motivated by the idea that the \emph{reach} of the interaction should be limited.

\begin{theorem} \label{thm:reach}
For a generic time-dependent charging procedure the achievable quantum advantage under constraint \emph{\textbf{C0}} is upper bounded as
\begin{gather}
\label{eq:theorem_3}
\Gamma_{\textrm{\textbf{C0}}} < \gamma \big(k^2(m-1) + k\big),
\end{gather}
where $k$ is the interaction order and $m$ is the maximum participation number, i.e., any one battery interacts with at most $m$ other batteries at any given time.
\end{theorem}

The proof of Theorem~\ref{thm:reach} is given in Sec.~\ref{s:co1} of the Appendix. For many physical systems, both $k$ and $m$ are limited: 2 or 3-body interactions are the norm for fundamental processes, and higher interaction orders are generally hard to engineer here~\cite{zoller, Boxio:07, napolitano2011interaction}.
The effective participation number, or reach, $m$ tends to be constrained by the spatial arrangement of systems and the fact that interaction strength often drops off with distance. Exceptions to this rule include the Dicke model~\cite{PhysRev.93.99} where collective coherence leads to superradiance, the Lipkin-Meshkov-Glick model~\cite{LIPKIN1965188}, where all particles interact with each other, and  the M{\o}lmer-S{\o}rensen interaction~\cite{MolmerSorensen1998}, in which an ensemble of ions are effectively coupled by a spatially uniform electromagnetic field. 

Note that these bounds are not tight; while a scaling of the power $P_\sharp$ with the number of batteries $N$ is surely not feasible in the context of $k$-body interactions, it is more likely that the quantum advantage is tightly limited by $k$. In fact, we conjecture that, for any choice of time-dependent $k$-body interaction Hamiltonian $H$, a conservative bound for the quantum advantage is given by $\Gamma_{\textrm{\textbf{C0}}} < \gamma k$:

\begin{conjecture}\label{conj:conj}
Theorem~\ref{thm:circuit} holds for any time-dependent $k$-body interaction Hamiltonian subject to constraint \emph{\textbf{C0}}.
\end{conjecture}
\noindent
We examine this particular statement in Sec.~\ref{s:conjecture} of the Appendix, anticipating that the result holds if a particular mathematical conjecture does too. 
We collected extensive numerical evidence to support our conjecture, calculating $\Gamma_{\textbf{C0}}$ for a large set Hamiltonians, obtained from unitaries sampled uniformly according to the Haar measure, for $(N,k)=(3,2)$, $(4,2)$, $(4,3)$ and $(6,2)$. Not a single instance of $\Gamma_{\textbf{C0}}\geq \gamma k$ has been recorded in this way, and we believe that similar conjectures should also hold for constraints \textbf{C1} and \textbf{C2}. However, this numerical evidence does not represent a proof of our conjecture since our sampling procedure is not uniform with respect to the constraint the we imposed, and due to the fact that there could be a measure-zero set of Hamiltonians that disproves our conjecture. 

\section{Chapter summary}
\label{s:chapter_summary_batteries}
In this Chapter we have applied the operational interpretation of time-energy uncertainty relations to study the limits on achievable power for many-body quantum systems. The results discussed in Sec.~\ref{s:enhancing}, and published in Ref.~\cite{Campaioli2017}, complement the strain of research into quantum thermodynamics and quantum speed limits, by deriving a concrete upper bound on the ratio between the maximum power of interacting and non-interacting driving between product states. In particular, we discussed the notion of collective quantum advantage for the charging power of quantum batteries, for which we calculated a bound that depends on the order of the interaction $k$, on the participation number $m$, and on the energetic constraint used to compare different charging methods.  

First, we applied the QSL to prove two fundamental upper bounds for the quantum advantage, given in Eq.~\eqref{conservativebounds}, each corresponding to a different energetic constraint on the charging Hamiltonian. The advantage of using global operations over local ones is interpreted as the result of rapid evolution through the space of quantum states. While, in the case of pure states, entanglement is a necessary consequence of these global operations, a fully separable evolution is still accessible for those states that live in the separable ball. 

A striking consequence of our results, which holds in general for mixed states, is that an enhanced charging power is available even for arbitrarily mixed states, in remarkable analogy to the case of quantum metrology. There, an enhancement in sensing is still available for highly mixed states lying inside the separable ball~\cite{Modi2011}.

We also showed analytically that a quantum advantage that grows with the number of batteries is not achievable with any physically reasonable Hamiltonian. By restricting the order of the interaction, we severely restrict the space of available charging procedures, which effectively defines a non-trivial many-body quantum optimal control problem, the solution of which is generally not known. 

An important remark is that the bound that we derived in Eq.~\eqref{eq:theorem_3} is not tight. Obtaining a tight bound for such dynamics requires showing that, for any choice of initial and final states, there is always a driving Hamiltonian that saturates it. 
This is equivalent to solving a constrained \emph{quantum brachistochrone problem}~\cite{Mostafazadeh2007,Assis2007,Carlini2008,Wang2015} to find the fastest evolution that connects the given initial and final states.
When we consider pure state and we are allowed to perform any global operation ($k=N$), or only local ones ($k=1$), solutions to the quantum brachistochrone problem are generally easier to obtain. However, as soon as we consider mixed states (see Ch.~\ref{ch:fast_state_preparation}) and we impose strong constraints on the generators of the evolution this problem becomes hard to solve. In this sense, the bound of Eq.~\eqref{eq:theorem_3} provides a fundamental result on the minimal time required to evolve between two multipartite states when the dynamics is restricted by the order of the interaction, for which a tight quantum speed limit is not known yet.
\newpage\leavevmode
\newpage\leavevmode
\chapter*{Conclusions \& Outlooks}
\markboth{Conclusions}{Conclusions \& Outlooks}
\label{ch:conclusions} 
\addcontentsline{toc}{chapter}{Conclusions \& Outlooks}
In this thesis we studied the operational interpretation of time-energy uncertainty relations, which set a bound on the minimal time of quantum evolution.
In particular, we aimed to solve some outstanding issues with the traditional formulation of QSLs, with emphasis on their tightness, feasibility, and significance. 
We addressed these tasks with a geometric approach, formulating the problem of time-optimal evolution as that of finding the shortest orbit between initial and final states. This allowed us to improve the framework used for the derivation of such bounds, constructively search for time-optimal solutions, and effectively apply these results to thermodynamic tasks, such as work extraction and deposition with quantum systems.

In Ch.~\ref{ch:qsl_mixed} we obtained a QSL for unitary evolution, and demonstrated is performance. This result was achieved using the generalised Bloch representation, and equipping the state space with a distance that corresponds with the angle between the vectors representing initial and final states. To the best of our knowledge, this QSL corresponds to the tightest bound for the unitary evolution of quantum systems, and we therefore used it in Ch.~\ref{ch:fast_state_preparation} in order to assess the performance of optimised Hamiltonians obtained by means of our iterative method, featured in Ref.~\cite{Campaioli2019}. A promising direction to further improve the QSL for unitary evolution is to look for a suitable metric for the space of mixed states with a given degeneracy structure, generally given by a flag manifold~\cite{Bengtsson2008}, which might yield an attainable bound.

The advantages of using the geometric framework offered by the generalised Bloch representation have been further emphasised in Ch.~\ref{ch:qsl_open}, where we generalised our QSL to the case of open evolution; a result that has been recently recognised in Refs.~\cite{Bukov2019,Brody2019,Yang2019}. Both for the case of unitary and open evolution, the main outlook is to incrementally improve the tightness of these QSLs, and provide efficient methods for their estimation. 

Another interesting research avenue is to address the converse problem, i.e., that of solving the time-optimal evolution of quantum systems. This is a notoriously hard task, that has received particular attention in the field of quantum optimal control~\cite{Wang2015}. The iterative method proposed in Ch.~\ref{ch:fast_state_preparation} addresses this problem for the case of unconstrained unitary evolution between arbitrary states. The method is based on the iterative suppression of the parallel \emph{ineffective} components of the generator of the evolution, and is interpreted as an optimisation over the geometric phases gained by the initial state during the evolution. While the efficacy of this method is supported by extensive numerical evidence, an interesting task is to apply group theoretic method to analytically bound its convergence properties. This iterative method can in principle be generalised to the case of open evolution, however, it is not clear how to effectively achieve such generalisation, or if a similar approach could be at all extended to the case of controlled systems. 

In Ch.~\ref{ch:quantum_batteries} we showed how QSLs can be applied in the context of quantum thermodynamics. There, we demonstrated the power advantage of using global operations over local ones for charging many-body quantum batteries. We then showed that this advantage often relies on highly non-local, genuinely many-body interactions, which are often hard (or impossible) to engineer in practice. We thus derived a new bound on charging power for physical systems, which accounts for limited interaction order and range. In this context, the main outlook is to further improve the QSLs for the case of many-body systems, with particular emphasis on their optimal control.

Another research perspective consists in progressing the study of quantum batteries, which has considerably advanced since the publication of the seminal paper by Alicki and Fannes~\cite{Alicki2013}. Several authors have proposed and studied practical implementations of quantum batteries, consisting of one-dimensional spin chains~\cite{Le2017}, cavity-assisted arrays of two-level systems~\cite{Ferraro2018,Zhang2018a}, modelled by means of superconducting pairs, and harmonic oscillators~\cite{Andolina2018,Zhang2018}. New insights have been obtained on the extractable work, on the role of quantum correlations, as well as on the difference between classical and many-body batteries~\cite{Andolina2018a,Andolina2019}, while the authors of Ref.~\cite{Friis2017} have studied cycle precision when charging is obtained by means of the feasible class of Gaussian unitaries. 
The most recent efforts have been devoted to the study of dissipative systems~\cite{Barra2019,garcapintos2019}. In this context, the author of this thesis has participated in the study of an energy-stabilization protocol based on the use of sequential measurements~\cite{Gherardini2019}. Further investigations should focus on the interplay between quantum batteries and their environment, aiming to improve energy stabilisation and efficient charging. 

In conclusions, the research conducted for this thesis consolidates the importance of QSLs, providing new results and methods with significance for theoretical and applied quantum mechanics. The geometric approach to QSLs adopted allowed us to fine-tune the balance expressed by time-energy uncertainty relations. As a result, the rather abstract notion of \textit{state space} assumed a much more tangible role, providing the variable to be optimised in order to achieve time-optimal evolution, and reaffirming the geometry of quantum mechanics as a powerful tool for future studies and applications.

\appendix
\chapter{Appendix}
\section{Proof of Theorem~\ref{th:qsl_tilde}}
\label{s:proof_phi}
\noindent
First, we prove that $\Phi$ is a distance that reduces to the Fubini-Study distance for the case of pure states. The function $\Phi(\rho,\sigma)\geq 0$ since $\tr[\rho\sigma]=\sum_{a,b}\lambda_a\lambda_b|\braket{r_a|s_b}|^2$ is positive and always smaller than or equal to $\tr[\rho^2] = \tr[\sigma^2] $. This can be proved using the Hilbert-Schmitdt distance $\tr[(A-B)(A^\dagger-B^\dagger)]/2\geq0$ for $A=\rho$ and $B=\sigma$, obtaining $\tr[\rho^2]-\tr[\rho\sigma]\geq 0$. For $\sigma=\rho$, $\Phi(\rho,\rho) =0 $, and if $\Phi$ is zero $\tr[\rho\sigma] = \tr[\rho^2] = \tr[\sigma^2]$ therefore $\sigma = \rho$. Symmetry holds due to the cyclic property of the trace and because $\rho$ and $\sigma$ have the same purity $\tr[\rho^2]$. Lastly,  $\Phi(\rho,\sigma)=\arccos(\sqrt{(1-(d-1)\lVert \bm{r}\rVert_2^2\; \hat{\bm{r}}\cdot\hat{\bm{s}})/(1-(d-1)\lVert \bm{r}\rVert_2^2 )})$ is monotonic in $\hat{\bm{r}}\cdot\hat{\bm{s}}$, thus respects the triangle inequality.  Ergo, $\Phi$ is a distance on the space of those states that can be unitarily connected. For the case of pure states, $\Phi$ reduces to the Fubini-Study distance, since, for $\rho =\ketbra{\psi}{\psi}$ and $\sigma =\ketbra{\phi}{\phi}$, $\sqrt{\tr[\rho\sigma]} = |\braket{\psi|\phi}|$ and $\tr[\rho^2]=1$.

The proof of Eq.~\eqref{eq:qsl_tilde} is identical to that for Eq.~\eqref{eq:qsl_gba}, expect for the fact that
\begin{equation}
    \Phi(\rho_t,\rho_{t+dt}) = \arccos\bigg(\sqrt{\frac{1-dt^2(\tr[\rho_t^2 H_t^2]-\tr[(\rho_t H_t)^2])}{\tr[\rho_t^2]}}\bigg).
\end{equation}
We expand $\arccos(\sqrt{1-c}) \sim \arccos(1-c/2) \sqrt{c}+\mathcal{O}(c)$ for small $c>0$, obtaining
$\int_0^\tau\Phi(\rho_t,\rho_{t+dt}) =  \tau Q_\Phi $ which leads to $\Phi(\rho,\sigma) \leq Q_\Phi \ \tau$. \hfill \qedsymbol

\section{Example: QSL for three-level systems}
\label{s:qutrits}
\begin{figure}[htbp]
    \centering
    \includegraphics[width=0.8\textwidth]{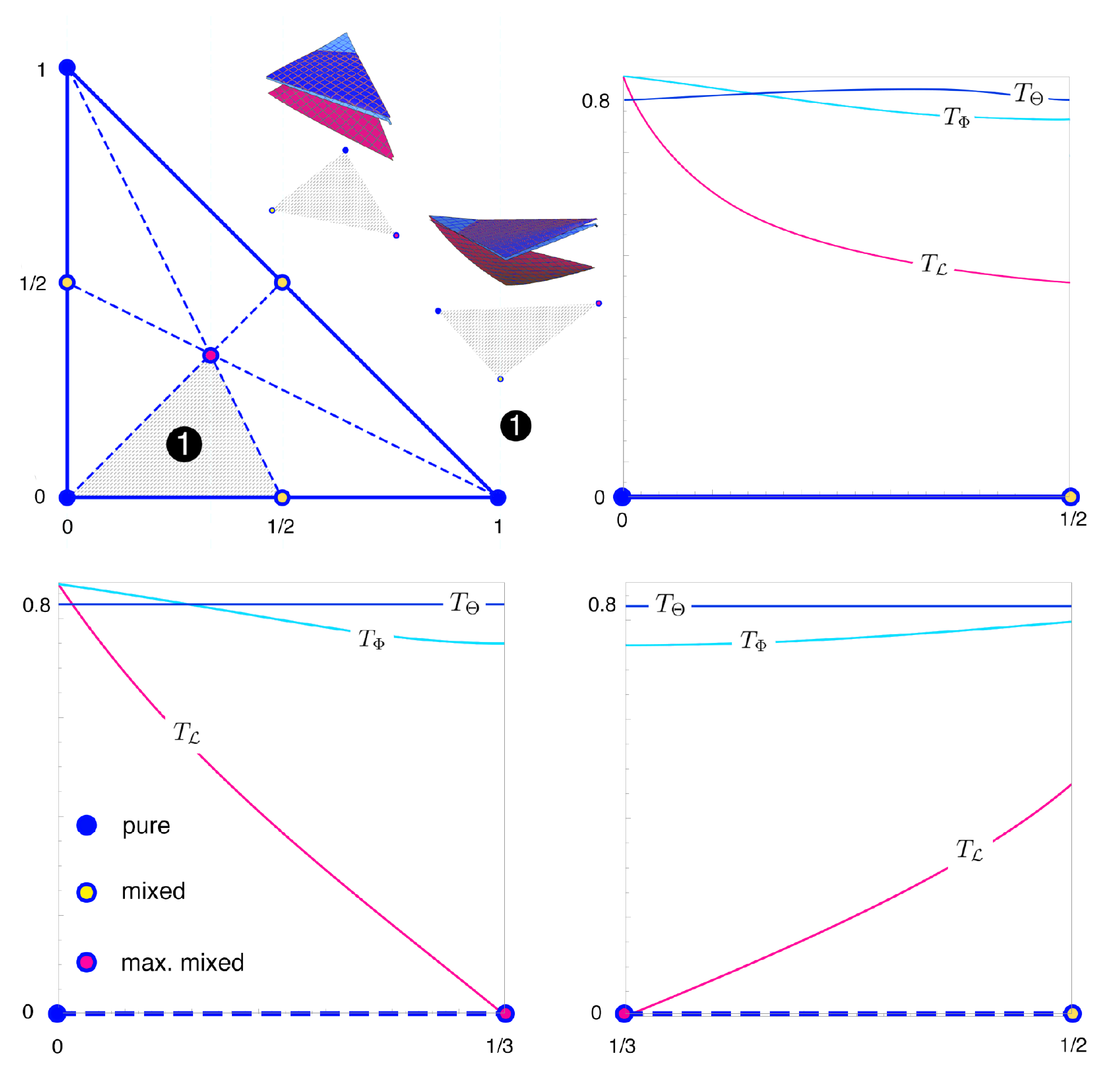}
    \caption{Bounds $T_{\mathcal{L}}$ (Eq.~\eqref{eq:QSL_mixed}, magenta line), $T_\Phi$ (Eq.~\eqref{eq:qsl_tilde}, light blue line),
    and $T_\Theta$ (Eq.~\eqref{eq:qsl_gba}, blue line), as a function of the eigenvalues $\lambda_1,\lambda_2,\lambda_3$, for a specific choice of mixed
    qutrit state $\rho$ and Hamiltonian $H$, for driving $\rho\to\sigma=O\rho O^\dagger$, with $O=\exp[-i H]$. Solid blue vertices correspond to pure states, yellow and magenta vertices to maximally mixed states of rank 2 and 1, respectively.}
    \label{fig:qutrits}
\end{figure}

\noindent
In this section we study the new bounds in Eqs.~\eqref{eq:qsl_gba} and~\eqref{eq:qsl_tilde} for the case of qutrits, and we compare them to the one in Eq.~\eqref{eq:QSL_mixed}. The spectrum $\{\lambda_1,\lambda_2,\lambda_3\}$ can be represented with the standard 2-simplex $\Delta_2$ (equilateral triangle), or by its projection onto the plane defined by $\lambda_1$ and $\lambda_2$, since the third component $\lambda_3$ has to be equal to $1-\lambda_1-\lambda_2$, and $\lambda_1+\lambda_2\leq1$. Within this space, we only need to consider the portion \circled{1} given by 
$0\leq \lambda_1\leq 1/2$, $\lambda_2\leq\lambda_1$ and $\lambda_2\leq 1-2\lambda_2$ (see Fig.~\ref{fig:qutrits})~\cite{Bengtsson2008}. This region is determined by the three vertices $(0,0,1)$ (solid blue dot), $(1/2,0,1/2)$ (yellow dot), and $(1/3,1/3,1/3)$ (magenta dot), that correspond to a pure state, a mixed state with two identical eigenvalues equal to $1/2$, and the maximally mixed state, respectively, and delimited by the segments that connect these vertices. These segments are characterized by two kinds of degeneracy: the one that connects the pure state to the mixed state associated with $(1/2,0,1/2)$ (given by $(\lambda_1,0,1-\lambda_1)$, solid blue line), is composed of fully non-degenerate mixed states, while the other two segments (dashed blue lines) contain mixed states with two identical eigenvalues (the same degeneracy structure as for pure states).

We generated random states and Hamiltonians, as details in {\fontfamily{phv}\selectfont\textbf{Box~\ref{tech:qsl_tightness}}}, and then studied the three different bounds in the region \circled{1} (see Fig.~\ref{fig:qutrits}). As described in the main text, $T_\Phi = T_{\mathcal{L}} > T_\Theta$ at the pure vertex. The bound $T_\Theta$ is constant along the dashed lines, while it can vary continuously along the solid solid blue line. For all of the generated Hamiltonians, $\max[T_\Theta,T_\Phi]-T_{\mathcal{L}}>0$. 
\newpage

\section{Comparison of QSL bounds}
\label{a:tightness}
\noindent
As mentioned in the letter, we have considered some significant bounds~\cite{Sun2015,DelCampo2013,Deffner2013b,Mondal2016} to test the performance of our bound $T_D$.
We analytically compare our bound to Sun's, Del Campo's, and Deffner's bounds. The last three bounds are given by
\begin{align}
    \label{eq:sun}
    &T_{\textrm{Sun}} = \frac{\bigg|1-\frac{\tr[\rho\sigma]}{\sqrt{\tr[\rho^2]\tr[\sigma^2]}}\bigg|}{2 \overline{\left({\lVert \dot{\rho}_t\rVert }/{\lVert \rho_t \rVert }\right)}},\\
    \label{eq:delcampo} 
    &T_{\textrm{Del Campo}} = \frac{\bigg|1-\frac{\tr[\rho\sigma]}{\tr[\rho^2]}\bigg|\lVert \rho \rVert ^2}{\overline{\lVert \dot{\rho}_t\rVert} },\\
    \label{eq:deffner}
    &T_{\textrm{Deffner}} = \frac{\sin^2\bigg[\arccos \bigg( \mathcal{F}(\rho,\sigma)\bigg) \bigg]}{\overline{\lVert \dot{\rho}_t\rVert} },
\end{align}
where $\mathcal{F}(\rho,\sigma) = \tr[\sqrt{\sqrt{\rho}\sigma\sqrt{\rho}}]$ is the quantum fidelity between $\rho$ and $\sigma$.
The orbit-dependent term of all of these bounds only depends on the strength of the generator $\lVert \dot{\rho}_t \rVert $, or can be bounded by some quantity that only depends on this term. This observation allows us to evaluate the relative tightness of these bounds and of $T_D$ just by comparing their orbit-independent terms. Let us assume that $\tr[\rho^2]\geq\tr[\sigma^2]$, without loss of generality, and introduce the enhanced bounds
\begin{align}
    \label{eq:over_sun}
    &T_{\textrm{Sun}}^{\;\star} = \frac{\bigg|1-\frac{\tr[\rho\sigma]}{\sqrt{\tr[\rho^2]\tr[\sigma^2]}}\bigg|\lVert \rho \rVert }{2 \overline{\lVert \dot{\rho}_t\rVert} }, \\
    \label{eq:over_deffner}
    &T_{\textrm{Deffner}}^{\;\star} = \frac{\sin^2\bigg[\arccos \bigg(\mathcal{E}(\rho,\sigma)\bigg) \bigg]}{\overline{\lVert \dot{\rho}_t\rVert} },
\end{align}
where $\mathcal{E}$ is the sub-fidelity
\begin{gather}
    \label{eq:sub_fidelity}
    \mathcal{E}(\rho,\sigma)=\sqrt{\tr[\rho\sigma]+\sqrt{2(\tr[\rho\sigma]^2 -\tr[\rho\sigma\rho\sigma]) }},
\end{gather}
which is a lower bound to $\mathcal{F}$~\cite{Miszczak2008}. Both enhanced bounds $T^{\;\star}_{\textrm{Sun}}$ and $T^{\;\star}_{\textrm{Deffner}}$ are larger than the respective bounds of Eqs.~\eqref{eq:sun}~and~\eqref{eq:deffner}. Therefore, whenever  $T_D$ is larger than the enhanced bounds it is also surely larger than the actual ones. Moreover, the enhanced bounds have orbit-independent terms that only depend on the following four parameters
\begin{align}
    \label{eq:x}
    &x: =\tr[\rho^2], \\
    \label{eq:y}
    &y:=\tr[\sigma^2], \\
    \label{eq:z}
    &z:=\tr[\rho\sigma], \\
    \label{eq:beta}
    &\beta:=\tr[\rho\sigma\rho\sigma],
\end{align}
where $x,y$ are bounded by $1/d$ from below and by $1$ from above, $z$ is bounded by $\sqrt{x y}$ from above, and $\beta$ is bounded by $z^2$ from above.
We proceed with the evaluation of the relative tightness of these bounds and of $T_D$ just by comparing their orbit-independent terms, obtaining
\begin{align}
    \label{eq:analytical_sun}
    &\bigg | 1 - \frac{z}{\sqrt{xy}} \bigg|\frac{\sqrt{x}}{2} \leq \sqrt{x+y-2z} \Rightarrow T^{\;\star}_{\textrm{Sun}} \leq T_D,
 \end{align}
and
\begin{gather}
    \bigg | 1 - \frac{z}{x} \bigg|x \leq \sqrt{x+y-2z}\Rightarrow T_{\textrm{Del Campo}} \leq T_D, 
 \end{gather}
for all $\rho,\sigma \in \mathcal{S}(\mathcal{H}_S)$ and all processes.

As mentioned in the main text, Deffner's bound is proven to be valid only when one of the two states is pure, i.e., for $\rho=\rho^2$ (or $\sigma=\sigma^2$)~\cite{Sun2015}, i.e., when $x=1$. Under this condition sub-fidelity, fidelity and super-fidelity all coincide~\cite{Miszczak2008} to be equal to $\sqrt{\tr[\rho\sigma]}$, and we obtain
\begin{gather}
\begin{split}
   &\sin^2\big[\arccos\big(\sqrt{z}\big)\big]\leq\sqrt{1+y-2z} \\
   &\Rightarrow T^\star_{\textrm{Deffner}}\leq T_D,
\end{split}
 \end{gather}
which proves our statement.
 
\section{Extending the validity of bound by Deffner et al.}
\label{a:deffner}
We will now show that our bound can be used to extend the validity of Deffner's bound~\cite{Deffner2013b} to the case of mixed initial states $\rho$, with $\tr[\rho^2]<1$.
As mentioned earlier, we can directly compare our bound $T_D$ to the enhanced bound $T^\star_{\textrm{Deffner}}$, by replacing the fidelity with the sub-fidelity $\mathcal{E}(\rho,\sigma)$,
given in Eq.~\eqref{eq:sub_fidelity}. The bound $T^\star_{\textrm{Deffner}}$ is always larger then the actual bound $T_{\textrm{Deffner}}$. Thus, anytime $T_D$ is larger then $T^\star_{\textrm{Deffner}}$,  then $T_D$ is also larger than $T_{\textrm{Deffner}}$, which is then guaranteed to be valid for such choice of initial and final states $\rho$ and $\sigma$.

Even though there is not a universal hierarchy between these two bounds, we can express a ranking between $T_D$ and $T^{\;\star}_{\textrm{Deffner}}$ using the following strategy:
We calculate the probability $P(T_D \geq T^*_{\textrm{Deffner}})$ of $T_D$ being larger than the upper bound on $T_{\textrm{Deffner}}$ in the space spanned by $z\in[0,\sqrt{x y}]$ and $\beta \in [0,z^2]$, as the ratio between the area where $T_D\geq T^*_{\textrm{Deffner}}$ and the area of the full space spanned by $z$ and $\beta$,
\begin{gather}
    \label{eq:probability_test}
    P(T_D\geq T^\star_{\textrm{Deffner}}) = \frac{\int_0^{\sqrt{xy}} \int_0^{z^2} \frac{\sgn(\Gamma (x,y,z,\theta))+1}{2} dz \;d\beta}{\int_0^{\sqrt{xy}} \int_0^{z^2} dz\; d\beta},
\end{gather}where $\sgn$ is the sign function, and
\begin{gather}
    \label{eq:performance_function}
    \begin{split}
    &\Gamma (x,y,z,\theta) = \sqrt{x+y-2z}\; + \\
    & \;\;- \sin^2\bigg[\arccos\bigg(\sqrt{z+\sqrt{2(z^2-\beta)}}\bigg)\bigg].
    \end{split}
\end{gather}

The probability $P(T_D\geq T^*_{\textrm{Deffner}})$ is a function of $x$ and $y$ measures how often $T_D$ is larger then Deffner in the space spanned by $z\in[0,\sqrt{xy}]$ and $\beta \in [0,z^2]$, given $x$ and $y$. As a result, we obtain a general \emph{rule of thumb} to decide which bound to use given the purity of initial and final states: For $y\geq 1- x$ bound $T_D$ is outperforms Deffner's (and vice versa for $y\leq 1-x$), as shown in the left panel of Fig.~\ref{fig:hierarchy_deffner}.
Additionally, we have directly compared our bound $T_D$ to $T_{\textrm{Deffner}}$ numerically, sampling $3\cdot10^6$ initial and final states from the Bures and the Ginebre ensembles. Our bound outperforms Deffner's for the vast majority of the cases, as shown in the right panel of Fig.~\ref{fig:hierarchy_deffner}.

\section{Relation between constraints}
\label{s:bounds}

\noindent
Here we show that $\overline{\Delta E}_\sharp \le \overline{\mathcal{E}}_\sharp$ and $\overline{E}_\sharp \le 2\overline{\mathcal{E}}_\sharp $ by direct calculation. In the first case we have
\begin{align}
    \overline{\Delta E}_\sharp & = \frac{1}{\tau_\sharp}\int_0^{\tau_\sharp} dt \sqrt{\tr[ H_\sharp^2(t)\rho_\sharp(t)] - \tr[ H_\sharp^2(t)rho_\sharp(t)]^2 }, \notag\\
    & \leq \frac{1}{\tau_\sharp}\int_0^{\tau_\sharp} dt \sqrt{\tr[ H_\sharp^2(t)\rho_\sharp(t)]} \notag\\
    & \leq \frac{1}{\tau_\sharp}\int_0^{\tau_\sharp} dt \, \lVert H_\sharp(t) \rVert_{op} = \overline{\mathcal{E}}_\sharp,
\end{align}
where we have used $\tr[ H_\sharp^2(t)\rho_\sharp(t)] \le \lVert H_\sharp^2(t)\rVert_{op} = \lVert H_\sharp(t) \rVert_{op}^2$ to get to the final line.

Similarly, the time-averaged energy
\begin{align}
    \overline{E}_\sharp & = \frac{1}{\tau_\sharp}\int_0^{\tau_\sharp} dt \,  \{\tr[ H_\sharp(t)\rho_\sharp(t)] - \omega_G(t)\} , \notag\\
    & \leq \frac{1}{\tau_\sharp}\int_0^{\tau_\sharp} dt \,  \{\lVert  H_\sharp(t) \rVert_{op} + |\omega_G(t)| \} , \notag\\
    & \leq \frac{2}{\tau_\sharp}\int_0^{\tau_\sharp} dt \, \lVert  H_\sharp(t) \rVert_{op}
     = 2\overline{\mathcal{E}}_\sharp. 
\end{align}
Thus, if the time-averaged operator norm of the Hamiltonian is bounded,  $\overline{\Delta E}_\sharp$ and $\overline{E}_\sharp$ are also bounded.

\section{Proof of Theorem~\ref{thm:circuit}}
\label{s:th2}
Consider $W(t)=\tr[H_0\rho^{(N)}(t)]-\tr[H_0\rho^{\otimes N}]$, the average work done on the system up to time $t$ during the charging process, where $\rho^{(N)}$ represents the collective state of the battery at time $t$ under the effect of the unitary generated by $H_\sharp(t)$. The instantaneous power is given by $P(t)=d_t W(t) = i \tr\{[H_\sharp(t) ,H_0] \rho^{(N)}(t)\}$. The strict inequality
\begin{gather}
    \label{eq:inequality_power}
    P_\sharp = \frac{1}{\tau_{\sharp}}\int_0^{\tau_{\sharp}} dt P(t) < \max_{H_\sharp(t)} \left\{ \lVert [H_\sharp(t),H_0] \rVert_{op} \right\} =: P^{\uparrow}
\end{gather}
follows from the fact that any unitary charging has to have vanishing instantaneous power for times $t=0$ and $t={\tau_\sharp}$. We now evaluate the commutator $[H_\sharp(t),H_0]$ in order to find an upper bound $P^{\uparrow}$ for the average power $P_\sharp$, remembering that we have $H_0=\sum_l H_0^{(l)}$ and $H_\sharp(t) = \sum_{\mu=1}^s h_{\mu}$. We will use the subscript $\bar{\mu}$ to indicate the set of battery indices that are not included in partition defined by $\mu$. Using the commutation relation between $h_{\mu}$ and $H_0^{(l)}$ we obtain
\begin{align}
   [H_\sharp(t),H_0] &= \sum_{\mu=1}^s  \left[ h_{\mu} \otimes \mathbb{1}_{\bar{\mu}} \,, \textstyle{\sum_{j=1}^k} H_0^{(\mu_j)} \otimes \mathbb{1}_{\bar{\mu}_j} \right] \nonumber \\
   &=\sum_\mu  \left[ h_\mu \,, \textstyle{\sum_{j=1}^k} H_0^{(\mu_j)} \otimes \mathbb{1}_{i\neq \mu_j \in \mu} \right] \otimes \mathbb{1}_{\bar{\mu}}.
\end{align}

Using the definition for $H_0$, it
follows from direct calculation that $\lVert \textstyle{\sum_{i=1}^k} H_0^{(\mu_i)} \rVert_{op} = \lambda_d k$. Let us define $\alpha_\mu := \lVert h_\mu  \rVert_{op}$ and introduce two normalized operators $X_\mu$ and $\iota_\mu$, as follows:
\begin{gather}
    X_\mu = \frac{h_\mu}{\alpha_\mu},
\quad    
    \iota_{\mu} = \frac{1}{\lambda_d k}{\sum_{i=1}^k H_0^{(\mu_i)}}.
\end{gather}
Using these, we can rewrite the commutator as
\begin{align}
   [H_\sharp(t),H_0] &= 2 \cdot \lambda_d k \sum_\mu \alpha_\mu\frac{1}{2} [X_\mu,\iota_{\mu}]\otimes\mathbb{1}_{\bar{\mu}} \nonumber \\
   \label{eq:power_y}
    & = 2\lambda_d k \sum_\mu \alpha_\mu Y_\mu  \otimes \mathbb{1}_{\bar{\mu}},
\end{align}
where $Y_\mu = \frac{1}{2} [X_\mu,\iota_{\mu}]$, such that $\lVert Y_\mu \rVert_{op} \leq 1$.

At any time $t$, the operator norm of the Hamiltonian $H_\sharp(t)$ is given by
\begin{gather}
    \label{eq:piecewise_norm}
    \lVert H_\sharp(t) \rVert_{op}  =   \bigg\lVert \sum_{\mu=1}^s h_\mu \bigg\rVert_{op} 
     = \sum_{\mu=1}^s \lVert h_\mu \rVert_{op}
     = \sum_{\mu=1}^s |\alpha_\mu |,
\end{gather}
where the equality holds due to the fact that, at each step in time, every term $h_\mu$ acts on a different $k$-partition of the Hilbert space. 
Accordingly, we obtain that
\begin{gather}
\label{eq:constraint_again}
    \frac{1}{\tau_{\sharp}} \int_0^{\tau_{\sharp}} dt \lVert H_\sharp(t) \rVert_{op}  =  \frac{1}{\tau_{\sharp}} \int_0^{\tau_{\sharp}} dt \sum_{\mu=1}^s |\alpha_\mu | \leq N\overline{\mathcal{E}}.
\end{gather}

Now we consider the expression given in Eq.~\eqref{eq:power_y}, to calculate the upper bound $P^{\uparrow}$. Once again, we use the fact that at each step in time there are $s$ terms acting on different $k$-partitions of the Hilbert space, such that the operator norm of $[H_\sharp(t),H_0]$ can be calculated exactly,
\begin{align}
    \lVert [H_\sharp(t),H_0] \rVert_{op} & = 2\lambda_d k  \bigg\lVert \sum_{\mu=1}^s \alpha_\mu Y_\mu \otimes \mathbb{1}_{\bar{\mu}} \bigg\rVert_{op} \nonumber\\
    & = 2\lambda_d k  \sum_{\mu=1}^s \lVert \alpha_\mu Y_\mu \otimes \mathbb{1}_{\bar{\mu}} \rVert_{op} \nonumber\\
    \label{eq:piecewise_norm_y}
    & \leq 2\lambda_d k  \sum_{\mu=1}^s |\alpha_\mu |,
\end{align}
where the inequality in line Eq.~\eqref{eq:piecewise_norm_y} holds due to the fact that $\lVert Y_\mu \rVert_{op} \leq 1$ by definition, and where the sum can be carried out of the operator norm thanks to the fact that, at any given time, the $s$ subgroups of $k$ batteries are not overlapping.
Using Eq.~\eqref{eq:constraint_again}, we obtain
\begin{align}
    \frac{1}{\tau_{\sharp}} \int_0^{\tau_{\sharp}} dt \lVert  [H_\sharp(t),H_0]  \rVert_{op}  &= 2\lambda_d k  \frac{1}{\tau_{\sharp}} \int_0^{\tau_{\sharp}} dt \sum_{\mu=1}^s |\alpha_\mu | \nonumber \\
    &\leq 2\lambda_d k N\overline{\mathcal{E}}.
\end{align}
Plugging this result back into Eq.~\eqref{eq:inequality_power}, we get 
\begin{align}
& \frac{1}{\tau_{\sharp}}\int_0^{\tau_{\sharp}} dt P(t)  < P^{\uparrow} 
= 2\lambda_d k N \overline{\mathcal{E}}.
\end{align}
We calculate the quantum advantage as in Eq.~\eqref{qadvantage}, where  $P_\|$ is given by the ratio between $W_\|$ and $\tau_\|=\bm{\mathsf{s}} \mathcal{L}_1/\min\{E,\Delta E\}$. 
Work $W_\|=N W$ is extensive, and in general $W=q 2\lambda_d$, where $0< q \leq 1$.
Thus, we obtain
\begin{align}
   \Gamma_{\rm C0}<\frac{2\lambda_d k N \overline{\mathcal{E}}}{2\lambda_d N q \frac{\min\{E,\Delta E\}}{\bm{\mathsf{s}}\mathcal{L}_1}} =  \frac{\bm{\mathsf{s}} \mathcal{L}_1 \overline{\mathcal{E}}}{q \: \min\{E,\Delta E\}} k,
\end{align}
as we intended to prove. \hfill $\blacksquare$

\section{Proof of Theorem~\ref{thm:reach}}
\label{s:co1}
Here our goal is to relate a generic unitary evolution to a circuit based charging procedure. We first note that the charging Hamiltonian can always be decomposed into a number $M$ of non-commuting terms:
\begin{gather}
\label{eq:neighbour}
   H_\sharp(t) =\sum_{j=1}^M H^{(j)}(t)
   \quad \mbox{with} \quad
   H^{(j)}(t)=\sum_{\mu=1}^s h^{(j)}_{\mu}(t),
\end{gather}
where $[h^{(j)}_{\mu}(t),h^{(j)}_{\mu'}(t)]=0$; this decomposition is in general different for different values of $t$.
The unitary evolution $U$ generated by this time-dependent Hamiltonian can always be approximated, using the Trotter-Suzuki decomposition~\cite{Poulin2011}, by the following product of unitary tranformations:
\begin{gather}
\label{eq:trotter}
    U_{\rm Trot}= \prod_{l=1}^L \prod_{j=1}^M\exp\left[ - i H^{(j)}(t)\left(\frac{l\tau_\sharp}{L}\right) \frac{\tau_\sharp}{L} \right].
\end{gather}
In the limit $L\rightarrow\infty$, $U_{\rm Trot} = U$; however, they do not correspond to the same implementation: The Hamiltonian generating $U_{\rm Trot}$ is piecewise time-independent and in the circuit form discussed in Sec.~\ref{s:th2}. Since each of the $M$ terms at each time step must be implemented sequentially, $U_{\rm Trot}$ takes $M$ times longer to run than $U$, with a corresponding drop in power  $P_\sharp= M P_{\texttt{Trot}}$.

Since we have an upper bound, from Theorem~\ref{thm:circuit}, on the quantum advantage for circuit model Hamiltonians, and the power for a more general Hamiltonian is at most $M$ times greater, it must be that $\Gamma_{\rm C0} < M\gamma k$ in this case. In order to complete the proof, we now need to consider how the minimum necessary value of $M$ scales with $k$ and $m$.

The quantity $m$ denotes the maximum number of other batteries any one can interact with. In order for the number of terms $M$ in Eq.~\eqref{eq:neighbour} to be sufficient for the required decomposition, it must at least equal the largest possible number of different $k$-partitions $\mu$ that have a non-trivial amount of indices in common, while containing the same index $\mu_i$ at most $m$ times. Let us provide a few simple examples to clarify the meaning of $M$, where we will assume that $N$ can be arbitrarily large.

\paragraph*{$(k=2, m=1)$} In this case $M$ is trivially equal to 1. A possible choice is given by the first 2-partition $(1,2)$, after which any other partition $(i,j)$ can contain neither 1 nor 2. This has to be true for any choice of other partitions, therefore $M=1$. In other words, in this case, the trotterization is not necessary and the unitary can be perfectly simulated with a piecewise unitary circuit. 

\paragraph*{$(k=2, m=2)$} Let us start with the first 2-partition $(1,2)$, followed by $(2,3)$ and $(1,3)$. Any other choice of two indices would form a partition that does not contain any element of at least one of the previous three, thus $M=3$. In this case the simulating circuit is at most 3 times slower than the actual unitary.

\paragraph*{$(k=3, m=2)$} Now the first 3-partition $(1,2,3)$ is followed by $(1,4,5)$, $(2,4,6)$ and $(3,5,6)$. Any other choice of three indices would form a partition that does not contain any element of at least one of the previous four, thus $M=4$.

In general, for a given $k$ and a given $m$, we could start -- without loss of generality -- from the first \emph{ordered} partition $(1,\dots,k)$. Remembering that each of those indices can appear at most $m$ times, we can construct $m$ sets containing $1$, followed by $m-1$ sets containing $2$, $3$, $4$ and so on until $m-1$ sets containing $k$, for a total of $k(m-1)+1$ terms. In the worst case scenario, all of these partitions have at least one element in common. However, any subsequent partition cannot contain any of the indices included in the first ordered partition $(1,\dots,k)$, thus $M\leq k(m-1)+1$.

Taking this most general, worst case scenario, we have a bound on the quantum advantage given by
\begin{gather}
\label{eq:conservative_bound_final}
    \Gamma_{\rm C0} <  (k(m-1)+1)\gamma k
    = \gamma \left( k^2 (m-1) + k\right),
\end{gather}
where $\gamma := \frac{\bm{\mathsf{s}} \mathcal{L}_1 \overline{\mathcal{E}}}{q \min\{E,\Delta E\}}$ is defined as in Theorem~\ref{thm:circuit}. \hfill $\blacksquare$
\vspace{5pt}

\section[A conjecture on achievable power limits]{A conjecture on achievable power limits}
\label{s:conjecture}
Let us consider a general time-dependent Hamiltonian that contains all the possible $k$-body interaction terms between the $N$ batteries that constitute the system, \textit{i.e.}, $H = \sum_\mu h_\mu$ contains $N!/k!(N-k)!$ terms in the sum. With the aim of obtaining an upper bound for the quantum advantage under the constraint \textbf{C0}, we follow the proof provided for Theorem 2, until Eq.~\eqref{eq:power_y}. We then find an explicit relation between the elements of $X_\mu$ and those of $Y_\mu$. Let us consider the product basis $\mathcal{B}_\mu:=\{|a\rangle_\mu\}$ for the subset of batteries defined by $\mu$. Each element $|a\rangle_\mu = \otimes_{i=1}^k |a_i\rangle_{\mu_i}$ is a product state of the partition of the Hilbert space associated with $\mu$. In this basis we can write
\begin{align}
    &\iota_{\mu} = \sum_a \eta_a \ketbra{a}{a}_\mu, \\
    \label{eq:x_mu}
    & X_\mu = \sum_{a<b} \Big(x_{ab}^\mu \ketbra{a}{b}_\mu + h.c.\Big) + \sum_{a}x^{\mu}_{aa} \ketbra{a}{a}_\mu.
\end{align}
By explicit calculation using this basis we obtain
\begin{gather}
\label{eq:y_mu}
    Y_\mu = \sum_{a<b} \bigg(\frac{\eta_b-\eta_a}{2}\bigg) \Big( x_{a,b}^\mu \ketbra{a}{b}_\mu - h.c. \Big),
\end{gather}
where $|\eta_a|<1$ and $0<(\eta_b - \eta_a)/2\leq 1$ for $a<b$, due to the structure of $\iota_{\mu}$. Our conjecture reduces to the following: 
\begin{gather}
\label{eq:conjecture}
  \lVert\textstyle{\sum_\mu}  \alpha_\mu Y_\mu  \otimes \mathbb{1}_{\bar{\mu}} \rVert_{op} \leq 
    \lVert \textstyle{\sum_\mu} \alpha_\mu X_\mu  \otimes \mathbb{1}_{\bar{\mu}} \rVert_{op},
\end{gather}
which is itself upper bounded by $N\overline{\mathcal{E}}$. If Eq.~\eqref{eq:conjecture} holds, then for any choice of time-dependent $k$-body interaction Hamiltonian $H$, subject to constraint \textbf{C0}, the average power $P_\sharp$ is upper bounded by $2\lambda_d k N \overline{\mathcal{E}}$, thus, $\Gamma_{\textrm{\textbf{C0}}} < \bm{\mathsf{s}} \mathcal{L}(\rho,\sigma) k$. An extensive numerical search failed to find any counterexamples to our conjecture.

The data collected for the quantity 
\begin{equation}
    P=\frac{\lVert \textstyle\sum_\mu  \alpha_\mu Y_\mu  \otimes \mathbb{1}_{\bar{\mu}} \rVert_{op}}{\lVert \textstyle\sum_\mu  \alpha_\mu X_\mu  \otimes \mathbb{1}_{\bar{\mu}} \rVert_{op}}
\end{equation}
is always smaller than the unit, as conjectured. Data has been collected sampling unitaries $u_\mu$ according to the Haar measure, and obtaining $h_\mu = i \log[u_\mu]$, where $\log[A]$ is the natural matrix logarithm of $A$; $P$ is evaluated explicitly for every sampled Hamiltonian. With a sample size $\nu = 10^5$ we ran the simulation for $(N,k)$ equal to $(3,2)$, $(4,2)$, $(4,3)$ and $(6,2)$. Not a single instance of $P>1$ has been recorded.
This numerical evidence does not represent a proof of our conjecture since there could be a measure zero set of Hamiltonians for which $P>1$.

\addcontentsline{toc}{chapter}{Bibliography}
\bibliographystyle{apsrev4-1}
%


\end{document}